\documentclass[twoside,12pt]{article}
\usepackage{epsfig}
\usepackage{psfig}

\def\Journal#1#2#3#4{{#1} {#2} (#4) #3 }

\def\NPA{{\em Nucl. Phys.} A}

\def\PRO{{\em Prog. Theor. Phys.}}
\def\NPB{{\em Nucl. Phys.} B}

\def\PLB{{\em Phys. Lett.} B}

\def\PRL{\em Phys. Rev. Lett.}
\def\PREV{\em Phys. Rev.}
\def\PREP{\em Phys. Rep.}

\def\PRD{{\em Phys. Rev.} D}
\def\PR{{\em Phys. Rev.}}
\def\PRC{{\em Phys. Rev.} C}

\def\ZPC{{\em Z. Phys.} C}

\def\ANNP{\em Ann. Phys. (N.Y.)}

\def\FDP{{\em Fortschr. Phys.}}
\def\INTA{{\em Int. J. Mod. Phys.} A}
\def\JHEP{{\em JHEP}}
\def\EPJC{{\em Eur. Phys. J.} C}
\def\EPJA{{\em Eur. Phys. J.} A}
\def\PRP{{\em Phys. Rep.} }
\def\PZETF{{\em Pisma Zh. Eksp. Teor. Fiz.}}
\def\PPNP{{\em Prog. Part. Nucl. Phys.} }
\def\SJNP{{\em Sov. J. Nucl. Phys.} }
\def\NPBPr{{\em Nucl. Phys. Proc. Suppl.} }
\def\JETPL{{\em JETP Lett.}}
\def\SJETP{{\em Sov. Phys. JETP}}
\def\JPG{{\em J. Phys. G}} 
\def\ARNPS{{\em Ann. Rev. Nucl. Part. Sci.}}
\def\PTP{{Prog. Theor. Phys.}}
\def\PREPC{{\em Phys. Rep. C}} 
\def\CJP{{\em Czech. J. Phys.}}
\def\PAN{{\em Phys. Atom. Nucl.}}
\def\MPLA{{\em Mod. Phys. Lett. A}}

\def\vp{\varphi}

\newcommand{\be}{\begin{equation}}
\newcommand{\ee}{\end{equation}}
\newcommand{\bea}{\begin{eqnarray}}
\newcommand{\eea}{\end{eqnarray}}

\newcommand{\eqb}{\begin{equation}}
\newcommand{\eqe}{\end{equation}}
\newcommand{\dmb}{\begin{displaymath}}
\newcommand{\dme}{\end{displaymath}}
\newcommand{\pd}{\partial}
\newcommand{\ep}{\varepsilon}
\newcommand{\eab}{\begin{eqnarray}}
\newcommand{\eae}{\end{eqnarray}}
\newcommand{\ra}{\right\rangle}
\newcommand{\la}{\left\langle}
\newcommand{\e}{\mbox{e}}

\newcommand{\bmE}{\mathbf E}

\newcommand{\bmx}{\mathbf x}
\newcommand{\bmy}{\mathbf y}

\begin{document}

\title{ \vspace{1cm} Operator Product Expansion and Quark-Hadron Duality:\\ Facts and Riddles}
\author{Ralf Hofmann\\
\\
Institut f\"ur Theoretische Physik, Universit\"at Heidelberg, Germany}
\maketitle
\begin{abstract} 

We review the status of 
the practical operator product expansion (OPE), when applied to two-point 
correlators of QCD currents which interpolate to 
mesonic resonances, 
in view of the violations of local quark-hadron duality. Covered 
topics are: a mini-review of mesonic QCD sum rules 
in vacuum, at finite temperature, or at 
finite baryon density, a comparison of model calculations of current-current 
correlation functions in 2D and 4D with the OPE expressions, 
a discussion of meson distribution amplitudes in the light of 
nonperturbatively nonlocal modifications of the OPE, and 
a reorganization of the OPE which (partially) resums powers 
of covariant derivatives. 
  
\end{abstract}

\newpage

\tableofcontents
\newpage

\section{Preface}

We start by giving a short outline of the present review article: 

In the next five sections we lay the foundations for reviewing 
(some of) the research on local quark-hadron duality violations being conducted 
within the last twenty years or so. We introduce the operator product expansion (OPE) as 
an expansion of a current-current correlator in powers of inverse Euclidean, 
external momenta and the strong coupling constant $\alpha_s$. 
We distinguish between the full OPE, which most probably is an asymptotic series, 
and truncations thereof used in practice. 
A definition of local quark-hadron duality, which conceptually rests on 
the OPE, is given. The concept of the OPE is 
applied to hadronic physics in the framework of the Shifman-Vainshtein-Zakharov (SVZ) sum rules 
associated with light- and heavy-quark mesons in vacuum and in a 
hot or baryon rich medium. We do not discuss 
the sum-rule program for baryonic resonances. We also review briefly the relation 
between power corrections and renormalons. Secs. 2 to 5 naturally have 
large overlaps with existing review articles and books on QCD sum rules 
and on theoretical as well as on practical aspects 
of the OPE \cite{QCDSRrev,Beneke1999}.

In Sec. 6 we take a closer look at 
violations of local duality. We first compare the predictions 
of analytically continued practical OPEs with the 
experimental data. Second, we review an instanton model calculation 
of the light-quark current-current correlator in Euclidean spacetime and compare the result with 
the practical OPE of this correlator. Third, we review an analysis of the 
current-current correlator in the 't Hooft model. The focus is on oscillatory 
components in the decay width of heavy mesons (as a function of the meson mass) 
when taking $O(1/N_c)$ corrections into account. 
The adopted point of view in the latter 
two investigations is that the model calculations are in some sense 
realistic, that is, they largely resemble the experimental 
situation. Disagreement between the prediction 
of hadronic spectra resting on the model calculation on the 
one hand and the OPE on the other hand thus are interpreted as violations 
of local quark-hadron duality. In Sec.\,\ref{sec:OPEA} we give an 
overview on experimental hadron spectra induced by the electromagnetic current and 
by $\tau$ decays. We also address $B_s$-$\bar{B}_s$ mixing: a problem which 
can be tackled by {\sl appealing} to local quark hadron duality. 
A rather large overlap of Sec.\,\ref{sec:QPIB}
and Sec.\,\ref{sec:2DM} with the reviews \cite{QHDrev} exists. 
In the present review the analysis in the 't Hooft model 
is presented in a more self-contained way than in \cite{QHDrev}.  

A discussion of the 
effects of nonperturbative nonlocality in modified practical OPEs is carried out in 
section 7 of the review. In a first step the relation between 
power corrections in an OPE and mesonic distribution amplitudes is discussed. Subsequently, 
the theoretical and phenomenological need for the inclusion 
of nonlocal condensates is pointed out. We also review the systematic 
inclusion of nonperturbatively nonlocal quantities in the OPE-based description 
of the hadron-to-hadron, hadron-to-vacuum, and vacuum-to-vacuum 
matrix elements of time-ordered current-current products. 
Phenomenologically consequences 
are explored in the latter case. To the best of the author's 
knowledge no review article exists which 
would have a substantial overlap with section 7.

The presentation is not very technical. Rather, 
we have tried to capture the {\sl basic} theoretical concepts and their applications. 
Most of the time our discussion aims at a direct comparison with the experimental situation. 
This review is by no means complete -- our apologies to all those authors whose contributions are 
not explicitly referred to.


\section{Introduction\label{sec:PF}}

Quantum Chromodynamics (QCD), the non-Abelian gauge theory of interacting quarks and 
gluons, is by now the widely accepted 
microscopic theory of strong interactions. 
Embedded into the Standard Model of particle physics it has 
passed various experimental tests. 
The expansion about the situation of asympotically free 
fundamental degrees of freedom at large momenta or particle masses is a conceptually and practically 
appealing feature \cite{GrossWilczek1973,Politzer1973,FritzschGell-MannLeutwyler1973}.
It has a wealth of applications: the perturbative calculability of the 
evolution of hadron structure functions \cite{GribovLipatov1972,
Dokshitzer1977,AltarelliParisi1977,AltarelliParisiPetronzio1977}, 
a perturbative description of quark or gluon induced 
jets in $e^+e^-$ annihilations \cite{StermanWeinberg1977}, the perturbative 
matching of the full theory to effective theories for heavy flavors 
\cite{CaswellLepage1986,BodwinBraatenLepage1995,EichtenHill1990,
Georgi1990,FalkGrinsteinLuke1991,Neubert1994R,PinedaSoto1998,BauerFlemingPirjolStewart2001} 
and the matching of the electroweak sector of the Standard Model involving perturbative QCD 
corrections with a low-energy effective theory describing the 
weak decays and mixings of heavy flavors, see for example 
\cite{GrinsteinSavageWise1989,Misiak1993} for $b\to s$ decays. Last but not 
least, at large external momenta asymptotic freedom guarantees the usefulness of an 
expansion of Euclidean correlation functions of hadron-interpolating 
currents starting from the parton model \cite{Bjorken1969,Feynman1969}. This is crucial for 
the method of QCD sum rules \cite{SVZ19791,SVZ19792} whose explanation is 
the starting point of the present review. 
  
While QCD is well understood and tested at large 
external momenta $p\gg 0.5\,$GeV the strongly coupled low-energy regime denies a 
perturbative treatment. 
The only presently known fundamental approach to learn about the 
dynamics of the relevant degrees of freedom populating the ground state of QCD at 
low energy and being responsible for the hadronic spectrum and 
hadron dynamics are lattice simulations \cite{Wilson1974}. Analytical approaches 
are not clear-cut -- not even the 
identification of the relevant degrees of freedom is unique --, 
and explanations of low-energy 
phenomena have only been found partially so far. For example, 
the instanton \cite{BelavinPolyakovShvartsTyupkin1975} gas 
provides a plausible, microscopic mechanism for dynamical chiral 
symmetry breaking \cite{BanksCasher1980} and 
the large mass of the $\eta^\prime$ \cite{'tHooft1976} but unless interactions between instantons 
are {\sl parametrized} as in the instanton-liquid model (see 
\cite{ShuryakSchafer1997} for a review) it does not explain color confinement. Dual models 
\cite{Suzuki1988} based on 
the idea that confinement is realized through the dual Meissner effect 
\cite{Mandelstam1976,'tHooft1981} explain confinement 
almost by definition and can not be considered fundamental. 
    
QCD sum rules provide a successful analytic way to 'scratch' 
into the nonperturbative regime. Many exhaustive reviews of the 
field exist \cite{QCDSRrev}, and therefore we will limit 
ourselves to a very basic introduction in the next section. 
The sum-rule approach is pragmatic in the sense that it does not aim 
at calculating nonperturbative ground-state parameters or hadronic wave functions from first principles. 
Rather, QCD sum rules extract universal nonperturbative parameters from limited experimental 
information to subsequently use them for predictions. 
The sum-rule method has been extended to investigate hadronic 
properties in the presence of a hot \cite{BochkarevShaposhnikov1985} or a dense 
hadronic medium (for a review see \cite{CohenFurnstahlGriegelJin1995}) 
which is of great relevance for ongoing experiments 
with colliding (ultra)relativistic heavy ions.

The basic object of investigation in QCD sum rules is 
an $N$-point correlator of QCD currents. The sum-rule method rests on analyticity in the external 
momentum variable(s) and on the assumption of quark-hadron 
duality. In its strong, local form, quark-hadron duality is the situation that a hadronic 
cross section can be related pointwise to a 
theoretical expression obtained in terms 
of quark and gluon variables. At low energy strong limitations 
on the accuracy of the theoretical expression exist. One possible theoretical approach is the 
analytical continuation of the OPE \cite{Wilson1969} which 
is an {\sl asymptotic} series in powers of the strong 
coupling $\alpha_s$ and in powers of the inverse external momentum $Q^{-1}$ (up to logarithms)
\cite{SVZ19791,ChibisovDikemanShifman1996}. To circumvent the problem associated with the 
{\sl asymptotic} series one appeals to analyticity in the external momentum variable 
to relate the incompletely known theoretical part to an {\sl average} over the 
hadronic cross section by means of a 
dispersion relation \cite{SVZ19791}. This pragmatic approach is rather 
fruitful.  

Different meanings are attributed to quark-hadron duality in the literature. 
For the present review, which is mainly concerned with its local form, we use the definition 
given by Shifman in \cite{Shifman12000} for a one-variable 
situation such as the $e^+e^-$ annihilation into hadrons. Shifman's definition is as follows: 
The current correlator, associated with the process of hadron creation out of the vacuum, is 
theoretically evaluated in terms of quark and gluon fields at 
an external momentum $q$ with $q^2=-Q^2<0$ (Euclidean region) using an OPE \cite{Wilson1969}.
Approximating this current correlator by a truncation of the purely perturbative part of the 
OPE at $\alpha_s^k$ and a truncation at 
power $Q^{-D}$ (not counting logarithms arising from anomalous dimensions of 
contributing operators), 
the uncertainty of the result should be of orders $\alpha_s^{(k+1)}$ and $Q^{-(D+1)}$ if the 
OPE represents a converging series. In practice one usually has $k=2$ or $k=3$ and $D=6$ or $D=8$ 
(practical OPE). The uncertainty in the truncated OPE would translate 
into an uncertainty of order $\alpha_s^{(k+1)}$ and $s^{-(D+1)/2}$ 
of the theoretically predicted 
spectral function $\rho_{\tiny{\mbox{theo}}}(s)$. The latter 
is obtained by calculating the imaginary part of a 
term-by-term analytical continuation to negative $Q^2$, $-Q^2\equiv s>0$, of the truncated OPE. 
If the experimentally measured spectral function 
$\rho_{\tiny{\mbox{exp}}}(s)$ coincides with $\rho_{\tiny{\mbox{theo}}}(s)$ 
up to the above uncertainty within a certain range of $s$ values 
we say that the quark-gluon prediction, resting on the OPE, is dual 
to the hadronic spectral function in this range. If this is not true we speak of a 
violation of local quark-hadron duality. While 
local quark-hadron duality is badly 
violated in practical OPEs at momenta close 
to the lowest resonance mass there usually is a window of Euclidean momenta $Q$ where the practical OPE is 
seen to be equal (or dual) to the dispersion {\sl integral} over the spectral function. We will refer to this kind 
of duality as global or sum-rule duality. It is known for a long time 
that in some exceptional channels QCD sum rules do not 
work \cite{NovikovShifmanVainshteinZakharov1981,ChetyrkinNarisonZakharov1999,
ChernodubGubarevPolikarpovZakharov2000,Zakharov2003}. Namely, a stability under the 
variation of quantities parametrizing our 
ignorance about the hadron spectrum and the vacuum structure does not 
exist in these channels.

\section{SVZ sum rules\label{sec:IntroM}}

To prepare for the discussion of duality violation in later chapters 
we briefly review the method of QCD sum rules of Shifman, 
Vainshtein, and Zakharov (SVZ)\cite{SVZ19791,SVZ19792} 
focussing on mesonic two-point correlators. A remark on the notation is in 
order: In this and subsequent sections multiple meanings are attributed to the symbol 
$T$. It may stand for a correlator or time-ordering or the 
temperature. The symbol $\bar{T}$ is the inverse Borel transform of $B[T]$. 
What is meant should become clear from the context. We apologize for the inconvenience.      

Let us start by giving a more specific characterization 
of the SVZ approach than in the last section: 
The idea is to express hadron parameters in 
terms of expectation values of gauge invariant operators which are composed of 
fundamental quark and/or gluon fields. 
In practice, the hadron parameters of the lowest states 
are studied. The lowest states are explicitly presented in models for the spectral functions. The higher 
resonances are absorbed into a continuum which is usually approximated 
by perturbative QCD. To separate the continuum from the lowest 
state an effective threshold $s_0$, which is a model parameter, is introduced. The parameter 
$s_0$ is to be determined by the sum rule by requiring hadron parameters to be least 
sensitive to its variation. Obviously, one does not assume {\sl local} quark-hadron 
duality here - only the {\sl integral} over the perturbative QCD spectrum is assumed 
to be equal to the {\sl integral} over the experimental spectrum down to the 
threshold $s_0$.

The basic object of investigation is the vacuum correlator of two gauge invariant, 
hadron-interpolating currents $j_{L}^{F}$ 
and $j_{L^\prime}^{F^\prime}$ in momentum space where $L,L^\prime$ and $F,F^\prime$ 
denote Lorentz and flavor quantum numbers, respectively 
\eqb
\label{correl}
T^{FF^\prime}_{LL^\prime}(q)\equiv i\int d^4x\ \e^{iqx}\la 0|\mbox{T}
\left(j_{L}^{F}(x)j_{L^\prime}^{F^\prime}(0)\right)|0\ra\ .
\eqe
Quantum numbers with respect to discrete symmetries are suppressed. The quantum 
numbers of a currents correspond to the probed hadronic resonance 
or the probed mixing of them. In Eq.\,(\ref{correl}) 
T refers to the time-ordering symbol. Notice that the mass-dimension of the 
correlator on the left-hand side of Eq.\,(\ref{correl}) is two 
if the currents are quark bilinears such as  
\eqb
\label{currenti}
j^i_\mu=\bar{\psi}\frac{\tau^i}{2}\gamma_\mu\psi\ ,\ \ \ \ \psi=(u,d)^T\ ,\ \ (i=1,...,3)\,.
\eqe
Here $\tau^i$ stand for SU(2) generators in the fundamental representation 
normalized to Tr$\tau^i\tau^j=2\ \delta^{ij}$. If a vector 
current $j^{F}_\mu$ is conserved then a current conserving (transverse) tensor
structure can be factored out of its correlator, and we have
\eqb
\label{transverse}
T^{FF}_{\mu\nu}(q)=(q_\mu q_\nu-q^2\,g_{\mu\nu})\ T^{FF}(q^2)\,.
\eqe
In the case of a quark-bilinear current this implies the scalar amplitude $T^{FF}(q^2)$ to 
be a dimensionless function of the square of the 
external momentum $q^2$. In the more general case of non or partially conserved currents, 
such as the axial current, the decomposition is into longintudinal 
and transverse tensor structures.   

\begin{figure}[tb]
\begin{center}
\begin{minipage}[t]{8 cm}
\epsfig{file=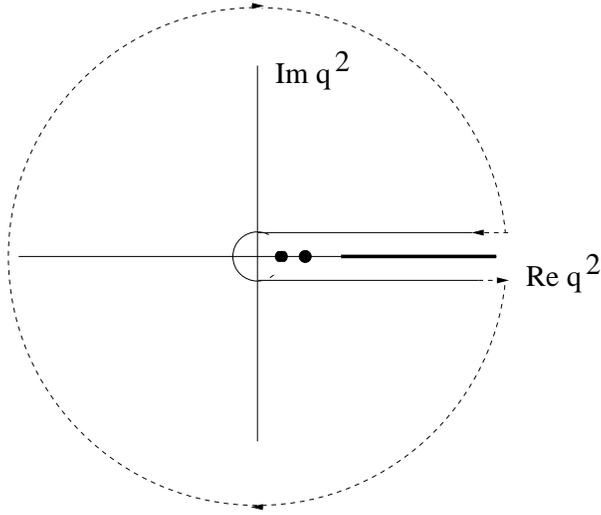,scale=.5}
\end{minipage}
\begin{minipage}[t]{16.5 cm}
\caption{The integration contour in the complex $q^2$ plane. The dashed line indicates the 
part at infinity. Dots denote isolated poles related to stable particles. 
The thick line is the continuous part of the spectrum. \label{Fig-analyticity}}
\end{minipage}
\end{center}
\end{figure}

Subtracted dispersion relations can be derived for the 
scalar amplitude $T^{FF}(q^2)$ assuming $T^{FF}(q^2)$ 
to be an analytic function of $q^2$ in the 
entire complex plane except for possible poles on and 
a cut along the positive, real axis starting at some threshold, see Fig.\,\ref{Fig-analyticity}
applied to $T^{FF}(p^2)/[(p^2)^N (p^2-q^2)]$, $(N\ge 0)$, and often to a 
circular origin-concentric contour of infinite 
radius deformed to spare out the positive real axis. Closing the circle 
at a finite radius $s_0<\infty$ leads to so-called finite-energy sum rules 
\cite{FESR1,FESR2} which we will not focus on here. We have
\eqb
\label{subtr}
T^{FF}(Q^2)=
\frac{(-Q^2)^N}{\pi}\int_{0}^{\infty}ds\ 
\frac{\mbox{Im}\ T^{FF}(s)}{s^N(s+Q^2)}-
 \frac{\left.d^{N}(T^{FF}(t)\times t)/dt^{N}\right|_{t=0}}{N!}\ (-Q^2)^{(N-1)}\,,
\eqe
where $Q^2=-q^2>0$, $N>0$ and it has been assumed that the integral along the circle vanishes, 
i.e. for a given $N$ the amplitude $T^{FF}$ has an according behavior at 
infinity. For $N=0$ the second term on the right-hand side of Eq.\,(\ref{subtr}) is omitted. 
The quantity $\frac{\left.d^{N}(T^{FF}(t)\times t)/dt^{N}\right|_{t=0}}{N!}$ 
is called subtraction constant. It is irrelevant if a so-called Borel transformation 
is applied to the dispersion relation Eq.\,(\ref{subtr}), see below.    

The discontinuity of $T^{FF}(q^2)$ across the cut and the residues of isolated poles are 
related to the hadronic spectral function $\rho^{FF}(s)$ which follows 
from an insertion of a complete set of hadron states 
inbetween the currents of Eq.\,(\ref{correl}). It is proportional to the total hadronic cross 
section in the considered channel. This is also known under the name 
Optical Theorem. For example, the total 
hadronic cross section $\sigma$ for the process
\eqb
e^+e^-\to \gamma\to \mbox{hadrons with isospin I=1}
\eqe
according to the Optical Theorem is related to $\mbox{Im}\ T^{33}$ (associated with 
the current of Eq.\,(\ref{currenti})) as \cite{PeskinSchroederBook,NarisonBook,TarrachBook}
\eqb
\label{cshadr}
\sigma(e^+e^-\to \gamma\to \mbox{hadrons with isospin I=1})=
\frac{16\pi^2\alpha_e^2}{s}\mbox{Im}\ T^{33}(s)\,,
\eqe
where $\alpha$ denotes the 
electromagnetic fine-structure constant. 
In sum rule applications one either substitutes a fit 
to the low-energy spectrum together with the perturbative QCD spectrum for $s$ 
greater than the threshold $s_0$ for $\mbox{Im}\ T^{FF}(s)\propto \rho^{FF}(s)$ 
to determine unknown vacuum parameters, see below, 
or one assumes the dominance of the lowest-lying hadron within the resonance region 
in a given channel and determines its properties from known vacuum parameters. 
This leads us to the SVZ approach to 
the left-hand side of Eq.\,(\ref{subtr}) employing the so-called 
Operator Product Expansion (OPE) in QCD. 

The OPE was originally proposed by 
K. Wilson \cite{Wilson1969} in the framework of a so-called 
skeleton theory for hadrons. The claim is that for a given theory 
a nonlocal product of composite operators $O_1(x)O_2(0)$ can usefully be expanded 
into a series 
\eqb
\label{OPE,coord}
O_1(x)O_2(0)=\sum_{D,{i_D}} c_{D}^{i_D}(x)\  O_{D}^{i_D}(0)\ .
\eqe
involving local operators $O_{D}^{i_D}(0)$ of increasing mass-dimension $D$ and 
c-number coefficients $c_{D}^{i_D}(x)$ - the so-called Wilson coefficients - as long as 
the (Euclidean) distance $|x|$ is small compared to the inverse of the 
highest dynamical mass scale in the theory.   
The OPE can be shown to exist in perturbation theory, 
see for example \cite{NovikovShifmanVainshteinZakharov1981}. In 
Eq.\,(\ref{OPE,coord}) the index $i_D$ runs over all possible, 
independent operators of dimension $D$. Taking the vacuum 
average of Eq.\,(\ref{OPE,coord}), only scalar contributions survive as 
a consequence of the Poincar\'e invariance of this state. 

Applying the OPE to a correlator of gauge
invariant currents in QCD, 
the scalar amplitudes in a given current correlator 
are expanded into a series of the form Eq.\,(\ref{OPE,coord}). 
The participating operators, 
such as ${\bf 1}$, $\frac{\alpha}{\pi}G^a_{\mu\nu}G^{\mu\nu}_a$ and $m(\bar{u}u+\bar{d}d)$,  
are gauge invariant, and the associated Wilson coefficients are perturbatively 
expanded in powers of $\alpha_s$. By definition, the expectation in the 
perturbative vacuum of the expansion singles out the unit operator 
${\bf 1}$.

In general, Wilson coefficients and operator 
averages depend on a normalization scale $\mu$ at which the 
perturbatively {\sl calculable} short-distance effects contained in the former are 
separated from the {\sl parametrized} long-distance behavior 
induced by nonperturbative fluctuations in the latter. 
If the currents of interest are conserved then 
the associated correlator should not depend on any factorization convention. 
In this case a residual $\mu$ dependence of the OPE is an artefact 
of the truncation of the $\alpha_s$ series in each coefficient function. An improved 
scale dependence of the Wilson coefficients can be obtained by solving perturbative 
renormalization group equations. At higher order in $Q^{-1}$ operators mix 
and anomalous dimension matrices must be diagonalized to 
decouple the system of evolution equations \cite{SVZ19791}, see also 
\cite{PeskinSchroederBook}. 
Vacuum averages of nontrivial QCD operators defined at $\mu$ are universal, 
that is, channel-independent 
parameters. They are the QCD condensates. These parameters have to be extracted 
from experiment, and they induce (up to logarithm's arising from anomalous operator dimensions) 
(negative) power corrections in $Q$. Already in the seminal paper \cite{SVZ19791} it 
was realized that the OPE in QCD can at best be an asymptotic expansion since small-sized 
instantons spoil the naive power counting $\left(\frac{\Lambda_{QCD}}{Q}\right)^{D}$ 
for operators of the form $G^{D/2}$ starting at the critical dimension $D=12$. 
This estimate was obtained from a dilute gas approximation 
for the instanton ensemble. Ever since, the discussion 
about the validity of the OPE in the light of its possible asymptotic nature, 
nonperturbative, nonlocal effects and the tightly related issue of 
local quark hadron duality has never faded away (see for example 
\cite{NovikovShifmanVainshteinZakharov1981,ChibisovDikemanShifman1996,
IsgurJeschonnekMelnitchoukvanOrden,Shifman12000,
Hofmann12001,Hofmann22001,Hofmann32001,BalitskyBraun1989,BalitskyBraun1990}). 
No definite conclusions have been reached although valuable insights were 
obtained in model approaches. To shed light on this discussion is in the 
main purpose of the present review. It is stressed here, however, 
that the phenomenological success of many 
QCD sum-rule applications using 'practical' OPEs, typically truncated at $D=6,8$, 
supports the pragmatic approach of Shifman, Vainshtein and Zakharov of `scratching' 
into the nonperturbative regime by ignoring questions of OPE convergence. For example, 
the prediction of the mass of the $\eta_c$ at 3.0\,GeV by a sum rule in the 
pseudoscalar $\bar{c}c$ channel matches almost precisely the result 
of a later experiment \cite{CB1980}.
  
Let us now discuss how Wilson coefficients are computed in practice. 
The perturbative part of a mesonic current correlator is the 
usual vacuum polarization function. In the case of 
electromagnetic currents this function was calculated to 4-loop accuracy in 
\cite{SurguladzeSamuel1991,GoroshniiKataevLarin1991}. 
For the calculation of the Wilson coefficients appearing 
in power corrections two methods are known. The so-called 
background-field method \cite{NovikovShifmanVainshteinZakharov1984} uses a 
Wick expansion of the time-ordered product of currents and 
interaction Lagrangians. Quark propagation is considered in a 
gauge-field background and the Fock-Schwinger or fixed-point 
gauge $x^\mu A_\mu(x)=0$ is used.  In the background-field 
method the gauge field 
and the quark field away from the origin are expressed in 
powers of adjoint and fundamental covariant derivatives 
acting on the field strength and the quark field at the origin, 
respectively. It is tacitly assumed in all practical applications that the 
gauge invariance of nonlocal operator products, 
as they arise in the Wick expansion, is ensured by {\sl straight} Wilson lines
\eqb
\label{Wilsonline}
W(z_i)={\cal P}\exp\left[ig\int_0^{z_i} dy^\mu\,A_\mu(y)\right]\,,\ 
\ \ y_\mu=\xi z_\mu\,,\ \ \  0\le\xi\le 1\,,
\eqe
which connect a field at $z_i$ with a field at the base point zero (see 
\cite{NikolaevRadyushkin1983} for an exhaustive discussion). The symbol ${\cal P}$ 
in Eq.\,(\ref{Wilsonline}) denotes the path-ordering prescription. 
The gauge invariance of the resulting nonlocal operators is at 
the price of allowing for gauge parallel transport only along 
straight lines connecting to the base point
zero. The method does not consider fluctuating gluons and therefore the 
computation of radiative corrections in Wilson 
coefficients is out of reach. It has been applied to the 
computation of Wilson coefficients for the gluonic 
operators of the form $g^3G^3$, $g^4G^4$, and $g^2(DG)^2=g^4j^2$ 
which induce the relevant power corrections in heavy quark 
correlators \cite{NikolaevRadyushkin11982,
NikolaevRadyushkin21982,NikolaevRadyushkin1983}. A second method, which 
allows for the consideration of radiative corrections, works as follows. 
Project out a particular Wilson coefficient in an OPE of interest 
by sandwiching the associated T product of currents with (hypothetical) 
external quark and/or gluon states, 
separate off the structure belonging to the corresponding operator, and take the 
limit of vanishing external momenta. The calculation can be carried out at any
order in perturbation theory - radiative corrections can be considered. 
This method was used in \cite{SVZ19791} and in 
\cite{BroadhurstGeneralis1983,BroadhurstGeneralis1984} together 
with the background method.  

Having discussed the two sides of the dispersion relation 
(\ref{subtr}) we will now review refinement procedures for the practical
evaluation of sum rules. Typically, precise 
information about the spectral function in a given channel 
is limited to the lowest lying resonance. It can thus be 
important to apply a weight function to the spectral function 
which suppresses the high-energy tail 
to sufficiently reduce the error in the spectral integral. A 
Borel transformation, which acts on the sum rule (\ref{subtr}) 
by an application of the operator
\eqb
\label{Borel}
{\bf L}\equiv \lim_{{Q^2\to\infty,\ N\to\infty}\atop Q^2/N
\equiv M^2,\ M^2\ \mbox{{\tiny fixed}}} 
\frac{(-1)^N}{(N-1)!}\ (Q^2)^N \left(\frac{\pd}{\pd Q^2}\right)^N\,,\ \ 
\ \ \ \ \ \ \ \ \ \ \ \  (Q^2=-q^2)
\eqe
is designed to do precisely this. On the spectral side it 
generates an exponential $\sim \exp[-s/M^2]$ as opposed to the
power suppression $\sim 1/(s+Q^2)$ at high energy and on the OPE side 
power corrections in the Borel parameter $M$ are 
factorially suppressed in the mass-dimension. A physical parameter, which 
is extracted from the sum rule, should be independent of $M$ 
within the so-called stability window. This region of values for $M$ is interpreted 
as the interval where the spectral side of the sum rule indeed approximates the 
OPE expression. Hadron parameters are extracted in this region. 

It sometimes proves useful to 
suppress the high-energy tail of the spectral function 
by a higher integer power of $s$ and then look at the insensitivity of the sum rule under 
variations in this power. In this case one 
considers various derivatives\footnote{In contrast to $T^{FF}(Q^2)$ 
the first derivative $\pd_{Q^2} T^{FF}(Q^2)$, the Adler function, 
needs no subtraction of the divergence related to the 
fermion loop at $O(\alpha_s^0)$.} of the sum rule (\ref{subtr}) 
with respect to the external momentum $Q^2$. This leads to 
a set of {\sl moment} sum rules.   

\section{Seminal sum-rule examples\label{sec:SSRE}}

Violations of local quark-hadron duality are believed to have a mild effect on QCD sum rules. 
We review specific QCD sum rules for (axial)vector mesons in this section. 
In view of future and ongoing experiments at facilities like the large hadron collider (LHC) and 
the relativistic heavy ion collider (RHIC), respectively, we also discuss the 
in-medium approach to QCD sum rules for (axial)vector current-current correlators.

\subsection{Borel sum rules for light (axial) vector meson currents \label{sec:BSR}}

We consider the following currents, which interpolate to the $\rho,\omega$, and $A_1$ resonances:
\eqb
\label{intcur}
j_\mu^{\rho,\omega}\equiv\frac{1}{2}(\bar{u}\gamma_\mu u\mp\bar{d}\gamma_\mu d)\,,\ \ \ 
j_\mu^{A_1}\equiv {\cal TP}\frac{1}{2}(\bar{u}\gamma_\mu\gamma_5 u-\bar{d}\gamma_\mu\gamma_5 d)\,.
\eqe
The operation ${\cal TP}$ projects the axial current 
onto its conserved part which allows to discard the pion 
pole contribution in the spectral function of its correlator.
After separating off a transverse tensor structure $q_\mu q_\nu-q^2 g_{\mu\nu}$ 
the OPEs for the current-current 
correlators $T^{\rho,\omega}$, and $T^{A_1}$ in vacuum are given as\cite{HatsudaKoikeLee1993}:
\eab
\label{OPEs(axialvector)}
T^{\rho,\omega}&=&-
\frac{1}{8\pi^2}\left(1+\frac{\alpha_s(\mu^2)}{\pi}\right)\log\left(\frac{Q^2}{\mu^2}\right)+
\frac{m_q}{2Q^4}\la\bar{u}u+\bar{d}d\ra+\frac{1}{24Q^2}\la\frac{\alpha_s}{\pi}G^a_{\mu\nu}G^a_{\mu\nu}\ra-\nonumber\\ 
&&\frac{\pi\alpha_s}{2Q^6}\la\left(\bar{u}\gamma_\mu\gamma_5\lambda^a u\mp
\bar{d}\gamma_\mu\gamma_5\lambda^a d\right)^2\ra(\mu)-\frac{\pi\alpha_s}{9Q^6}\la\left(\bar{u}\gamma_\mu\lambda^a u+
\bar{d}\gamma_\mu\lambda^a d\right)\sum_{q=u,d,s}\bar{q}\gamma_\mu\lambda^aq\ra(\mu)\,,\nonumber\\ 
T^{A_1}&=&-\frac{1}{8\pi^2}\left(1+\frac{\alpha_s(\mu^2)}{\pi}\right)\log\left(\frac{Q^2}{\mu^2}\right)-
\frac{m_q}{2Q^4}\la\bar{u}u+\bar{d}d\ra+\frac{1}{24Q^2}\la\frac{\alpha_s}{\pi}G^a_{\mu\nu}G^a_{\mu\nu}\ra-\nonumber\\ 
&&\frac{\pi\alpha_s}{2Q^6}\la\left(\bar{u}\gamma_\mu\lambda^a u-
\bar{d}\gamma_\mu\lambda^a d\right)^2\ra(\mu)-\frac{\pi\alpha_s}{9Q^6}\la\left(\bar{u}\gamma_\mu\lambda^a u+
\bar{d}\gamma_\mu\lambda^a d\right)\sum_{q=u,d,s}\bar{q}\gamma_\mu\lambda^a q\ra(\mu)
\,,
\eae
where a one-loop correction is included in the perturbative part, the Wilson 
coefficients in the power corrections are given without radiative corrections, 
$\mu$ denotes the normalization point, degenerate 
light-quark masses $m_q=m_u=m_d$ have been assumed, and $\lambda^a$ are SU(3) 
generators in the fundamental representation normalized to 
tr$\lambda^a\lambda^b=2\delta^{ab}$. In contrast to dimension four 
the operator averages at $D=6$ depend on the 
normalization point $\mu$ logarithmically \cite{SVZ19791}. For simplicity and because effective 
anomalous operator dimensions are small, we neglect 
this dependence here.

\begin{figure}[tb]
\begin{center}
\begin{minipage}[t]{8 cm}
\epsfig{file=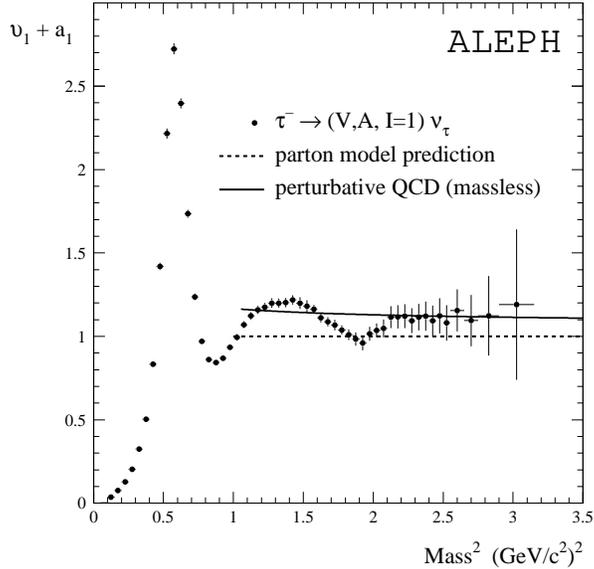,scale=.5}
\end{minipage}
\begin{minipage}[t]{16.5 cm}
\caption{The sum of $\rho$ and $A_1$ spectral functions as 
measured by the Aleph collaboration in 
$\tau$ decays. Plot taken from \cite{Aleph1998}.\label{Fig-aleph-v+a}}
\end{minipage}
\end{center}
\end{figure}

\begin{figure}[tb]
\begin{center}
\begin{minipage}[t]{8 cm}
\epsfig{file=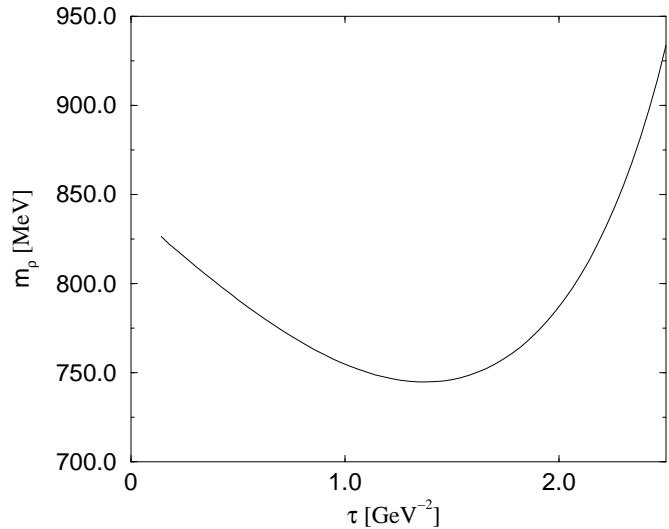,scale=.5}
\end{minipage}
\begin{minipage}[t]{16.5 cm}
\caption{The dependence of $m_\rho$ on the Borel parameter $\tau$.\label{Fig-mrho-tau}}
\end{minipage}
\end{center}
\end{figure}

On the spectral side, we use a zero-width model for the 
lowest resonance and approximate higher resonances by a 
(channel-dependent) perturbative continuum starting at 
threshold $s^{mes}_0$ (see for example \cite{NarisonBook}). 
The resonance part involves a meson-to-vacuum element. In the case of vector mesons it 
is parametrized as
\eqb
\label{coupling-to-current}
\la 0|j_\mu^{mes}|mes(p)\ra\equiv f_{mes} m_{mes}\epsilon_\mu\,
\eqe
where $f_{mes}$ $(mes=\rho,\omega)$ is a decay constant and 
$\epsilon_\mu$ the meson's polarization vector. The contributions of the pion continuum to the spectral functions 
for $s\le s_0$ are for practical reasons neglected here\footnote{At finite temperatures $T$ 
this contributions becomes important for $T\ge m_\pi$.}.
In Fig.\,\ref{Fig-aleph-v+a} the experimental data for the sum of 
$\rho$ and $A_1$ spectral functions is shown for center-of-mass 
energies up to $\sqrt{s}=\sqrt{3}$\,GeV. The two resonance peaks at $m^2_\rho\sim (0.77)^2\,\mbox{GeV}^2=
0.59\,\mbox{GeV}^2$ and $m^2_{A_1}\sim (1.26\,\mbox{GeV})^2=
1.59\,\mbox{GeV}^2$ are clearly visible. A similar situation holds in the $\omega$ channel. 
After Borel transformation we obtain the following sum rules
\eqb
\label{Borelsrs}
T^{\rho,\omega,A_1}(M^2)=
\frac{1}{M^2}\left[f_{mes}^2\e^{-m_{mes}^2/M^2}+\frac{1}{8\pi^2}
(1+\frac{\alpha_s}{\pi})\e^{-s_0/M^2}\right]\,,\nonumber\\ 
\eqe
where the left-hand side of Eq.\,(\ref{Borelsrs}) can is represented by 
the Borel transforms of the respective OPEs 
(applying the operator (\ref{Borel}) term by term). 
To extract information on the meson parameters one solves the sum rule, 
Eq.\,(\ref{Borelsrs}), for the resonance piece $R\equiv f_{mes}^2\e^{-m_{mes}^2/M^2}$. 
Assuming the coupling constants $f_{mes}$ and the resonance masses $m_{mes}$ 
to be sufficiently insensitive\footnote{This is a a bootstrap-like approach to the sum rule. 
We first assume insensitivity of hadron parameters to variations in the Borel parameter and show 
{\sl subsequently} that this assumption is self-consistent. For $m_{\rho}$ this is shown 
in Fig.\,\ref{Fig-mrho-tau}. The value of $s_0$ below 
is chosen such that the size of the window of insensitivity is maximized.} to changes in the 
Borel parameter $\tau\equiv\frac{1}{M^2}$, one can solve for $m_{mes}$ 
by performing a logarithmic derivative
\eqb
\label{logder}
m_{mes}^2=-\frac{\partial}{\partial \tau} \mbox{log}\ R\,.
\eqe
Using the following numerical values for the condensates
\footnote{The quark condensate is 
determined by the pion decay constant $f_\pi$, the pion mass $m_\pi$, the light current-quark mass by virtue of the 
Gell-Mann-Oakes-Renner relation \cite{Jamin}, and the gluon condensate can be 
extracted from ratios-of-moments in the $J/\Psi$ channel, 
see \cite{SVZ19791} and next section.}   
\eab
\label{condval}
\la\bar{q}{q}\ra\ (\mu=1\ \mbox{GeV})&=&-(250 \ \mbox{MeV})^3\,,  \nonumber\\  
\ \la\frac{\alpha}{\pi} 
G^a_{\mu\nu}G_a^{\mu\nu}\ra\ (\mu=1\ \mbox{GeV})&=&0.012\ \mbox{GeV}^4\,,
\eae
a continuum threshold $s_0^\rho=1.5\,\mbox{GeV}^2$, and 
approximating the four-quark condensate by squares of 
chiral condensate by assuming {\sl exact} vacuum saturation \cite{SVZ19791}
(taking place in the limit of a large number of colors, $N_c\to\infty$), 
one obtains a dependence of $m_{\rho}$ on $\tau$ as shown in Fig.\,\ref{Fig-mrho-tau}. 
We have neglected the contribution of the 
quark condensate at D=4 since it is numerically 
small compared to the gluon-condensate $m_q/\Lambda_{QCD}\ll 1$.
Fig.\,\ref{Fig-mrho-tau} clearly indicates 
that the sum rule for the $\rho$ meson 
is stable in the Borel mass for values 
around $M^2=0.8\,\mbox{GeV}^2$. One obtains qualitatively similar results 
in the $\omega$ and $A_1$ channels \cite{SVZ19792}.

\subsection{Quarkonia moment sum rules\label{sec:J/psi}}

So far we have considered light-quark channels using Borel 
transformed sum rules. In the case of mesons containing a heavy-quark pair moment sum rules turn out to be useful 
\cite{NovikovOkunSVVZ1978}. The methods of Ref.\,\cite{NovikovOkunSVVZ1978} 
have been developed 
further over the years and often applied to 
determine the mass of the bottom quark (for a 
review see \cite{ReviewMomentSRs}).
To be sensitive to the value of the quark mass the focus in the past was mainly 
on large moments since these suppress the perturbative continuum 
\cite{SVZ19792,NikolaevRadyushkin1983L,NikolaevRadyushkin1983}. In general, moment sum rules 
require a precise analysis of the quarkonium threshold and an according 
definition of the quark mass \cite{CorcellaHoang2003}. We briefly introduce the 
method here since we will rely on it 
in Sec.\,\ref{sec:DOE}. 

One starts with the correlator (\ref{correl}) where now light-quark 
currents of definite SU(3)$_F$ quantum numbers 
are replaced by heavy-quark currents such 
as $j_\mu^{h}=\bar{h}\gamma_\mu h$, $h$ denoting one of the 
heavy-quark fields $c,b,t$. 
Since this current is conserved we may again separate 
off a transverse structure $(q_\mu q_\nu-q^2g_{\mu\nu})$ and then only 
consider the scalar amplitude $T^{\bar{h}h}$. In contrast to the light-quark case, 
where the OPE essentially is an expansion in $\frac{\Lambda_{QCD}}{Q}$, the scale that 
power-suppresses nonperturbative corrections is naturally given by the 
heavy-quark mass $m_h$. One therefore expands both sides of 
the sum rule in powers of the dimensionless parameter $z\equiv Q^2/(4m_h^2)$, 
\eqb
\label{Cexp}
T^{\bar{h}h}(Q^2)=\frac{3Q_c^2}{16\pi^2}\sum_{n\ge0} C_n z^n\,,
\eqe
where $Q_c$ denotes the electric charge of the heavy quark, 
and compares coefficients $C_n$. These can be expressed in terms of the 
moments ${\cal M}_n$, defined as
\eqb
\label{moments}
{\cal M}_n=\left.\frac{1}{n!}\left(-\frac{d}{dQ^2}\right)^n 
T^{\bar{h}h}(Q^2)\right|_{Q^2=0}\,,
\eqe
as follows
\eqb
\label{coefficientsn}
{\cal M}_n=\frac{Q_c^2}{16\pi^2}
\left(\frac{-1}{4m_h^2}\right)^n C_n\,.
\eqe
The perturbative part of the coefficient $C_n$ is known up to 
order $\alpha_s^2$ and $n=8$ \cite{ReindersRubinsteinYazaki1981,ChetyrkinSteinhauser1996,
ChetyrkinSteinhauser1997}. It can be written as
\eqb
\label{Cnpert}
C_n=C_n^{(0)}+\frac{\alpha_s(\mu)}{\pi}\left(C_n^{(10)}+
C_n^{(11)}l_{m_h}\right)+
\left(\frac{\alpha_s(\mu)}{\pi}\right)^2\left(C_n^{(20)}+
C_n^{(21)}l_{m_h}+C_n^{(22)}l^2_{m_h}\right)\,,
\eqe
where $l_{m_h}$ is a short-hand for $\log(m_h^2/\mu^2)$ 
and the coefficients $C_n^{(\cdots)}$ 
are listed up to $n=8$ in \cite{KuhnSteinhauser2001}. 
The nonperturbative part of the moment ${\cal M}_n$, 
induced by the condensates $\la g^2 G^a_{\mu\nu}G^a_{\mu\nu}\ra, 
\la g^3 f^{abc}G^a_{\mu\nu}G^b_{\nu\lambda}G^c_{\lambda\mu}\ra$, and 
$\la g^4j_\mu^a j_\mu^a\ra$, $j_\mu$ being the light-flavor singlet current, 
was calculated in \cite{NikolaevRadyushkin21982,
NikolaevRadyushkin1983} by using the background-field method. 
In our convention, Eq.\,(\ref{moments}), the 
corresponding expressions are listed in \cite{HoangHofmann2003}. Low-$n$ moments are more sensitive to the spectral 
continuum, $large$-n moments to the resonance part of the spectrum. The former probe 
the relativistic part of the spectrum, and an expansion in $\alpha_s$ 
is appropriate. The latter probe the nonrelativistic physics in the quarkonium 
threshold region. The expansion parameter $\alpha_s$ is modified by the velocity of the heavy quark 
and given as $\sqrt{n}\alpha_s$. For, say,  $n>4$ a 
resummation of the spectrum to all orders in $\sqrt{n}\alpha_s$ by a Schroedinger 
equation for the bound state should be performed (for a review see \cite{Hoang2002}).

On the phenomenological side the correlator is expressed in terms of a 
dispersion integral. In terms of the heavy-quark pair cross 
section $\sigma_{e^+e^-\to c\bar{c}+X}$ and 
muon pair cross section $\sigma_{e^+e^-\to\mu^+\mu^-}$ in 
$e^+e^-$ annihilations this 
leads to the following expression for the moments
\eqb
\label{momentsExp}
{\cal M}^{exp}_n 
\, = \,
\frac{1}{12\pi^2\,Q_c^2}\,\int \frac{ds}{s^{n+1}}\,
\frac{\sigma_{e^+e^-\to h\bar
    h+X}(s)}{\sigma_{e^+e^-\to\mu^+\mu^-}(s)}
\,.
\eqe
It is obvious from Eq.\,(\ref{momentsExp}) that with increasing $n$ the 
hadron spectrum is probed at lower and lower 
center-of-mass energy $\sqrt{s}$. To reliably extract the quark-mass parameter $m_h$ 
at high $n$ therefore needs a high-precision treatment of the spectral function close to 
the threshold of quarkonium production. When $m_c$ was first estimated on parton level 
only the lowest four moments were considered\cite{SVZ19792}. 

If instead of analyzing the parameter $m_h$ by using low moments one wants to estimate 
power corrections it is better to consider the 
ratio $r_n\equiv\frac{{\cal M}_n}{{\cal M}_{n-1}}$ 
of adjacent moments since, contrary to the moments 
themselves, this quantity does not contain high 
powers of $m_h$ for large $n$. Large $n$ are
needed to be sensitive to the nonperturbative 
information of the resonance region. The ratio-of-moment analysis 
thus is less sensitive to the error in the extraction 
of $m_h$ and therefore more reliable. Historically, the value of 
the gluon condensate $\la \frac{\alpha_s}{\pi} 
G^a_{\mu\nu}G^a_{\mu\nu}\ra\sim 0.012\,\mbox{GeV}^4$ 
was first estimated using a stability analysis of $r_n$ for $n=2\cdots 10$ 
in the J/$\psi$ channel including resonances up to $\psi(4400)$ 
and the usual perturbative 
continuum model with threshold $s_0=(4.2\,\mbox{GeV})^2$. In Sec.\,\ref{sec:CSRRGC} 
we will have to go beyond the dimension four power
correction when we analyze the `running' of the gluon condensate 
in the framework of a reshuffled OPE containing nonlocal nonperturbative information.  

\subsection{QCD sum rules in the (axial)vector meson channel at $T,\mu_B>0$ \label{sec:Tmu}}

In this section we address the case of a QCD current correlator in the 
(axial)vector-meson channel and its OPE inside a hot 
or dense medium. On the one hand, the high-temperature situation is of relevance 
since the properties of mesonic resonances (width and peak positions) 
in such an environment are measured in ongoing and future experiments 
at RHIC and LHC, respectively. To do this, 
the measurement of the, almost unperturbed by the nuclear environment, 
invariant mass spectrum of dileptons is carried out. Dileptons, produced in the early stages 
of a relativistic heavy-ion collision by vector meson decay, 
tell us about the mass and the width of the primary. 
A sudden `melting' of the spectrum should unambiguously indentify the deconfinement transition. 
QCD sum rules are predestined to make predictions of the 
$T$ dependence of spectral parameters \cite{BochkarevShaposhnikov1985,DeyEletskyIoffe1990,Leutwyler1992,HatsudaKoikeLee1993,
EletskyIoffe1993,HofmannGutscheFaessler2000,Mallik1997,MallikMukherjee1998,MallikSarkar20021,
MallikSarkar20022,MarcoHofmannWeise2002}. On the other hand, there are 
interesting effects inside a cold baryon-rich nuclear medium which can be probed by heavy ion colliders 
operating at lower center-of-mass energy. For example, isospin 
resonance mixing can be induced by an 
isospin-asymmetric environment. This is possibly accessible to a 
theoretical treatment using QCD sum rules and assuming the nuclear environment to be modelled by 
a {\sl dilute} gas of nucleons \cite{HatsudaLee1992,AsakawaKo1993,JinLeinweber1995,HatsudaLeeShiomi1995,
LeupoldPetersMosel1998,LeupoldMosel1998,Leupold2001,ZschockePavlenkoKampfer2002,ZschockePavlenkoKampfer2003,
CohenFurnstahlGriegelJin1995,HatsudaShiomiKuwabara1996,DrukarevRyskinSadovnikova2001,MishraReinhardtStockerGreiner2002,
Dutt-MazumdarHofmannPospelov2000}.

The main focus of our brief 
review of the (axial)vector-meson sum rules in medium 
is on the occurrence of a twist expansion\footnote{defined as mass dimension minus Lorentz spin} 
when replacing the vacuum average over a bilocal 
current product by a {\sl Gibbs} average
\eqb
\label{VactoGibbs}
\la0|\cdots|0\ra \to \la\la\cdots\ra\ra\equiv Z^{-1}
\mbox{Tr}\e^{-\beta (H-\mu_B Q_B)}\cdots\,,
\eqe
where $Z\equiv \mbox{Tr}\e^{-\beta (H-\mu_B Q_B)}$ denotes 
the grand-canonical partition 
function for the QCD Hamiltonian $H$, 
$\mu_B$ defines a baryon chemical potential, and $Q_B\equiv\int d^3x\,j_0^B$ is the
associated, conserved baryon charge. A resummation of such an 
expansion genuinely takes nonperturbative nonlocalities 
in the associated hadron states into account.  
At finite temperature $T$ a small expansion parameter 
$T/Q$ exists ($T<\Lambda_{QCD}, Q\sim 1\,\mbox{GeV}$), 
and such a resummation is apparently not needed. The situation is quite different at finite nucleon 
density, where the expansion parameter is $m_p/Q$ ($m_p$ the proton mass) and a (partial) 
resummation of the twist expansion is imperative. The present section serves as a 
prerequiste to Sec.\,\ref{sec:PhenNN}, where it is pointed out
that the consideration of nonperturbative nonlocalities in {\sl vacuum} OPEs 
is necessary for a good description 
of certain hadron properties. 

\subsection{Light (axial)vector mesons at $T>0$\label{sec:AVM}}

The pioneering work in formulating QCD sum rules for the 
case of finite temperature and/or 
baryon density was launched by A. I. Bochkarev and M. E. 
Shaposhnikov in 1985 \cite{BochkarevShaposhnikov1985}. 
In particular, the $\rho$-meson channel at finite temperature, as it
was first treated in this paper, has been revisited several 
times over the years \cite{DominguezLoewe1989,DeyEletskyIoffe1990,DominguezLoewe1991,EletskyIoffe1993,
EletskyEllisKapusta1993,DominguezLoeweRojas1993,HatsudaKoikeLee1993,
EletskyIoffe1994,Mallik1997,MallikMukherjee1998,HofmannGutscheFaessler2000,
MarcoHofmannWeise2002,MallikSarkar20021,MallikSarkar20022,ZschockePavlenkoKampfer2002} 
because a number of substantial points 
were overlooked in \cite{BochkarevShaposhnikov1985} on both sides of the sum rule.  
 
Let us first discuss the basic points of the approach in 
\cite{BochkarevShaposhnikov1985} and its 
application to the $\rho^0$ channel. Instead of using a causal $T$ product 
the formulation relies on a the retarded ordering of currents 
because of its more adequate analytical properties. In either case, 
the correlator of conserved currents $T_{\mu\nu}$ 
can be decomposed into two invariants $T_1(q_0^2=(u_\mu q^\mu)^2,q^2)$ and 
$T_2(q_0^2=(u_\mu q^\mu)^2,q^2)$ 
since through the presence of a preferred rest frame 
a new covariant $u_\mu$=(1,0,0,0) - the four-velocity of the heat bath -  exists besides 
the external momentum $q_\mu$, and therefore
\eqb
\label{twoinvariants}
T_{\mu\nu}(q,u)=(q_\mu q_\nu-q^2\,g_{\mu\nu})T_1+
(u_\mu-\omega \frac{q_\mu}{q^2})(u_\nu-\omega \frac{q_\nu}{q^2})T_2\,,
\eqe
where $\omega=u^\mu q_\mu$. In the limit $\vec{q}\to 0$, however, 
$T_1$ and $T_2$ depend on each other, and it suffices to consider 
only one of them. By truncating the trace in the Gibbs average (\ref{VactoGibbs}) 
to the vacuum and one-particle pion states (1-$\pi$ states) (dilute pion-gas approximation), 
for $T\le 160\,$MeV this is well 
justified due to the chiral gap in the
spectrum ($m_\pi\sim 140\,$MeV, $m_\rho\sim 770\,$MeV), a Borel 
sum rule for the $\rho^0$ channel at finite temperature was derived in 
\cite{BochkarevShaposhnikov1985} in the limit $\vec{q}\to 0$. 
Evaluating the spectral side in this approximation, 
there appears a zero-temperature 
2-$\pi$ continuum weighted by a function $\tanh(\sqrt{s}/4T)$, 
usually neglected in vacuum sum rules, and, in addition, a 
scattering contribution to the spectral density,
\eqb
\label{scatteringcont}
\propto \delta(s)\theta(\omega^2-4m_\pi^2)n_B(\omega/2T)\,,
\eqe
accounting for a 1-$\pi$ intermediate state scattering off the 
current into a heat-bath pion and vice versa \cite{DeyEletskyIoffe1990}. 
The $\rho$-meson contribution to the spectral function 
is as in vacuum since the 1-$\pi$-to-$\rho$ matrix element of the current vanishes 
in the Gibbs-trace. At $s>s_0$ the spectrum is 
approximated by thermal QCD perturbation theory. There is a quark-antiquark continuum 
and a scattering term. The latter arises from the interaction of the current with {\sl quarks} in the heat 
bath. At $T\sim 130\,$MeV it amounts to about three times the corresponding pion contribution. 
This shows the relative suppression of the pion scattering term in the hadronic part of the
spectrum. 

On the OPE side, the $T$ dependence of the dimension-six (four-quark) 
operators in Eq.\,(\ref{OPEs(axialvector)}) 
that contribute 
to the vacuum correlator was treated in 
\cite{BochkarevShaposhnikov1985} by applying the 
fluctuation-dissipation theorem, that is, by expressing the 
Gibbs averages of local operators through spectral functions in 
the respective channels
\eqb
\label{fluctuatrion-dis}
\la\la A(0)B(0)\ra\ra=\pi\int\frac{d^4p}{(2\pi)^4}
\coth(\omega/2T)\rho(\omega,\vec p,T)\,,
\eqe
where $\rho$ is the absorptive part of the retarded correlator of the 
two currents $A$ and $B$. In order to make sense of such a spectral 
function the original current product has to be Fierz rearranged into products of 
gauge invariant currents. Each of these products can then be evaluated 
using (\ref{fluctuatrion-dis}). For alternative ways of calculating 
pion averages over four-quark operators see below \cite{HatsudaKoikeLee1993} 
and \cite{DrukarevRyskinSadovnikaFaessler2002}. For dimension-four operators 
this does not apply and some not too solid 
arguments were used to conclude that the $T$ dependent part of their 
Gibbs averages is small \cite{BochkarevShaposhnikov1985} and 
therefore can be neglected. As a result of their Borel analysis 
Bochkarev and Shaposhnikov find that the both the $\rho$ 
mass parameter and the spectral continuum
threshold $s_0$ experience a drastic drop at $T\sim 150\,$MeV.

Three comments are in order: (i) On the spectral side in 
\cite{BochkarevShaposhnikov1985} 
the fact was not taken into account  
that at finite $T$ the axial and the vector channel mix, 
and consequently that both resonances $\rho$ and $A_1$ should appear in 
the spectral function of the $\rho^0$ channel. 
In the dilute pion-gas approximation of \cite{BochkarevShaposhnikov1985} 
this was first shown in \cite{DeyEletskyIoffe1990} up to first 
order in the parameter $\epsilon\equiv\frac{T^2}{6f_\pi^2}$. Powers of $\epsilon$ arise 
by reducing 1-$\pi$, 2-$\pi$, 3-$\pi$, $\cdots$ states 
arising in the Gibbs-trace (\ref{VactoGibbs}) by means of the LSZ reduction formula, 
{\sl neglecting} their momenta, and using PCAC and current algebra in the chiral limit. 
Taking into account only these finite-$T$ corrections, the correlators 
$T^{\rho,A_1}_{\mu\nu}(q,T)$ are expressible as a
superposition of vector and axial-vector correlators 
$T^{\rho,A_1}_{\mu\nu}(q,0)$ at zero temperature. 
Up to order $T^2$ only the $\epsilon$ expansion contributes, 
and one obtains,
\eab
\label{EletskyandDey} 
T^{\rho}_{\mu\nu}(q,T)&=&(1-\epsilon)T^{\rho}_{\mu\nu}(q,0)+
\epsilon T^{A_1}_{\mu\nu}(q,0)\,,\nonumber\\ 
T^{A_1}_{\mu\nu}(q,T)&=&(1-\epsilon)T^{A_1}_{\mu\nu}(q,0)+
\epsilon T^{\rho}_{\mu\nu}(q,0)\,.
\eae
Thus the resonance poles of the $\rho$ and $A_1$ mesons 
do not move as a function of $T$ (and would not to any order in $\epsilon$ if this was the only 
expansion for $T$ corrections). The relative weight of the 
$A_1$ meson in the spectral integral, however, 
increases with growing temperature. Interestingly, 
it was found in \cite{DeyEletskyIoffe1990} by a Borel sum-rule analysis 
applied to a 
single-resonance spectral function (erroneously, since the $\rho$ mass does not shift at order $\epsilon$) 
that the $\rho$ mass shifts towards 
higher values as $T$ 
grows. One still can interprete this result in terms of the spectral weight moving 
towards higher invariant mass-squared $s$ as $T$ increases (growing importance of 
the $A_1$ resonance, see Eq.\,(\ref{EletskyandDey})). 
Except for \cite{HofmannGutscheFaessler2000} 
none of the QCD sum rule analysis subsequently performed, 
which all assumes a single resonance in the spectrum of 
the (axial)vector channel, has reproduced this behavior. Going to order $T^4$, there 
is an order $\epsilon^2$ correction, arising 
from the zero-momentum 2-$\pi$ states, but also a correction of order 
$\left(T^2/Q^2\right)^2$ which originates 
from finite pion momenta in the 1-$\pi$ state. The latter was 
estimated in \cite{EletskyIoffe1994} by expressing the 
two invariants $T^{\pi}_{1,2}$, defined in analogy to Eq.\,(\ref{twoinvariants}) with 
$u_\mu\to p_\mu$ and $\omega\to\nu=p^\mu q_\mu$ in the 
1-$\pi$ matrix element of the current product, in terms of the measured pion 
structure functions $F_{1,2}(x,q^2)$, $x=Q^2/2\nu$, 
using dispersion relations. There is a $\left(T^2/Q^2\right)^2$ correction in both the 
Lorentz invariant and violating parts $T_{1,2}$, respectively. These terms 
are induced by {\sl nonscalar} condensates in the 
OPE leading us to the next comment on \cite{BochkarevShaposhnikov1985}.  

(ii) In 
\cite{BochkarevShaposhnikov1985} only scalar operators were considered 
in the OPE. The O(4)-invariance is, however, reduced to an O(3)- or rotational 
invariance by the presence of the heat bath, and 
thus a number of additional operators 
are allowed to contribute to the OPE. This was first noticed in 
\cite{HatsudaKoikeLee1993} where a systematic twist-expansion 
was used to identify the relevant, nonscalar operators. 
The gluonic stress, contributing at 
dimension four, was further investigated in 
\cite{EletskyEllisKapusta1993} and in \cite{Mallik1997,MallikMukherjee1998} 
in view of operator mixing under a change of the renormalization scale. 
In \cite{HatsudaKoikeLee1993} the 1-$\pi$ matrix elements of operators, 
such as parts of $\theta_{00}$ at dimension four
($\theta_{\mu\nu}$ denotes the QCD energy-momentum tensor) and operators of 
the type $\bar{q}\gamma_0 D^3_0q$ at $D=6$, with non-zero 
twist - all other operator averages were omitted because of 
non-calculability - were evaluated 
using pionic parton distribution functions in the leading-order scheme 
and the parametrization of \cite{GlueckReyaVogt1992} 
at the sum rule scale $\mu=1\,$GeV. As noticed above on general grounds, 
the thermal phase-space integrals over 
1-$\pi$ matrix elements of pure-quark scalar operators in the OPE 
are expressible in terms of zero-temperature 
condensates and expanded 
in powers of $\epsilon$ only 
by applying the LSZ reduction formula, PCAC, and 
current algebra \cite{HatsudaKoikeLee1993}. The 1-$\pi$ matrix elements 
of the operator $\alpha_s/\pi\, G^2$ can be calculated by using the QCD 
trace-anomaly \cite{QCDTA}, $\theta^\mu_\mu=-1/8\pi(11-2/3N_F)\alpha_sG^2+\sum_qm_q\bar{q}q$. 
As a result, the matrix element is proportional to the pion 
mass and thus vanishes in the chiral limit. 
Even for realistic pion masses the $T$-induced shift of the 
gluon condensate is negligible \cite{HatsudaKoikeLee1993}. 
Assuming a single, narrow resonance plus scattering term plus 
continuum model for the spectral side, 
the Borel analysis of \cite{HatsudaKoikeLee1993} 
indicates a drastic {\sl decrease} of the $\rho$ mass and the continuum 
threshold $s_0$ at $T\sim 160-170\,$MeV. Notice that with the choice 
$\sqrt{s_0(T=0)}\sim 1.3\,$GeV$>m_{A_1}(T=0)=1.26\,$GeV in 
\cite{HatsudaKoikeLee1993} the $A_1$ resonance can effectively be viewed 
as a part of the perturbative continuum. A mixing of the 
$A_1$ and $\rho$ channels was noticed in \cite{HatsudaKoikeLee1993} on the OPE 
side, in accord with the general result in \cite{DeyEletskyIoffe1990}. 
Similar results were obtained for the $\omega$ and $A_1$ 
channels.    

\begin{figure}[tb]
\begin{center}
\begin{minipage}[t]{8 cm}
\epsfig{file=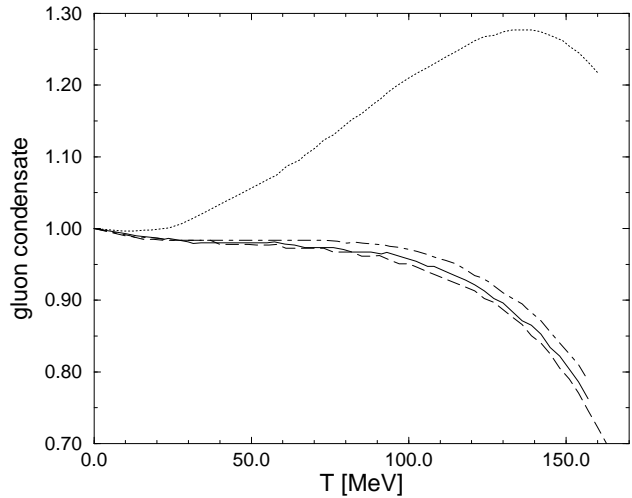,scale=.5}
\end{minipage}
\begin{minipage}[t]{16.5 cm}
\caption{The ratio of the $T$ dependent to the vacuum gluon condensate as 
obtained in \cite{HofmannGutscheFaessler2000}. 
The dot-dashed, solid, and long-dashed curves correspond to continuum
thresholds $s_0(0)=1.2\,$GeV$^2$, $s_0(0)=1.5\,$GeV$^2$, and $s_0(0)=1.8\,$GeV$^2$, 
respectively. The dotted 
curve shows the effect of perturbative renormalization-group evolution on 
the vacuum part of four-quark operators when allowing for an 
effective normalization scale $\sqrt{s_0(T)/s_0(0)}Q$. 
Taken from \cite{HofmannGutscheFaessler2000}.\label{Fig-gkT}}
\end{minipage}
\end{center}
\end{figure}

\begin{figure}[tb]
\begin{center}
\begin{minipage}[t]{8 cm}
\psfig{bbllx=70, bblly=232, bburx=485, bbury=610, 
file=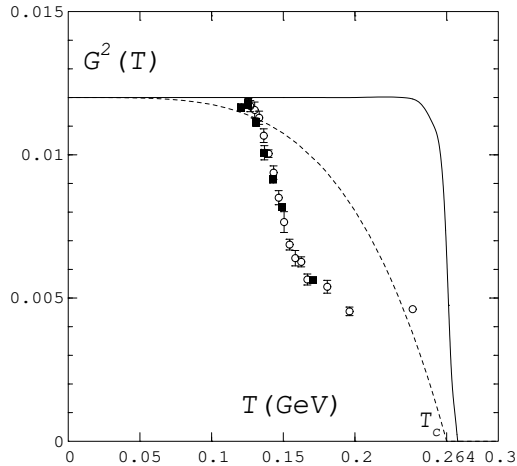, scale=.5 
}
\end{minipage}
\begin{minipage}[t]{16.5 cm}
\caption{The $T$ dependence of the gluon condensate as obtained on the lattice. Open circles and squares
denote the result obtained for QCD with light and heavy dynamical 
quarks. The dashed line shows the ideal-gas situation 
(a formula derived from a scaling argument in \cite{Miller2000}),
and the solid line is a fit to the lattice result for the $T$ dependence of the 
gluon condensate. Both lines refer to SU(3) Yang-Mills theory. Taken from \cite{Miller2000}.  
\label{Fig-gkT-Lat}}
\end{minipage}
\end{center}
\end{figure}

(iii) No attempt was made in \cite{BochkarevShaposhnikov1985} to consider a $T$ dependence of 
the vacuum state itself. This is suggestive since 
parameters like $f_\pi$, entering the average over the pion state in Eq.\,(\ref{VactoGibbs}), 
have a $T$ dependence which, in the case of $f_\pi$, is calculable in 
thermal chiral perturbation theory \cite{Leutwyler1992}. Chiral perturbation theory also predicts the $T$ dependence
of the quark condensate \cite{GerberLeutwyler1989}. For an investigation of the Gell-Mann-Oakes-Renner relation at
finite temperature and the calculation of the 
$T$ dependence of the pion mass relying on finite-energy sum rules see 
\cite{DominguezFeteaLoewe1996}.
Nothing, however, is known about the $T$ dependence of 
parton distribution functions. Assuming that a $T$ dependence of the vacuum state 
is the dominating dependence of the matrix elements in the Gibbs average 
(and thus that the $T$ dependence of 1-$\pi$ matrix elements can be neglected - the 
effect would anyway be of higher order in $T$) and assuming that such a 
dependence arises only implicitly through a 
$T$ dependence of the continuum threshold $s_0$ (related to a 
$T$ dependence of the QCD scale $\Lambda_{QCD}$), a scaling of the vacuum averages of operators 
with powers of their mass dimension $2n$, $(s_0(T)/s_0(0))^{n}$, 
was introduced in \cite{HofmannGutscheFaessler2000}. A $T$ dependence of the 
vacuum state is, indeed, suggested by the condensation of 
center vortices in the confining phase of QCD. These topological objects can be viewed as coherent 
thermal states themselves, hence the (mild) $T$ dependence of the ground state. 
Condensed center vortices are apparently responsible for quark confinement and chiral 
symmetry breaking \cite{CenterVortex}. The scaling with powers in 
$s_0(T)/s_0(0)$ introduced in \cite{HofmannGutscheFaessler2000} is an effective, 
phenomenological way of considering this effect. Note that this 
scaling does not capture the small effects of omitted, 
higher hadronic resonances in the Gibbs average. As a result, 
a {\sl positive} ``mass shift'' similar to the one 
in \cite{DeyEletskyIoffe1990} and, as a byproduct, 
a moderate drop of the gluon condensate around $T=160\,$MeV, which at least qualitatively is in accord 
with lattice results \cite{Miller2000}, was obtained, compare Figs.\,\ref{Fig-gkT} and \ref{Fig-gkT-Lat}. Notice 
that this decrease of about $30$\% (for $s_0(T=0)=1.5\,$GeV$^2$) 
as compared to the value at $T=0$ is practically entirely due to the vacuum average and 
not due to 1-$\pi$ matrix elements.

(iv) The usefulness of thermal, i.e. on-shell quarks (the scattering term), in \cite{BochkarevShaposhnikov1985} 
is quite questionable at low temperatures. 

We have seen that thermal, practical OPEs of current-current 
correlators allow for additional, O(3)-invariant, operators to appear. 
These operators arise in the Gibbs average 
from matrix elements over 1-$\pi$ states with 
nonvanishing spatial momentum. An expansion in 
powers of $T^2/Q^2$ arises in the chiral limit. 
Temperature induced corrections in Gibbs averages over 
{\sl scalar} operators can be organized as expansions in two parameters, $T^2/6f_\pi^2$ and 
$T^2/Q^2$ (in the chiral limit). The former arises from the (repeated) use 
of the LSZ reduction formula, PCAC and current algebra treating 
the pion as a noninteracting, elementary particle in the soft limit; the latter arises 
from the structure of finite-momentum pions. 
Since practical vacuum OPEs are, roughly speaking, expansions in $\Lambda_{QCD}^2/Q^2$ 
we conclude that the ``convergence'' of the expansion is not 
threatened by finite-temperature effects. 

\subsection{Vector mesons in nuclear matter\label{sec:NME}}

The treatment of current correlation involving (axial)vector mesons 
in a cold and dense environment using QCD sum rules 
is technically analogous to the case of finite temperature. Much work has been devoted to the calculation 
of the change of the mass and width of light vector mesons in a baryon-rich 
environment \cite{HatsudaLee1992,AsakawaKo1993,JinLeinweber1995,
HatsudaLeeShiomi1995,KlinglKaiserWeise1997,
KlinglWeise1999,LeupoldPetersMosel1998,LeupoldMosel1998,
Leupold2001,ZschockePavlenkoKampfer2002,ZschockePavlenkoKampfer2003} 
(for summaries see \cite{CohenFurnstahlGriegelJin1995,
HatsudaShiomiKuwabara1996,DrukarevRyskinSadovnikova2001}) 
since these should 
be measurable in terms of the invariant-mass spectra of dileptons 
emitted in the course of a heavy-ion collision, for an analysis within 
the Walecka model see \cite{MishraReinhardtStockerGreiner2002}, at a facility 
like SIS18 (GSI) with the HADES detector. For an adaption of vector-axialvector mixing 
to the situation of pions in a nuclear environment 
see \cite{Krippa1998}. A sum-rule analysis of the $\rho$-$\omega$ mixing 
induced by an isospin asymmetric nucleon density $\rho_N=\rho_p+\rho_n$ 
was performed in \cite{Dutt-MazumdarHofmannPospelov2000}. \cite{HatsudaHenleyMeissnerKrein1994}, 
respectively. This effect occurs in vacuum due to the breaking of the SU(2)$_{\tiny\mbox{F}}$ 
symmetry by the different electric charges and masses 
of up- and down-quarks \cite{SVZ197913}. For a sum-rule
analysis of the off-shell situation see \cite{HatsudaHenleyMeissnerKrein1994}. 
It was shown in \cite{Dutt-MazumdarHofmannPospelov2000} that by 
an appropriate and realistic choice of the isospin asymmetry $\alpha_{np}$ of the nucleonic environment, 
defined as $\alpha_{np}\equiv\frac{\rho_p-\rho_n}{\rho_p+\rho_n}$, the vacuum mixing can either 
be compensated or enhanced. Most of the above-mentioned works 
are technically rather involved. A simple and beautiful discussion 
of the $\rho$ meson (positive) mass shift in nuclei is, however, given in \cite{EletskyIoffe1997}. 
New developments 
concerning the treatment of nucleon matrix elements,in particular the ones 
of four-quark operators, deserve a review article 
in their own right, for a recent publication relying on the perturbative chiral quark model 
see \cite{DrukarevRyskinSadovnikaLGF2003}. Here we focus on interesting OPE 
aspects at finite density which hint on a fundamental manifestation of 
strong interactions at purely {\sl Euclidean} external momenta in 
both vacuum and hadron properties: the occurrence of strong 
nonperturbative correlations characterized 
by mass scales considerably larger than the 
perturbative scale $\Lambda_{QCD}$. Let us now briefly 
review some technical aspects of finite density sum rules. 

On the spectral side, a linear-density or dilute-gas approximation for the Gibbs average 
in (\ref{VactoGibbs}), which consists of taking into account only the 1-nucleon state 
besides the vacuum, again leads to the occurrence of a scattering term 
in the spectral function which is due to the scattering of a bath-nucleon off the current into an 
intermediate-state nucleon and vice versa. The question whether a treatment of in-medium 
resonance physics relying on the linear-density approximation is reliable 
for nucleon densities larger than the saturation density is open. 
Moreover, the consideration 
of finite vector-meson width in a {\sl pure} sum-rule treatment of the resonance 
seems to be problematic \cite{LeupoldPetersMosel1998}. The sum rule apparently contains 
too few information to predict both the density dependence of 
the resonance mass and the width. On the other 
hand, consistency of a spectral function calculated in the
framework of an effective chiral theory with the in-medium OPE of the 
correlator of the associated currents 
was obtained in \cite{KlinglKaiserWeise1997}.  

On the OPE side, O(3)-invariant operators 
contribute and can be organized in a twist expansion. Their 
nucleon averages are expressed in terms of integrals 
over nucleonic quark parton distributions, and a new expansion parameter, $\frac{m^2_p}{Q^2}$, 
emerges. In practice, one omits twist-four and also mixed operators due to the 
very limited information about their nucleon averages. The nucleon average over $\bar{q}q$ and 
$\frac{\alpha}{\pi}G^2$ are determined by the nucleon $\sigma$ term and 
by using the QCD trace anomaly, respectively, see \cite{CohenFurnstahlGriegelJin1995}. 
The treatment of nucleon averages over scalar four-quark 
operators is not as straight-forward as in the pionic case where 
chiral symmetry fixes these matrix elements in terms of vacuum averages. One way of proceeding is 
a mean-field like approximation\footnote{We use the same terminology as in \cite{HatsudaLee1992}.} 
(MFA) adjusted to the linear-density treatment \cite{HatsudaLee1992,HatsudaLeeShiomi1995}. For a treatment 
beyond the linear-density approximation methods have been worked out in 
\cite{CohenFurnstahlGriegelJin1995,DrukarevRyskinSadovnikova2001}. 
The status of the MFA is quite obscure (for a recent discussion see 
\cite{ZschockePavlenkoKampfer2002,ZschockePavlenkoKampfer2003} where 
the strong sensitivity of the in-medium mass-shifts of $\rho$ and $\omega$ mesons 
on the value of the in-medium four-quark condensates is stressed). 

The evaluation of the Borel sum rules in the 
$\rho$ channel yields a decrease of the $\rho$ mass 
\cite{HatsudaLee1992,AsakawaKo1993,JinLeinweber1995,HatsudaLeeShiomi1995} with 
increasing density. As for a the behavior of width {\sl and} mass no definite 
conclusion is possible \cite{LeupoldMosel1998,Leupold2001}. The old results for the 
$\omega$ channel in \cite{HatsudaLee1992,HatsudaLeeShiomi1995,JinLeinweber1995}, 
where in comparison to the $\rho$ channel an enhancement of the screening term by a factor of 9 was overlooked 
\cite{KlinglKaiserWeise1997,Dutt-MazumdarHofmannPospelov2000} and a negative 
shift of the resonance mass was obtained, are in clear contradiction to 
more recent analysis \cite{Dutt-MazumdarHofmannPospelov2000,ZschockePavlenkoKampfer2003} 
which points towards a {\sl positive} mass shift. Note, however, 
that the calculation of the $\omega$ mass shift in \cite{KlinglKaiserWeise1997}, 
which is based on a chiral, effective theory, also indicates a negative sign. Consistency with 
the OPE in this case was reached by applying nuclear 
ground-state saturation in the same way as in the vacuum:
\eqb
\label{qssnuc} 
\la\Omega(\rho)|(\bar{q}\gamma_\mu\gamma_5\lambda^a q)^2|\Omega\ra=
-\la\Omega(\rho)|(\bar{q}\gamma_\mu\lambda^a q)^2|\Omega\ra=\frac{16}{9}\kappa(\rho)
\la\Omega(\rho)|\bar{q}q)|\Omega\ra^2\,.
\eqe
This approximation is different from the MFA. In Eq.\,(\ref{qssnuc}) a density 
dependence of the correction factor $\kappa$ is allowed for. 

Let us make some summarizing comments on practical OPEs at finite nucleon density. (i) 
As we have seen, a new expansion parameter, $\frac{m^2_p}{Q^2}$, 
arises in the Gibbs averages over finite-twist, nonscalar operators. 
Recalling that $m_p\sim 940\,$MeV and that the external momentum $Q$ 
(or the Borel parameter $M$) should be not much larger than $\sim\,$1 GeV 
to be sensitive to resonance physics and associated 
power corrections in the OPE, we must conclude 
that a naive expansion is hardly controlled. 
However, as it was shown in \cite{FrimanLeeKim1999}, a summation of the twist-two 
correction to all orders in $\frac{m^2_p}{Q^2}$ appears to resolve this problem. Such a successful, 
partial summation of powers of $\frac{m^2_p}{Q^2}$ 
stresses the need to take nonperturbative nonlocalities in nucleon 
matrix elements into account. That this is not only true for the nucleon or, more generally, 
for any sufficiently stable hadron will be shown in detail in Sec.\,\ref{sec:PhenNN} where 
nonlocalities in vacuum matrix elements are imperative for a good description 
of certain hadronic properties. A very thorough discussion of the limitations 
of a local expansion of current correlators in the framework of 
nucleonic sum rules at isospin-symmetric finite baryonic density and of possible ways of 
improvement is performed in 
\cite{DrukarevRyskin1994}. This discussion rests on the pioneering work 
\cite{DrukarevLevin1988} on QCD sum rules for the nucleon at finite 
baryonic density. (ii) The screening term can dominate the 
density dependent part of the sum rule 
(for example in the $\omega$ channel or for the 
mixed $\rho$-$\omega$ correlator with mean-field treatment of four-quark operator averages) 
\cite{Dutt-MazumdarHofmannPospelov2000}. 
We can take this as a general indication 
that most of the density dependence of the resonance parameters is induced 
by the hadronic model for the rest of the spectral function and 
not by QCD parameters. So the situation is reversed as compared to vacuum sum-rules, where information 
on the lowest resonance is obtained in terms of QCD parameters and not in terms of extra 
hadronic information. 
(iii) The status of the mean-field treatment of nucleonic matrix over 
four-quark operators \cite{HatsudaLee1992,HatsudaLeeShiomi1995} 
is unclear. It was shown in \cite{ZschockePavlenkoKampfer2003} 
how a change by a factor of four in the contribution of 
four-quark operators can already change the sign of the mass-shift of the $\omega$ resonance 
at nuclear saturation density.

\section{OPE and Renormalons\label{sec:Renorm}}

In renormalized perturbation theory the divergent large-order behavior in 
correlators like the one in (\ref{correl}) can be related to power corrections of these objects  
\cite{GrossNeveu1974,Lautrup1977,tHooft1977,Parisi1978,Parisi1979,David1984,Mueller1985}. In this section 
we very briefly discuss the origin of this phenomenon and applications in QCD. We strongly draw upon the 
review by Beneke \cite{Beneke1999} which contains the relevant references up to the year 1999. 
We will explicitly refer to only some of the subsequent developments in applications of renormalons.
 
In a power-in-$\alpha_s$ perturbative expansion up to order $N$ of, say, a two-point current correlator
\footnote{In what follows the nonexistence of a constant term in 
Eq.\,(\ref{pertcurr}) is inessential.} 
\eqb
\label{pertcurr}
T(\alpha_s)=\sum_{n=0} r_n\alpha_s^{n+1}
\eqe
certain classes of diagrams, which we {\sl assume} to dominate the expansion in $\alpha_s$ in QCD, 
are associated with factorially-in-$n$ increasing coefficients 
$r_n\sim K a^n n^b n!$, (a,b,K constants), at large $n$ . In this case the expansion would be 
asymptotic, that is, there exists a truncation $N^*<\infty$ which {\sl minimizes} 
the truncation error. In gauge theories like QCD no proof is available 
for this asymptotic behavior. 

To have a sensible definition of a divergent series with factorially growing 
coefficients $r_n$ it is useful to first look 
at the {\sl Borel transform} of this series. For the series in Eq.\,(\ref{pertcurr}) it is defined as
\eqb
\label{Borelper}
B[T](t)=\sum_{n=0}^\infty r_n\frac{t^n}{n!}\,.
\eqe
For a $B[T](t)$, which has no non-integrable divergences on the positive, real $t$ axis, and which does not increase 
too strongly for $t\to\infty$, one can define the {\sl Borel integral} as
\eqb
\label{Borelint}
\bar{T}(\alpha_s)=\int_0^\infty dt\, \e^{-t/\alpha_s}B[T](t)\,.
\eqe
If $\bar{T}(\alpha_s)$ exists then it defines the {\sl Borel sum} 
of the original series $T(\alpha_s)$. 
If $B[T](t)$ has poles, which would then be a map of the 
diverging behavior of the series $T(\alpha_s)$, in the domain $t\ge 0$ then one can still define a Borel integral for 
$B[T](t)$ by deforming the integration path in the complex $t$ plane such that 
these singularities are circumvented. As a result, the Borel sum 
$\bar{T}(\alpha_s)$ usually acquires an imaginary part. There is, however, no unique 
deformation prescription - poles can be 
circumvented by deforming to positive or negative imaginary values of $t$ - 
which could be obtained from first principles in QCD perturbation theory. The difference 
between the two possible prescriptions embodies an 
ambiguity of the Borel integral which generically can be removed 
by adding {\sl exponentially} small terms $\sim \e^{-1/(a\alpha_s)}$ 
to the power series $T(\alpha_s)$. One refers to the poles on 
the real $t$ axis, which originate form factorially 
diverging coefficients in the perturbative expansion, as {\sl renormalon} poles.  

Following the presentation in \cite{Beneke1999} let us now look 
more specifically at how such singularities arise. 
We consider the Adler function, which is defined as
\eqb
\label{adlerfunction}
D(Q^2)=4\pi^2 Q^2\frac{d}{dQ^2} T(Q^2)\,,
\eqe
because it is free of divergences related to the outer 
fermion loop. In Eq.\,(\ref{adlerfunction}) $T(Q^2)$ is defined 
as in Eq.\,(\ref{transverse}).

\begin{figure}[tb]
\begin{center}
\hspace{-3cm}
\begin{minipage}[t]{8 cm}
\epsfig{file=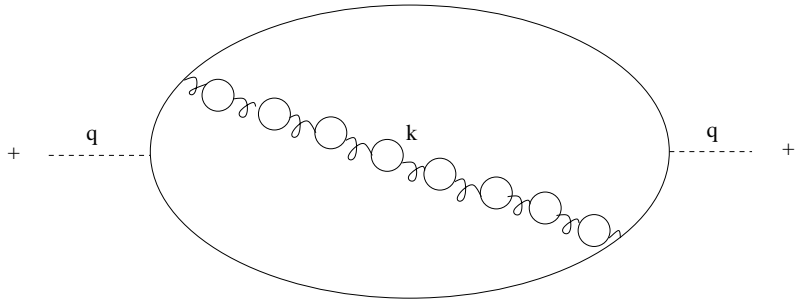,scale=.5}
\end{minipage}
\begin{minipage}[t]{16.5 cm}
\caption{A bubble-chain diagram which contributes to the 
perturbative expansion of a two-point current correlator.\label{Fig-Renormalon}}
\end{minipage}
\end{center}
\end{figure}

More specifically 
we are only interested in contributions 
arising from chains of fermion bubbles as 
in Fig.\,\ref{Fig-Renormalon}. At each order in $\alpha_s$ 
these contributions are gauge invariant by themselves. 
The QCD renormalized fermion bubble 
leads to the following fermion-bubble-chain induced 
expression for the Adler function 
\eqb
\label{adlerfunctionren}
D(Q^2)=\sum_{n=0}^\infty\alpha_s\int_0^\infty \frac{d\xi}{\xi^2}F(\xi)
\left[\beta_{0f}\alpha_s\log\left(\xi^2\frac{Q^2\e^{-5/3}}{\mu^2}\right)\right]^n\,,
\eqe
where $\xi\equiv-k^2/Q^2$, $k$ denoting the momentum flowing through the chain. 
The fermionic contribution to the (scheme independent) one-loop QCD $\beta$ function is defined as 
$\beta_{0f}\equiv \frac{1}{6\pi}N_f>0$, $\mu\sim Q$ denotes the normalization point, and the internal 
fermion-loop subtraction has been performed in the $\overline{\mbox{MS}}$ scheme. 
The function $F(\xi)$ is known exactly. It implies that 
for large $n$ the integrand in Eq.\,(\ref{adlerfunctionren}) 
is dominated by $\xi\ll 1$ and $\xi\gg 1$. In the former case $F(\xi)\sim\frac{2}{\pi}\xi^4$ and the latter 
$F(\xi)\sim\frac{4}{9\pi}\xi^{-2}\left(\log\xi^2 +\frac{5}{6}\right)$. 
This leads to the following approximate 
(the low-$n$ contributions are not well approximated) expansion in $\alpha_s$ of the Adler function
\eqb
\label{adlerfuntionapp}
D(Q^2)=\frac{1}{\pi}\sum_{n=0}^\infty\alpha_s^{n+1}\beta^n_{0f}\left[\left(\frac{Q^2}{\mu^2}\e^{-5/3}\right)^{-2}
(-2)^{-n} n!+\frac{4}{9}\frac{Q^2}{\mu^2}\e^{-5/3}n!\left(n+\frac{11}{6}\right)\right]\,.
\eqe
The first (sign alternating since $\beta^n_{0f}>0$) and second (sign non-alternating since $\beta^n_{0f}>0$) 
terms in the square brackets in Eq.\,(\ref{adlerfuntionapp}) 
are due to the $\xi\ll 1$ and $\xi\gg 1$ contributions to the 
integral in Eq.\,(\ref{adlerfunctionren}), respectively. The Borel transform of 
Eq.\,(\ref{adlerfuntionapp}) reads
\eqb
\label{adlerfunctionBT}
B[D](v)=\frac{2}{\pi}\left(\frac{Q^2}{\mu^2}\e^{-5/3}\right)^{-2}\frac{1}{2-v}+
\frac{4}{9\pi}\frac{Q^2}{\mu^2}\e^{-5/3}\left[\frac{1}{(1+v)^2}+
\frac{5}{6}\frac{1}{1+v}\right]\,,
\eqe
where $v\equiv=-\beta_{0f}t$. The pole at $v=2$, which is related to the 
behavior at small chain momenta, $\xi\ll 1$, is called first 
infrared (IR) renormalon whereas the single and the double pole at $v=-1$, 
which originated from large chain momenta, $\xi\gg 1$, is called first 
ultraviolet (UV) renormalon. According to Eq.\,(\ref{Borelint}) only the 
latter makes a contribution to the Borel integral and generates a {\sl negative linear} power 
correction in $Q^{-2}$. This is in contradiction to what we expect from the OPE where the 
leading power in $Q^2$ is $-2$ arising in the chiral limit from the gluon condensate 
$\la\frac{\alpha}{\pi}G^a_{\mu\nu}G^{\mu\nu}_a\ra$. What went wrong? 
The problem can be traced back to the fact that in 
considering only (gauge invariant) fermionic bubble chains and consequently only looking at the fermion 
contribution to the full QCD $\beta$ function we actually computed renormalon poles 
which are close to mimicking the large-$n$ behavior of an Abelian theory. Working in a covariant gauge, 
one could naively add the gluon and ghost bubble chains. The result, however, 
would be gauge dependent. A gauge invariant prescription to 
incorporate non-Abelian effects into our large-order investigations is to simply 
replace $\beta_{0f}$ by the full one-loop coefficient $\beta_0$ of the QCD $\beta$ function. 
This prescription includes also non-bubble-chain diagrams. Since the sign of $\beta_0$ is opposite to 
the one of $\beta_{0f}$ the IR renormalon pole moves to the negative (or positive) $v$ (or $t$) axis 
and thus contributes to the Borel integral whereas the UV renormalon 
pole ceases to make a contribution. As a result, the lowest nonperturbative correction 
is a power $Q^{-4}$ and induced by an IR renormalon. Replacing $\beta_{0f}$ by $\beta_{f}$ in 
Eq.\,(\ref{adlerfunctionren}) and performing the sum first yields
\eqb
\label{renormskel}
D(Q^2)=\int_0^\infty\frac{d\xi^2}{\xi^2}\,F(\xi^2)\alpha_s(\xi\e^{-5/6})\,,
\eqe
where $\alpha_s$ is the running coupling at one-loop. Thus the effect of a gauge invariant sum of 
diagrams including the fermion bubble-chain is to replace the chain by a single gluon line 
which couples to the external fermions via the one-loop {\sl running} coupling $\alpha_s(\xi\e^{-5/6})$. 
Although our prescription $\beta_{0f}\to\beta_{f}$ seems to be 
ad hoc it can be justified diagrammatically that 
renormalon poles are located at integer values of $v$ (or values of $t$ that are 
multiples of $\frac{1}{\beta_0}$.
Using Eq.\,(\ref{renormskel}), it is easy to see that the first 
IR renormalon contribution to $D(Q^2)$ is $\sim {\Lambda_{QCD}^4}{Q^4}$ with an ambiguous but $\mu$-independent 
numerical factor where the scale $\Lambda_{QCD}$ is a typical hadron scale. 
All this matches nicely with the OPE approach where lower power corrections are forbidden by the absence of 
the corresponding gauge invariant operators, and the operator $\frac{\alpha}{\pi}G^a_{\mu\nu}G^{\mu\nu}_a$ 
is renormalization group invariant at one loop. One may then say that the first IR renormalon is 
factored into the condensate and is associated with chain momenta $k\sim\Lambda_{QCD}\ll \mu$ 
while the Borel summable UV renormalons, corresponding to momenta $k\sim Q\gg\mu$, 
do contribute to the Wilson coefficients in an unambiguous way.   

As we have seen, the renormalon approach offers 
some insight into the {\sl structure} of power corrections even though the 
coefficients of the power corrections are ambiguous. At present it is not clear whether the OPE is 
asymptotic or not, and IR renormalons can not be used to predict 
the convergence properties of the OPE as an expansion in 
powers of $Q^{-2}$ itself. 
In fact, they only indicate the very limited set of power corrections which 
are related to large-order perturbation theory. However, one may think of more possibilities for the 
generation of power corrections, namely, power corrections which are entirely beyond the 
reach of perturbation theory or power corrections arising 
in Wilson coefficients from short distances. From a comparison of an analytical continuation 
to $Q^2<0$ with experimental spectral functions it is obvious 
that the so-called practical OPE violates local quark-hadron duality in the sense that 
we have defined it in Sec.\,\ref{sec:PF}. The construction and phenomenological test of 
reorderings of the OPE, which contain summations of $Q^{-2}$ powers to 
all orders and yet allow for a factorization of the large momenta regimes as in ordinary OPEs, 
is discussed in Sec.\,\ref{sec:ResOPE}.

In the remainder of this section we list two modern phenomenological applications 
of renormalons. For event shape variables and fragmentation cross sections in lepton-pair annihilation into hadrons, 
which are not described by an OPE, the identification 
of power corrections is not clear cut. Resorting to the so-called large $\beta_0$ approximation for 
the perturbative expansion, power corrections 
to the logarithmic scaling violations in these quantities were 
treated using renormalon resummations \cite{fragmentation}. For current correlators associated with 
lepton-pair annihilation and $\tau$ decay the relative strength of power corrections 
in their respective OPEs can be predicted from the 
corresponding residues of IR renormalon poles in a given scheme 
assuming that renormalons are the only source for these
corrections. Comparing the large $\beta_0$ approximation, in which this program is carried out, 
with known, low-order exact results only provides a partial justification for this 
approximation. This approach is an interesting model for 
power corrections and provides semi-quantitative insights, for a excellent discussion of this issue 
see \cite{Beneke1999} and references therein. 

\section{Violation of local quark-hadron\label{sec:VLQHD}}

In this second part a review of the experimentally measured violation of local 
quark-hadron duality in inclusive processes is given. Attempts to understand 
this violation using the model of current correlation with quarks 
propagating in an instanton 
background are reviewed. Finally, we discuss 
the issue in the framework of the 't Hooft model in the 
limit $N_c\to\infty$, where the model is exactly solvable, 
and also for large, but finite $N_c$.  

\subsection{\it Experimental facts and lattice results\label{sec:OPEA}}

Despite the practical successes of the use of the OPE in the framework 
of QCD sum rules, recall the moment analysis 
in the $J/\psi$ channel (Sec.\,\ref{sec:J/psi}). The theoretical 
status of this expansion in general field theories has never reached a 
satisfactory level, see \cite{GuptaQuinn1982,David1982} for a discussion of 
{\sl scalar} field theories with unstable vacua. In fact, 
it was even claimed in \cite{David1983} as a result of an analysis of the 2D O(N) nonlinear 
$\sigma$ model that no cancellation between IR renormalons in the perturbative part 
of the OPE of the propagator with IR renormalons present in 
the condensate part takes place. This means that 
the definition of local condensates is ambiguous.  

In 4D QCD there is not yet an analytical way to decide 
on the role of perturbative contributions to the vacuum condensates. 
Direct calculations of current or field correlators were performed in (suitable limits of) various 
field-theory models and compared with the OPE 
\cite{NovikovShifmanVainshteinZakharov1985}, and it was found that 
the amount of perturbative contribution varies from model to model. 
Pragmatically assuming that the local condensates in QCD are dominated 
by nonperturbative effects, as it is done 
in any sum-rule application of the OPE, a clarification of the nature of this 
expansion in negative powers of the external, Euclidean momentum $Q$ is still needed. 
This is, in particular, pressing in applications where analytical 
continuations of the OPE to the Minkowskian signature are needed as we will see below.   

Let us gather some experimental evidence that the 
inclusive spectra, which correspond to certain current-current 
correlators, do deviate substantially from the analytical continuation of their practical 
OPEs in the resonance region. 
We will consider three processes. (i) $e^+e^-$ annihilation into hadrons, 
(ii) axial-vector mediated, $\tau$ decays into hadrons, and (iii) width 
difference in the decays of $B_s$ and $\bar{B}_s$ mesons.

\begin{figure}[tb]
\begin{center}
\hspace{-3cm}
\begin{minipage}[t]{8 cm}
\epsfig{file=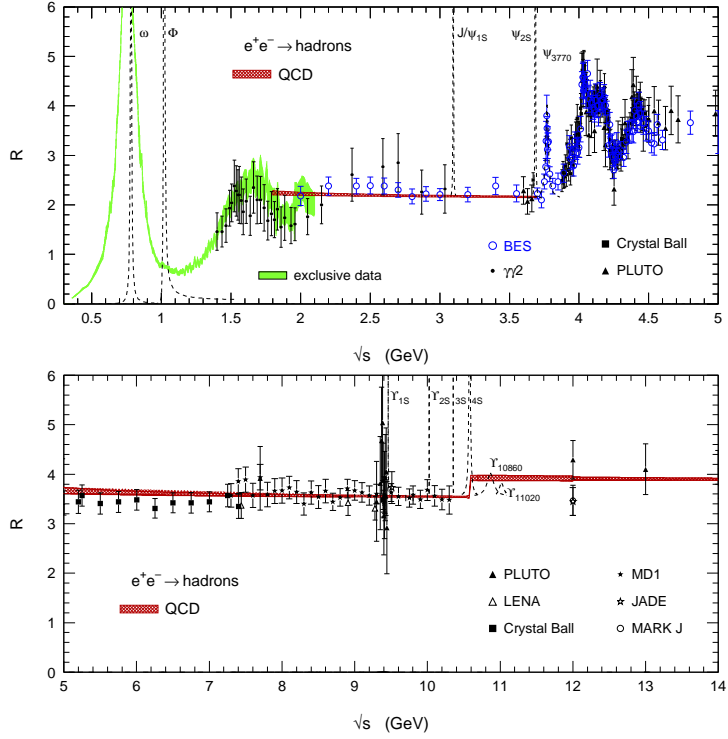,scale=.5}
\end{minipage}
\begin{minipage}[t]{16.5 cm}
\caption{The spectrum of $e^+e^-$ annihilation into hadrons up to $\sqrt{s}=14\,\mbox{GeV}$. 
The cross-hatched band are the results of perturbation theory which agree nicely with 
the data in the continuum regions $2\,\mbox{GeV}<\sqrt{s}<3.1\,\mbox{GeV}$, 
$4.6\,\mbox{GeV}<\sqrt{s}<9.1\,\mbox{GeV}$, and 
$2\sqrt{s}> 11.2\,\mbox{GeV}$. The resonances in the region $\sqrt{s}<2\,\mbox{GeV}$ and in 
the vicinity of the $c\bar{c}$ and $b\bar{b}$ thresholds 
are out the reach of perturbation theory. Taken from \cite{DavierEidelmanHockerZhang2003}.
\label{Fig-ree}}
\end{minipage}
\end{center}
\end{figure}

(i): In Fig.\,\ref{Fig-ree} the (electromagnetic) vector-current induced 
spectrum of $e^+e^-$ annihilation, 
\eqb
\label{Rpred}
R=\frac{\sigma_{tot}(e^+e^-\to \mbox{hadrons})}{\sigma (e^+e^-\to\mu^+\mu^-)}\,,
\eqe
into hadrons is shown for $\sqrt{s}$ up to $14\,\mbox{GeV}$. 
It is obvious from the figure that a spectrum calculated from a continuation of the practical 
OPE, $R=-\frac{1}{\pi}\mbox{Im}(Q^2=-s-i0)$, 
violates local quark-hadron duality considerably within the resonance regions 
since the contribution of power corrections arising from operators 
with anomalous dimensions alter the perturbative result in 
Fig.\,\ref{Fig-ree} only in a smooth way at finite $\sqrt{s}$.

\begin{figure}[tb]
\begin{center}
\hspace{1cm}
\begin{minipage}[t]{8 cm}
\epsfig{file=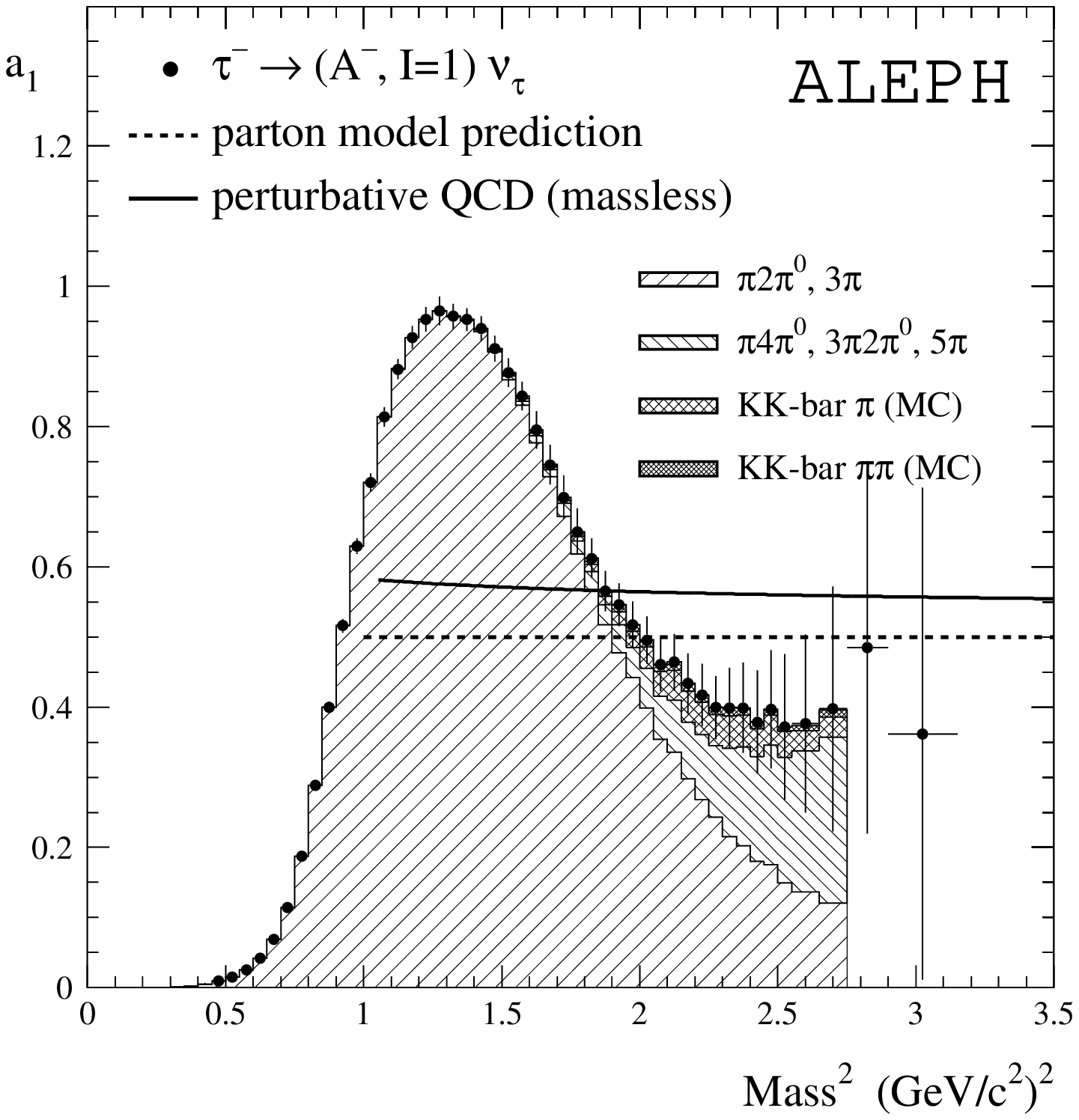,scale=.5}
\end{minipage}
\begin{minipage}[t]{16.5 cm}
\caption{Spectrum of axialvector-induced, $\tau$ decay into hadrons. 
Taken from \cite{Aleph1998}.
\label{Fig-A1}}
\end{minipage}
\end{center}
\end{figure}

(ii): The isovector-axialvector-induced spectrum (labelled by $A^{-}$) of $\tau$ decay into non-strange hadrons 
can be expressed as follows \cite{Aleph1998}
\eab
\label{A_1spectr}
R_{A^{-}}(s)&\equiv& \frac{M_\tau}{6|V_{ud}|^2S_{EW}}
\frac{B(\tau^{-}\to A^-\nu_\tau)}{B(\tau^{-}\to e^-\bar{\nu}_e\nu_\tau)}\times\nonumber\\ 
&&\frac{dN_{A^-}}{N_{A^-}ds}
\left(\left(1-\frac{s}{M_\tau^2}\right)\left(1+\frac{2s}{M_\tau^2}\right)\right)^{-1}\,
\eae
where $|V_{ud}|=0.9752\pm 0.0007$ is the CKM matrix element, 
$S_{EW}=1.0194\pm 0.004$ accounts for radiative 
electroweak corrections, $\frac{dN_{A^-}}{N_{A^-}ds}$ denotes the normalized 
invariant mass-squared distribution, and $M_\tau\sim 1.77\,$GeV is the mass of 
the $\tau$ lepton. In Fig.\,\ref{Fig-A1} the spectrum $R_{A^{-}}$ of $\tau$ 
decay as measured at LEP by the ALEPH collaboration \cite{Aleph1998} is shown.
Again, there is no way for a naive OPE continuation into the Minkowskian domain 
to generate the behavior of the spectrum around the $A_1$ resonance. 

(iii): Experimental information on $B_s$-$\bar{B}_s$ mixing is 
not yet available, but it will be investigated by CDF in the near future \cite{AzfarLyons2001}. 
A theoretical prediction resting on the assumption of local quark-hadron duality 
in the OPE approach exists at next-to-leading order in $\alpha_s$ 
\cite{BenekeBuchallaGreubLenzNierste1999}. Following \cite{Nierste2001} we briefly 
give some theoretical background on why 
$B_s$-$\bar{B}_s$ mixing can be a testing ground for the violation of 
local quark-hadron duality. Since $B_s$ may mix with its 
antiparticle $\bar{B}_s$ the two mass eigenstates $B_{H,L}$, which are linear combinations of 
$\bar{B}_s$ and $B_s$, have different masses, $\Delta m=M_H-M_L\not=0$, and different 
inclusive decay widths, $\Delta \Gamma=\Gamma_H-\Gamma_L\not=0$. $\Delta m$ and 
$\Delta \Gamma$ can be related to
the dispersive (absorptive) part of the $B_s$-$\bar{B}_s$ 
mixing amplitude, $M_{12}$ ($\Gamma_{12}$), 
as follows
\eqb
\label{adBB}
\Delta m=2|M_{12}|\,,\ \ \ \ \ \ \ \Delta \Gamma=2|\Gamma_{12}|\cos\phi\,.
\eqe
Practically, we have $\Delta \Gamma=2|\Gamma_{12}|$ since the CP 
violating phase $\phi$ is very small in the Standard Model. By means of the optical 
theorem the Standard-Model expression for $\Delta \Gamma$ is 
\eqb
\label{DelatGamma}
\Delta\Gamma=
\left|\frac{1}{M_{B_s}}\mbox{Im}\la\bar{B}_s|i\int d^4x\,T\,{\cal H}_{eff}(x){\cal H}_{eff}(0)|B_s\ra\right|\,
\eqe
where ${\cal H}_{eff}$ denotes the effective Hamiltonian mediating transition between 
$\bar{B}_s$ and $B_s$ in the Standard model after the heavy vector 
bosons $W^\pm$ and $Z_0$ have been integrated
out using renormalization-group improved perturbation theory. The operators appearing in the decomposition of ${\cal H}_{eff}$ are 
normalized at a scale $\mu_1=O(m_b)$. The variant of the OPE, which is used 
to estimate the right-hand side of Eq.\,(\ref{DelatGamma}), is an expansion in inverse 
powers of the $b$-quark mass $m_b$ - the so-called heavy quark 
expansion (HQE)\cite{VoloshinShifman1985}. Notice that this expansion relies on the validity of 
the HQE along the discontinuity in the Minkowskian domain. An experimental detection of 
a sizable CP asymmetry in the decays of $\bar{B}_s$ and $B_s$ would signal new physics. 
To test the Standard Model it is thus extremely important to have a good understanding of 
nonperturbative QCD effects and in particular of the validity of 
local quark-hadron duality in the use of HQE. To lowest order in $1/m_b$ one has 
\eab
\label{dispOPE}
&&\left|\mbox{Im}\la\bar{B}_s|i\int d^4x\,T\,{\cal H}_{eff}(x){\cal H}_{eff}(0)|B_s\ra\right|=\nonumber\\ 
&=&\frac{G_F^2m_b^2}{12\pi}|V_{cb}^*V_{cs}|^2\left|F\left(\frac{m_c^2}{m_b^2}\right)\la\bar{B}_s|Q|B_s\ra+
F_S\left(\frac{m_c^2}{m_b^2}\right)\la\bar{B}_s|Q_S|B_s\ra\right|\,.
\eae
In Eq.\,(\ref{dispOPE}) $G_F$ denotes the Fermi constant, $F$ and $F_S$ are 
the imaginary parts of the Wilson coefficients in the 
leading-order in $\alpha_s$ expansion, and $Q$($Q_S$) are the $\Delta B=2$ operators 
$\bar{s}_i\gamma_\mu(1\pm\gamma_5)b_i\bar{s}_j\gamma^\mu(1\pm\gamma_5)b_j$. The matrix 
elements of $Q$ and $Q_S$ are parametrized as
\eqb
\label{MEQ}
\la\bar{B}_s|Q(\mu_2)|B_s\ra=\frac{8}{3}f_{B_s}^2M_{B_s}^2 B(\mu_2)\,,\ \ \ \ 
\la\bar{B}_s|Q_S(\mu_2)|B_s\ra=-\frac{5}{3}f_{B_s}^2M_{B_s}^2
\frac{M_{B_s}^2}{(m_b(\mu_2)+m_s(\mu_2))^2}B_S(\mu_2)\,
\eqe
where $f_{B_s}$ and $M_{B_s}$ are the decay constant and the mass of the $B_s$ meson, and 
the masses $m_b(\mu_2)$ and $m_s(\mu_2)$ are defined in the $\overline{\mbox{MS}}$ scheme, and the 
normalization scale $\mu_2=O(m_b)$. In the 
vacuum-saturation approximation the ``bag'' factors $B(\mu_2),B_S(\mu_2)$ are equal to one.

At next-to-leading order in $\alpha_s$ there are already {\sl seven} operators\footnote{Dimensional 
regularization with anticommuting $\gamma_5$ matrices and the $\overline{\mbox{MS}}$ scheme is 
used (scheme dependence of the Wilson coefficients 
cancels against scheme dependence of the associated operator averages) 
in \cite{BenekeBuchallaGreubLenzNierste1999}.} which 
describe nonlocal contributions to the transition. Including $m_B^{-1}$ corrections
\cite{BenekeBuchallaDunietz1996}, the final answer for the quantity $\Delta\Gamma/\Gamma$, 
$\Gamma\equiv 1/2(\Gamma_L-\Gamma_H)$, reads \cite{Nierste2001}
\eqb
\label{DeltaGamma}
\frac{\Delta\Gamma}{\Gamma}=\left(\frac{f_{B_s}}{245\,\mbox{MeV}}\right)^2\,
[(0.234\pm0.035)B_S(m_b)-0.080\pm0.020]\,
\eqe
where $m_b(m_b)+m_s(m_b)=4.3\,$GeV ($\overline{\mbox{MS}}$ scheme) 
and $m_c^2/m_b^2=0.085$ have been used. Due to its tiny numerical value 
the contribution $\propto B$ has been neglected in Eq.\,(\ref{DeltaGamma}). 
With the result $f_{B_s}=(245\pm30)$\,MeV (for a QCD sum rule determination see for 
example \cite{JaminLange2002}) of an unquenched 
lattice calculation \cite{Hashimoto2000} (two dynamical fermion flavors)  
and the result $B_S(m_b)=0.87\pm 0.09$ of a quenched lattice calculation 
\cite{Yamada2000} one obtains (lattice errors added linearly)
\eqb
\label{DeltaGammaLat}
\frac{\Delta\Gamma}{\Gamma}=0.12\pm 0.06\,.
\eqe
In the limit of $N_c\to\infty$, where vacuum saturation is exact, 
and for $\Lambda_{QCD}\ll m_b-2m_c\ll m_b$ one can show that local duality 
holds exactly \cite{Aleksan1993}. In this case the result is 
$\frac{\Delta\Gamma}{\Gamma}=0.18$ which is just in the 
upper error limit of Eq.\,(\ref{DeltaGammaLat}). It is thus clear that a future experimental 
detection ($N_c=3$) of violations of local quark-hadron duality 
in the HQE for $\frac{\Delta\Gamma}{\Gamma}$ and/or of New Physics 
needs a much more precise (lattice)determination of 
$f_{B_s}^2B$ and $f_{B_s}^2B_S$, see for an unquenched, $N_f=2$ calculation of $B_{S}$ 
\cite{JLQCD2001} where the error (and the central value) 
are reduced in comparison to the result of the quenched calculation in 
\cite{Yamada2000}.

\subsection{\it Quark propagation in an instanton background\label{sec:QPIB}}

Since no full, analytical solution of QCD exists, which would allow for a direct comparison 
of the OPE with the exact result (and make the OPE superfluous), one has to resort to 
models of the current-current correlator. 
It was proposed in \cite{ChibisovDikemanShifman1996} that 
in analogy to the cancellation of renormalon 
ambiguities, arising from a factorial growth of coefficients in 
the $\alpha_s$ expansion, by exponentially small terms $\e^{1/\alpha_s}$, 
the OPE may (at best) be asymptotic in the according 
expansion in $\sim\frac{\Lambda_{QCD}}{Q}$. 
To make sense of it, one would then have to add exponentially 
small terms of the form $\e^{Q/\Lambda_{QCD}}$ 
which possibly would cure the violation of 
local quark-hadron duality by practical OPEs.  

A seemingly reasonable approach to see whether 
there is some truth in this proposal is to consider the quark 
propagation \cite{BrownCarlitzCreamerLee1978}, 
inherent in a given correlator of currents with massless quarks, 
in a dilute-gas instanton-antiinstanton background \cite{CarlitzLee1978,AndreiGross1978} or a 
general, (anti)selfdual background \cite{DubovikovSmilga1981} (like a dilute gas of 
multiinstantons and multiantiinstantons). 

Let us briefly review the calculation of the correlator of electric currents in a dilute-gas 
instanton-antiinstanton background as it was performed 
in \cite{AndreiGross1978} putting the (anti)instanton into 
the singular gauge. In the dilute-gas approximation 
it is only necessary to regard a single quark flavor $q$ of electric 
charge $Q$ - a sum over flavors can be performed at the end of the calculation. 
We consider the two-point correlator of the conserved 
current $j_\mu\equiv Q\bar{q}\gamma_\mu q$ in Euclidean position space
\eqb
\label{instcurcor}
T_{\mu\nu}(x)=\la T j_\mu(x)j_\nu(0)\ra\,.
\eqe
Considering, in a first step, 
quark-propagation in the background of a single (anti)instanton and 
disregarding radiative corrections, the Lorentz-trace $T(x)=T^{\mu}_{\mu}(x)$ 
is simply given as
\eqb
\label{instcurcorA}
T_{\pm}(x,\Omega_{\pm})=-\sum_F \mbox{Tr}\,Q_F^2\gamma^\mu S_{\pm}^F(x,0,\Omega_{\pm})\gamma_\mu 
S_{\pm}^F(0,x,\Omega_{\pm})\,
\eqe
where $S_{\pm}^i$ denotes the quark-propagator in the (anti)instanton background, 
the trace is over Dirac and color indices, the sum is over light quark flavors, and 
$\Omega_{\pm}$ denotes the collective parameters of the (anti)instanton. The propagators 
are expanded in mass $m_i$ around $m_i=0$
\eqb
\label{prop-in-inst}
S_{\pm}(x,y,\Omega)=-\frac{\Psi_0(x)\Psi_0^\dagger(y)}{m}+\sum_{\lambda\not=0}
\frac{\Psi_\lambda(x)\Psi_\lambda^\dagger(y)}{\lambda}+
m\sum_{\lambda\not=0}
\frac{\Psi_\lambda(x)\Psi_\lambda^\dagger(y)}{\lambda^2}+O(m^2)\,
\eqe
where $\lambda$ is a nonvanishing eigenvalue of the Dirac operator $i\gamma^\mu D_\mu$, 
and the subscript `0' refers to the zero-mode contribution. It is important to keep 
the term linear in $m$ when calculating $T(x)$ in the limit $m\to 0$. 
The zero-mode part in Eq.\,(\ref{prop-in-inst}) is given in terms of 
\eqb
\label{zero-mode}
(\Psi_0(x))_{\alpha,t}=\left(\frac{2}{\pi^2}\right)^{1/2}\frac{\rho_{\pm}}
{((x-x_{\pm})^2+\rho_{\pm}^2)^{3/2}}\left(i\gamma_\mu \hat{x}^\mu\gamma_2
\frac{1}{2}(1\pm\gamma_5)\right)_{\alpha,t}\,
\eqe
where $\hat{x}$ denotes a 4D unit vector, and $\alpha (t)$ is a Dirac index 
(fundamental SU(2) color index). The $O(m^0)$-part 
in Eq.\,(\ref{prop-in-inst}), $S^0_{\pm}(x,y,\Omega_{\pm})$, 
can be written as \cite{BrownCarlitzCreamerLee1978}
\eqb
\label{O(m^0)} 
S^0_{\pm}(x,y,\Omega_{\pm})=\gamma^\mu D^x_\mu\Delta_{\pm}(x,y,\Omega_{\pm})\frac{1}{2}(1\pm\gamma_5)+
(\Delta_{\pm}(x,y,\Omega_{\pm})D^x_\mu\gamma^\mu)\frac{1}{2}(1\mp\gamma_5)
\eqe
where $\Delta_{\pm}(x,y,\Omega_{\pm})$ denotes the propagator of a scalar, color-triplet 
particle, and $(\Delta_{\pm}(x,y,\Omega_{\pm})D_\mu\gamma^\mu)$ means that the 
covariant derivative acts from the right onto $\Delta_{\pm}(x,y,\Omega_{\pm})$. 
This propagator is explicitly known \cite{BrownCarlitzCreamerLee1978}, in singular gauge it reads
\eab
\label{sc.prop.sing}
\Delta_{\pm}(x,y,\Omega_{\pm})&=&-\frac{1}{4\pi^2(x-y)^2}
\left(1+\frac{\rho_{\pm}^2}{(x-x_\pm)^2}\right)^{-1/2}\times\nonumber\\ 
&&\left(1+\frac{\rho^2_{\pm}\sigma^{\mp}_\mu(x-x_{\pm})^\mu\sigma^{\pm}_\nu (y-x_{\pm})^\nu}
{(x-x_{\pm})^2(y-x_{\pm})^2}\right)\left(1+\frac{\rho_{\pm}^2}{(y-x_\pm)^2}\right)^{-1/2}\,
\eae
where $\sigma^{\pm}_\mu=(R_{ab}\,\sigma^b,\mp i)_\mu$, $R_{ab}\in\,$SO(3) is a (constant) rotation matrix 
in adjoint SU(2) color space, $\sigma^b\,,(b=1,2,3),$ denoting the 
Pauli matrices, $\rho_{\pm}$ the (anti)instanton radius, and $x_{\pm}$ is the center of the 
(anti)instanton. The $O(m)$ contribution in Eq.\,(\ref{prop-in-inst}), $S^1_{\pm}(x,y,\Omega)$, 
can simply be expressed as
\eqb
\label{O(m)}
S^1_{\pm}(x,y,\Omega)=m\int d^4z\,S^0_{\pm}(x,z,\Omega_{\pm}) S^0_{\pm}(z,y,\Omega_{\pm})\,.
\eqe
Inserting the zero-mode expression (\ref{zero-mode}), and the 
zeroth- (first)- order in $m$ expressions Eq.\,(\ref{O(m^0)}) (Eq.\,(\ref{O(m)})) 
into Eq.\,(\ref{instcurcorA}) and only considering the part, 
which survives the limit $m\to 0$, averaging over the color orientations of the 
instanton embedding into SU(3), subtracting the free current correlator $T_0$, performing 
the integration over (anti)instanton centers and radii over 
the remainder, and taking into account the contribution from 
instantons and antiinstantons in this part, one arrives at 
the following expression \cite{AndreiGross1978}
\eqb
\label{Tbeforerho}
\delta T(x)=(\sum_F Q_F^2)\frac{36}{\pi^2}\int \frac{d\rho}{\rho^5} D(\rho) 
\frac{\rho^4}{x^4}\pd_{x^2}\left(\frac{1}{x^2}\frac{1}
{(1+4\rho^2/x^2)^{1/2}}\,\log\frac{(1+4\rho^2/x^2)^{1/2}+1}{(1+4\rho^2/x^2)^{1/2}-1}\right)\,
\eqe
where $D(\rho)$ denotes the instanton density at one-loop perturbation 
theory (only gluonic fluctuations),
\eqb
\label{instdens}
D(\rho)\equiv\frac{0.1}{\rho^5}\left(\frac{8\pi^2}
{g^2(\Lambda_{QCD}\rho)}\right)^6\exp\left(\frac{-8\pi^2}{g^2(\Lambda_{QCD}\rho)}\right)\,
\eqe
which can be interpreted as the number of instantons of size between 
$\rho$ and $\rho+d\rho$ per unit space-time volume.
  
After separating off a factor $\frac{3}{4\pi^2} q^2$ (arising from the 
transverse tensor structure) in the Fourier transform of the entire correlator $T_0+\delta T(x,\rho)$ and 
after accounting for the Gaussian integration over fermionic fluctuations 
around the (anti)instanton 
we have 
\eab
\label{FTTbeforerho}
\bar{T}(Q^2)&=&-\frac{4\pi^2}{3Q^2}T(Q^2)\nonumber\\ 
&=&(\sum_F Q_F^2)\left(\log\frac{Q^2}{\mu^2}+16\pi^2\int\frac{d\rho}{\rho}
D(\rho)\Delta(\rho)\left[\frac{1}{3(Q\rho)^4}-\frac{1}{(Q\rho)^2}\int_0^1 
du\,K_2\left(\frac{2Q\rho}{\sqrt{1-u^2}}\right)\right]\right)\, 
\eae
where $\Delta(\rho)$ denotes the associated (dimensionless) fermion 
determinant \cite{'t Hooft1976}, and $K_2$ is a McDonald
function. The integral over $\rho$ is cut off at small $\rho$ by 
the fermion determinant. For large $\rho$ it is ill-behaved which signals that the 
dilute-gas approximation as well as the one-loop 
perturbative treatment of fluctuations breaks 
down. One usually introduces an upper cutoff $\rho_c\sim\Lambda_{QCD}^{-1}$ 
by hand. 

Clearly, there is a dimension-four power correction 
arising from the first part 
of the integrand in Eq.\,(\ref{FTTbeforerho}) 
which can be associated with the gluon condensate. The second part 
of the integrand, however, is not a power correction. For asymptotic 
momenta $Q\to\infty$ it falls off exponentially and can be taken 
as an indication for the searched-for exponentially small 
terms needed to cure the OPE. The reader may wonder why there is 
only a dimension-four power correction in Eq.\,(\ref{FTTbeforerho}). 
To answer this, let us recall that the appearance of 
dimension-six condensates is associated with 
radiative corrections - the prefactor $\alpha_s$ before the 
four-quark operators refers to a gluon exchange initiated by the current-induced 
quarks. On our above treatment of 
quark propagation, however, we did only consider radiative corrections to the background-field 
but not to the quark propagation in this background.

One may now continue Eq.\,(\ref{FTTbeforerho}) to the Minkowskian domain and determine 
its imaginary part to give a prediction for the ratio $R=-\frac{1}{\pi}\mbox{Im}(Q^2=-s-i0)$ 
in Eq.\,(\ref{Rpred}).

\begin{figure}[tb]
\begin{center}
\hspace{-4cm}
\begin{minipage}[t]{8 cm}
\epsfig{file=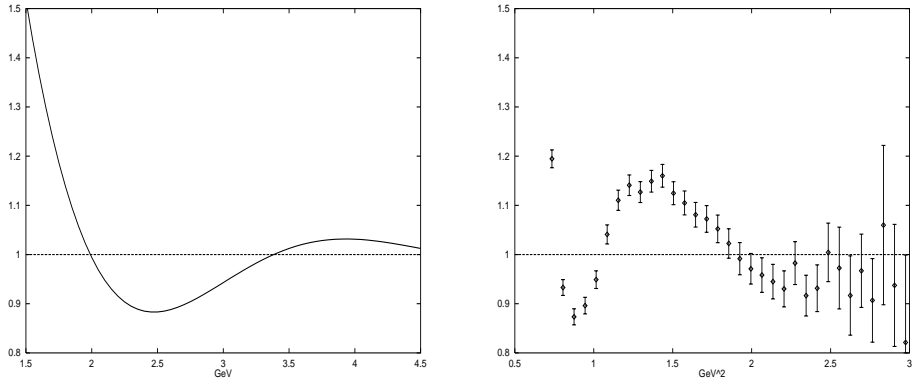,scale=.5}
\end{minipage}
\begin{minipage}[t]{16.5 cm}
\caption{Experimental results for $\bar{R}(E)$ (right panel) and the result of Eq.\,(\ref{continst}) 
obtained in the instanton model (left panel). The dashed line 
corresponds to the parton-model result. Plots taken from \cite{ChibisovDikemanShifman1996}. 
\label{Fig-R(E)}}
\end{minipage}
\end{center}
\end{figure}

The result is
\eab
\label{continst}
\bar{R}(E)&=&\frac{1}{\sum_F Q_F^2}\left( R_0(E)+R^I(E)\right)\nonumber\\ 
&=&1+\frac{16\pi^2}{3}\int\frac{d\rho}{\rho}
\frac{D(\rho)}{(E\rho)^2}\Delta(\rho)\left[\frac{1}{\rho^2}\delta(E^2)+
\frac{3}{2(E\rho)^2}\int_0^1 du\, J_2\left(\frac{2E\rho}{\sqrt{1-u^2}}\right)\right]\,
\eae
where $E\equiv\sqrt{s}$, $R_0(E)$ and $R^I(E)$ 
denote the free particle and the instanton induced parts, respectively, 
and $J_2$ is a Bessel function. 
In \cite{ChibisovDikemanShifman1996} the product $\frac{D(\rho)}{(E\rho)^2}\Delta(\rho)$ 
was approximated by the simplest possible form 
\eqb
\label{simpldens}
\frac{D(\rho)}{(E\rho)^2}\Delta(\rho)=d_0\rho_0\,\delta(\rho-\rho_0)\,
\eqe
where a value $\rho_0=1.15\,\mbox{GeV}^{-1}$ and $d_0=9\times10^{-2}$ was adopted
\footnote{These numbers are obtained by requiring that the instanton induced 
contribution to the semileptonic width of D-meson decay are 50\% of the 
parton-model prediction \cite{ChibisovDikemanShifman1996}.}. 
A comparison of the 
function $\bar{R}(E)$ in (\ref{simpldens}) and the 
experimental results for $\bar{R}(E)$ is presented in Fig\,\ref{Fig-R(E)}. 
It is obvious that the part in Eq.\,(\ref{continst}) not contained in 
the practical OPE is responsible for the 
resonance-like behavior at low $E$. Although quantitatively the two 
plots in Fig.\,\ref{Fig-R(E)} differ\footnote{Unfortunately, we have energy on the 
x-axis in the left panel and energy squared on the 
x-axis in the right panel. Even though this makes a direct comparison more cumbersome 
the author of the present review chose not to adapt the 
figures in \cite{ChibisovDikemanShifman1996}.}  - after all it is clear that an {\sl incomplete 
dilute-gas} approximation, recall the bold choice of the instanton weight 
in Eq.\,(\ref{simpldens}), is not a good approach - there is at least 
some qualitative agreement.    

\subsection{\it Duality analysis in the 't Hooft model \label{sec:2DM}}

QCD in two dimensions (QCD$_2$) 
considered in the limit $N_c\to\infty$ with $g^2 N_c$ fixed - 
the so-called 't Hooft model \cite{'t Hooft1974} - 
is exactly solvable. At finite but large $N_c$ a well controlled expansion in powers of $1/N_c$ is available. 
For this reason it is the ideal testing ground for questions on 
local quark-hadron duality, namely, at large external momenta the practical OPE of some 
polarization operator (current correlator) can directly be 
compared with the asymptotically exact result, and duality violating contributions 
can be identified. A vast literature exists on the subject, see for example 
\cite{CallanCooteGross1976,Einhorn1976,
EinhornNussinovRabinovici1976,Zhitnitsky1985,Burkardt1992,GrinsteinLebed1997,GrinsteinLebed1999,
BlokShifmanZhang1998,BigiShifmanUraltsevVainshtein1999,
BigiUraltsev1999,Grinstein2001,RozowskyThorn2001,BerrutoGiustiHoelbingRebbi2002}, and not all contributions 
can be explicitly referred to here. A review, which also discusses the string interpretation 
of the 't Hooft-model results, exists \cite{AbdallaAbdalla1996}. During the ten years or so 
the interest in the 't Hooft model was boosted by questions of duality-violations in the weak 
decay of heavy quark flavors, see for example \cite{GrinsteinLebed1997,
BigiShifmanUraltsevVainshtein1999,GrinsteinLebed1999,BigiUraltsev1999,
BigiShifmanUraltsevVainshtein1999,Grinstein2001}, 
by the necessity to check the reliability of lattice calculations, see 
\cite{RozowskyThorn2001,BerrutoGiustiHoelbingRebbi2002}, 
by the need to estimate higher-twist corrections to
parton distribution functions \cite{Burkardt1992}. A dynamical 
understanding of chiral symmetry breaking in two dimensions was obtained relatively early 
\cite{Zhitnitsky1985}. In this section we are mainly concerned 
with duality violations in spectral functions based on the OPEs of current correlators when 
allowing for $1/N_c$ corrections.

\subsubsection{Prerequisites}

Before going into the technical details of duality violations 
in the 't Hooft model we will here give a brief introduction 
into this model \cite{'t Hooft1974}. 

\begin{figure}[tb]
\begin{center}
\hspace{-2.5cm}
\begin{minipage}[t]{8 cm}
\epsfig{file=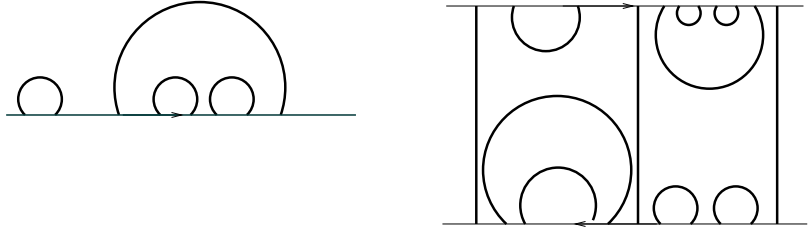,scale=.5}
\end{minipage}
\begin{minipage}[t]{16.5 cm}
\caption{Planar diagrams contributing to the quark self-energy and the quark-antiquark scattering amplitude. 
Thick lines denote gluon propagators (not dressed in the limit $N_c\to\infty$, $g^2 N_c$ fixed) and 
thin, arrowed lines propagators refer to quark propagators.
\label{Fig-Plan}}
\end{minipage}
\end{center}
\end{figure}

\begin{figure}[tb]
\begin{center}
\hspace{0.0cm}
\begin{minipage}[t]{8 cm}
\epsfig{file=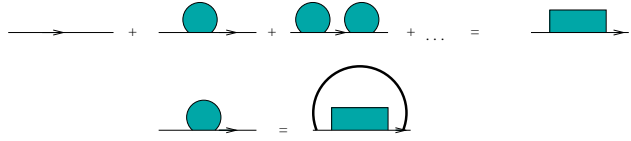,scale=.5}
\end{minipage}
\begin{minipage}[t]{16.5 cm}
\caption{The Dyson series for the quark self-energy (upper figure) and the equation which 
determines it in the limit $N_c\to\infty$, $g^2 N_c$ fixed. 
\label{Fig-Prop}}
\end{minipage}
\end{center}
\end{figure}

One considers the usual QCD Lagrangian 
\eqb
\label{QCD2}
{\cal L}=-\frac{1}{4}G^a_{\mu\nu}G_a^{\mu\nu}+\sum_F \bar{q}^F(i\gamma^\mu D_\mu-m_F)q^F\,.
\eqe
Spacetime is two dimensional, the gauge group is U($N_c$) instead of SU($N_c$), and the gauge coupling $g$ has 
the dimension of a mass. Conveniently, one works in light-cone coordinates 
$x^{\pm}=\frac{1}{\sqrt{2}}\left(x^1\pm x^0\right)$ and 
$A_{\pm}=\frac{1}{\sqrt{2}}\left(A_1\pm A_0\right)$ where $x^1=x_1$ and $x^0=-x_0$. 
Imposing the (ghost-free) light-cone gauge $A_{-}=A^+=0$, one 
has $G_{+-}=-\partial_{-}A_{+}$ and Eq.\,(\ref{QCD2}) 
reduces to
\eqb
\label{QCD2gf}
{\cal L}=\frac{1}{4}(\pd_{-}A_{+}^a)^2+
\sum_F \bar{q}^F(i\gamma^\mu \pd_\mu-m_F+g\gamma_{-}A_{+})q^F\,.
\eqe
Taking $x^+$ as the new time direction, the field $A_{+}^a$ has no time-derivatives, 
and thus it is not dynamical. It will provide for a static Coulomb force between the quarks. 
Since $\gamma_{-}^2=\gamma_{+}^2=0$, 
$\gamma_{+}\gamma_{-}+\gamma_{-}\gamma_{+}=2$, and since the 
vertex in Eq.\,(\ref{QCD2gf}) comes with a $\gamma_{-}$ one can eliminate the gamma 
matrices from the Feynman rules. Suppressing color indices, the 
gluon propagator is $1/k^2_{-}$, the quark propagator is $k_{-}/(2k_{+}k_{-}-m_F^2+i\epsilon)$, 
and the vertex is $2g$. It was shown in \cite{'t Hooft19742} that in the limit 
$N_c\to\infty$ with $g^2 N_c$ fixed only {\sl planar} diagrams with no fermion loops like the 
ones in Fig.\,\ref{Fig-Plan} survive in the calculation 
of any amplitude. 
Due to this extreme simplification 
the equation for the quark self-energy (the rectangular blob in Fig.\,\ref{Fig-Prop}) 
can be written in untruncated 
form. 

\begin{figure}[tb]
\begin{center}
\hspace{-1cm}
\begin{minipage}[t]{8 cm}
\epsfig{file=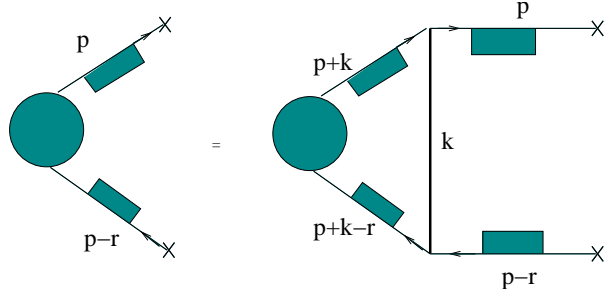,scale=.5}
\end{minipage}
\begin{minipage}[t]{16.5 cm}
\caption{Homogeneous Bethe-Salpeter equation which determines the spectrum of meson states 
in the limit $N_c\to\infty$, $g^2 N_c$ fixed. 
\label{Fig-Wavefunction}}
\end{minipage}
\end{center}
\end{figure}

To solve this equation requires, in intermediate steps, 
the introduction of a symmetric ultra-violet cutoff as well 
as an infra-red cutoff. The former is a consequence of the strong gauge fixing and 
has no physical interpretation. The removal of the latter in the final result, 
which does not depend on the ultra-violet
cutoff, shifts the 
pole of the quark-propagator to infinity - one concludes that the spectrum has no 
single quark state. To determine the spectrum of 
quark-antiquark bound states one has to look at the homogeneous 
Bethe-Salpeter equation as depicted in Fig.\,\ref{Fig-Wavefunction}. 

Exploiting that the Coulomb force is instantaneous to separate the loop integrals 
and introducing the following dimensionless 
quantities (compare with Fig.\,\ref{Fig-Wavefunction})
\eqb
\label{dimless}
\gamma^2\equiv m^2/\mu^2\,,\ \ \ \ \mu_n^2\equiv m_n^2/\mu^2\,,\ \ \ \ x\equiv p_{-}/r_{-}\,,
\eqe
where $\mu\equiv (g^2N_c)/\pi$ and $m_n$ is the mass of the $n$th meson, one obtains 
the 't Hooft equation for the mesonic wave functions $\phi_n(x)$
\eqb
\label{tHoofteq}
\mu_n^2\phi(x)=\frac{(\gamma^2-1)\phi_n(x)}{x(1-x)}-P\int_0^1 dy\, \frac{\phi_n(y)}{(x-y)^2}\,
\eqe
with $P$ denoting principle-value integration, $P\left[1/(x-y)^2\right]\equiv\lim_{\epsilon\to 0}
\frac{1}{2}[1/(x-y+i\epsilon)^2+1/(x-y-i\epsilon)^2]$. In writing Eq.\,(\ref{tHoofteq}) 
we have assumed the masses of the participating quark 
and antiquark to be equal, $m_F=m_{\bar{F}}=m$.
It was shown in \cite{'t Hooft1974} that the ``Hamiltonian'' 
defined by the right-hand side of Eq.\,(\ref{tHoofteq}) is hermitian and positive 
definite (finite quark mass, $\gamma>0$) on the Hilbert space of 
functions which vanish at x=0,1 like $x^{\beta}, (1-x)^{\beta}$ where $\beta$ is a root of 
$\pi\beta\cot(\pi\beta)=1-\gamma$. The spectrum is discrete and is for large $n$ approximated 
by $\phi_n\sim\sqrt{2}\sin(\pi nx)$ and $\mu_n^2=\pi^2 n$. For $\gamma=0$ the lowest meson 
mass vanishes, the associated wave function is $\phi_0=1$.

Since the spectrum is real and positive definite resonances do not 
decay into one another - their width is zero. At large $s$ (or $n>n_0$) 
the spectral function of a correlator $T$ of currents, 
which couple to all meson states equally, is therefore well approximated by 
equidistant, zero-width spikes:
\eqb
\label{sfQCD2}
\rho(s)=\mbox{Im}\,T=\mbox{const}\frac{N_c}{2}
\sum_{n>n_0}^\infty \delta\left(\frac{s}{\pi^2\mu}-n\right)\,.
\eqe

\subsubsection{Current-current correlator and spectral function at $O(1/N_c)$}

Our discussion of local duality violation of current-current correlators in QCD$_2$ 
beyond the limit $N_c\to\infty$ 
relies on work \cite{BlokShifmanZhang1998} which uses the older results in 
\cite{CallanCooteGross1976,Einhorn1976,
EinhornNussinovRabinovici1976}. 

At finite $N_c$ the quark-antiquark bound states of the 't Hooft model are 
unstable. For a decay $a\to b+c$ the width $\Gamma_a$ at $O(1/N_c)$ 
is given as \cite{CallanCooteGross1976,Einhorn1976,EinhornNussinovRabinovici1976}
\eqb
\label{width QCD2}
\Gamma_a=\frac{1}{8m_a}\sum_b\sum_c\frac{g_{abc}^2}{\sqrt{I(m_a,m_b,m_c)}}\,,
\eqe
where $I(m_a,m_b,m_c)=1/4[m_a^2-(m_b+m_c)^2][m_a^2-(m_b-m_c)^2]$ and the meson 
coupling is given as
\eqb
\label{mesoncouplQCD2}
g_{abc}=32\mu^2\sqrt{\frac{\pi}{N_c}}\left[1-(-1)^{\sigma_a+\sigma_b+
\sigma_c}\right](f_{abc}^{+}+f_{abc}^{-})\,.
\eqe
The parity of the $a$th 
meson is $\sigma_{a}$. 
The quantities $f_{abc}^{\pm}$ are constants for on-shell decay. They are given 
by overlap integrals between the meson wave functions $\phi_{a,b}$ and 
the Bethe-Salpeter kernel, see \cite{BlokShifmanZhang1998}. 
For lack of better analytical knowledge Eq.\,(\ref{width QCD2}) was used in 
\cite{BlokShifmanZhang1998} with the 
asymptotic spectrum, even for $a,b,c\le n_0$. For small $a$ one 
should not trust this approximation. For an investigation of duality violating components at 
large $Q$ it is, however, justified. A numerical evaluation of 
Eq.\,(\ref{width QCD2}) and a subsequent fit to a square-root dependence 
yields the following estimate
\eqb
\label{widthfitQCD2}
\Gamma_n=\frac{(A=15\pm 1.5)\mu}{\pi^2N_c}\sqrt{n}\left[1+O(1/n)\right]\,.
\eqe
This knowledge of the dependence of width on $n$ can 
be exploited to estimate the correlator $T(q^2)=i\int d^2x\,\e^{iqx}\la Tj(x)j(0)\ra$ 
of the scalar current $j=\bar{q}q$. In the limit $N_c\to\infty$ it is given as 
\eqb
\label{corQCD2}
T(q^2)=-\sum_{n=0}^\infty\frac{g_n^2}{q^2-m_n^2+i\epsilon}\,.
\eqe
Requiring this to match the leading perturbative 
order for $-q^2=Q^2\to\infty$ (duality in mesonic and quark 
description of asymptotic freedom in 
$T(Q^2\to\infty)$), one obtains \cite{CallanCooteGross1976}
\eqb
\label{gnQCD2}
g_n^2=N_c\pi\mu^2\,, \ \ \ \ \ \ \ \ (n\ \mbox{odd})\,.
\eqe
\begin{figure}[tb]
\begin{center}
\hspace{-1cm}
\begin{minipage}[t]{8 cm}
\epsfig{file=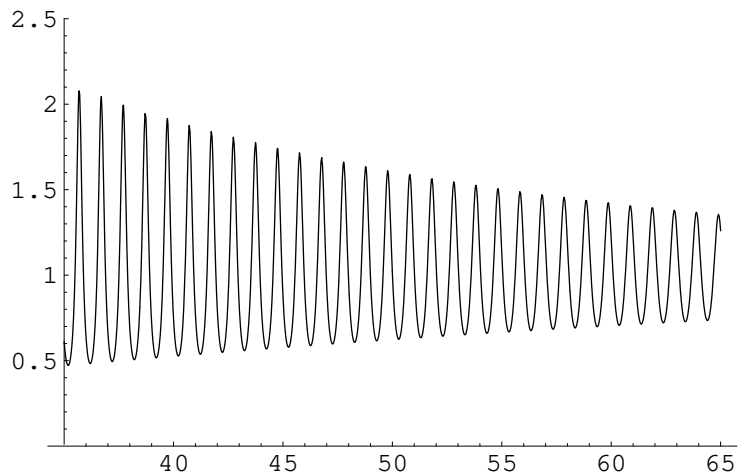,scale=1.}
\end{minipage}
\begin{minipage}[t]{16.5 cm}
\caption{The spectral density $\mbox{Im}\,T(s)$ in the 't Hooft model 
with the leading $1/N_c$ contribution to the resonance widths included. The center-of-mass 
momentum $s$ is given in units of $2\pi^2\mu^2$. Plot taken from \cite{BlokShifmanZhang1998}. 
\label{Fig-sdQCD2}}
\end{minipage}
\end{center}
\end{figure}
Inserting Eq.\,(\ref{gnQCD2}) into 
Eq.\,(\ref{corQCD2}) and taking the imaginary part, 
one arrives at an expression like in Eq.\,(\ref{sfQCD2}). 
By means of a dispersion relation one can define Im\,$T(q^2>0)$ to $T(q^2)$ everywhere 
in the complex $q^2$-plane, up to a constant. This gives \cite{BlokShifmanZhang1998} 
\eqb
\label{dispTQCD2}
T(q^2)-T(0)=-\frac{N_c}{2\pi}\,\psi(\sigma)\,,\ \ \ 
\ \sigma=\frac{Q^2}{2\pi^2\mu^2}+\frac{1}{2}\,
\eqe
where $\psi(\sigma)$ denotes the logarithmic 
derivative of Euler's Gamma function. So far we have discussed 
the narrow-width case ($N_c\to\infty$). The finite-width case is treated in 
the Breit-Wigner approach 
by replacing $q^2-m_n^2+i\epsilon\to q^2-m_n^2+\Sigma(q^2)$ 
in Eq.\,(\ref{corQCD2}). Since 
$\mbox{Im}\,\Sigma(q^2=m_n^2)=m_n\Gamma_n=\frac{Am_n^2}{\pi^3N_c}$ 
one may take $\Sigma(Q^2)=\frac{A}{\pi^4N_c}Q^2\log\frac{Q^2}{2\pi^2\mu^2}$. 
As a result, the correlator with $1/N_c$ corrections in the widths included reads
\eqb
\label{dispTQCD21/N} 
T(q^2)-T(0)=-\frac{1}{1-A/(\pi^4N_c)}\frac{N_c}{2\pi}\,\psi(\tilde{\sigma})\,, \ \ \ \ 
\tilde{\sigma}=\left(\frac{Q^2}{2\pi^2\mu^2}\right)^{-A/(\pi^4 N_c)+1}+\frac{1}{2}\,.
\eqe
Fig.\,\ref{Fig-sdQCD2} shows the spectral density $\mbox{Im}\,T(s)$. 
The broadening of the resonances with increasing $s$ is 
clearly visible. How does this result compare 
with the practical OPE? The OPE corresponding 
to Eq.\,(\ref{dispTQCD21/N}) is known to be the following (asymptotic) expansion 
\cite{BlokShifmanZhang1998}
\eqb
\label{OPEdispTQCD21/N}  
T(q^2)-T(0)=-\frac{1}{1-A/(\pi^4N_c)}\frac{N_c}{2\pi}\,\left[\log\tilde{\sigma}+\frac{1}{\tilde{\sigma}}-
\sum_{n=1}(-1)^{n-1}\frac{B_n\tilde{\sigma}^{-2n}}{2n}\right]\,.
\eqe
The Bernoulli numbers behave at large $n$ as $B_n\sim (2n)!$ 
which shows that the expansion is asymptotic. The variable 
$\tilde{\sigma}$ can be obtained from $\sigma$ 
by replacing $Q^2\to z$. Consequently, the powers in $Q^{-2}$ 
in Eq.\,(\ref{OPEdispTQCD21/N}) are slightly
displaced at finite $N_c$ as compared to their 
integer values at $N_c\to\infty$. Expanding in this deviation $\alpha=\frac{A}{\pi^4N_C}$, 
this introduces logarithmic corrections: 
\eqb
\label{logsQCD2}
\left(1/Q^2\right)^{2n-\alpha}
\to\left(1/Q^2\right)^{2n}(1+\alpha\log Q^2+\cdots)\,. 
\eqe
The logarithms in Eq.\,(\ref{logsQCD2}) lead 
to {\sl smooth} contributions to the spectral density at {\sl finite} $s$ 
which, however, vanish in the 
limit $N_c\to\infty$. A direct (not using the asymptotic expansion (\ref{OPEdispTQCD21/N})) 
calculation of the spectral density reveals an oscillating component 
$2\exp[-\alpha s/\mu^2]\cos[s/(\pi\mu^2)]$ for $\alpha s\gg \mu^2$ 
which, however, is not suppressed by $1/N_c$. This contribution 
signals the violation of local quark-hadron
duality.

\subsubsection{Decays of heavy mesons}

As a final testing ground for the study of the 
violation of local quark-hadron duality in the 't Hooft model we 
will now discuss the weak decay of heavy-light mesons. There has been a rather high research 
activity within the last few years investigating the issue of both global, see for example 
\cite{GrinsteinLebed1997}, and local duality, 
see for example \cite{BigiShifmanUraltsevVainshtein1999,BigiUraltsev1999,GrinsteinLebed1999,Grinstein2001}. 
This is natural since important lessons for the (3+1) dimensional case, 
which is targeted by present experiments at the $B$ factories, 
can be drawn. 

\begin{figure}[tb]
\begin{center}
\hspace{-2cm}
\begin{minipage}[t]{8 cm}
\epsfig{file=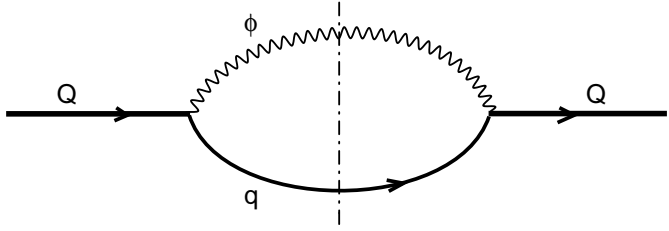,scale=0.7}
\end{minipage}
\begin{minipage}[t]{16.5 cm}
\caption{Lowest-order diagram for the transition amplitude in Eq.\,(\ref{DecGQCD2}). The wave line 
represents the propagation of the pseudoscalar massless particle which is 
composed of $\bar{\psi}_a$ and $\psi$. Taken from \cite{BigiShifmanUraltsevVainshtein1999}. 
\label{Fig-TransQCD2}}
\end{minipage}
\end{center}
\end{figure}

Our discussion mainly follows \cite{BigiShifmanUraltsevVainshtein1999}. 
To describe weak decays the 't Hooft-model Lagrangian has to 
be supplemented by a weak-interaction part. 
Effectively, we can take it to be of the 
current-current form
\eqb
\label{ccwiQCD2}
{\cal H}_{eff}^V=\frac{G}{\sqrt{2}}\left(\bar{q}\gamma_\mu Q\right)
\left(\bar{\psi}_a\gamma^\mu\psi_b\right)\,
\eqe
for two vector currents since the axial current reduces 
to the vector current in (1+1) dimensions. The constant $G$ is the 2D analogue of the Fermi coupling in 4D. 
By means of the optical theorem the 
inclusive decay with $\Gamma_{H_Q}$ reads
\eqb
\label{DecGQCD2}
\Gamma_{H_Q}=\mbox{Im}\frac{i}{M_{H_Q}}\int d^2x \la H_Q|T {\cal H}_{eff}^V(x){\cal H}_{eff}^V(0)|H_Q\ra\,
\eqe
where $H_Q$ denotes the state with the heavy meson at rest. 
Depending on whether we describe semileptonic or hadronic decay the fields ${\psi}_{a,b}$ in 
Eq.\,(\ref{ccwiQCD2}) are either leptonic fields or quark fields. We only discuss the case where 
$m_\psi=0$. As was shown in \cite{BigiShifmanUraltsevVainshtein1999}, 
the current $\bar{\psi}_a\gamma^\mu\psi_b$ can be substituted by 
$\epsilon^{\mu\nu}\pd_\nu\phi/\sqrt{\pi}$ where $\phi$ denotes a pseudoscalar, massless field
\footnote{In the leptonic as well as in the case where $\psi_{ab}$ are quark fields 
their polarization operator is exact at one-loop level owing to the facts that 
$\gamma^\alpha\gamma^\mu\gamma_\alpha=0$ and that 
a product over an odd number of gamma matrices reduces to one. There is a pole at 
$s=0$ corresponding to a massless particle. In the case of quark fields it is 
known that in the 't Hooft model the only nonvanishing meson-current coupling of the 
vector current is the one to the massless meson.}. The corresponding 
lowest-order diagram for the transition amplitude is shown in Fig.\,\ref{Fig-TransQCD2}. 
The parton-model result is
\eqb
\label{partonQCD2}
\Gamma_Q=\frac{G^2}{4\pi}\frac{m_Q^2-m_q^2}{m_Q}\,
\eqe
and can be obtained by replacing the heavy meson by the heavy 
quark to be able to {\sl calculate} the average 
over the operator $T^0=c^0_{\bar{Q}Q}\bar{Q}Q$ with $2\mbox{Im}\,c^0_{\bar{Q}Q}=\Gamma_Q$. 
Going beyond the parton-model approximation, the (spacetime integral of the) 
operator $\bar{Q}Q$ needs to be expanded in powers of $1/m_Q$. Up to $O(1/m_Q^3)$ one obtains 
\cite{BigiShifmanUraltsevVainshtein1999}
\eqb
\label{barQQQCD2}
\frac{\la H_Q|\bar{Q}Q|H_Q\ra}{2M_{H_Q}}=1-
\frac{1}{2m_Q^2}\frac{\la H_Q|\bar{Q}(-D_1^2)Q|H_Q\ra}{2M_{H_Q}}+\frac{g^2}{2m_Q^3}
\frac{\la H_Q|\bar{Q}\gamma_\mu t^a\sum_q\bar{q}\gamma^\mu t^a q|H_Q\ra}{2M_{H_Q}}+
O\left(\frac{1}{m^4_Q}\right)\,.
\eqe
Including first-order in $g^2$ radiative corrections to the coupling $Qq\phi$ 
simply amounts to a shift of the quark masses: 
$m_{Q,q}^2\to m_{Q,q}^2-\beta^2$ where $\beta^2=\lim_{N_c\to\infty}\frac{g^2}{2\pi}N_c<\infty$ 
denotes the 't Hooft coupling. This is actually true to all orders \cite{BigiShifmanUraltsevVainshtein1999}. 
As a consequence, we have to all orders that $2\,\mbox{Im}\,c_{\bar{Q}Q}=\frac{G^2}{4\pi}\frac{m_Q^2-m_q^2}{\sqrt{m_Q^2-\beta^2}}$. 
Under radiative corrections the expansion of $c_{\bar{Q}Q}\bar{Q}Q$ 
up to $O(1/m^4_Q)$ thus is entirely due to the one of the 
operator $\bar{Q}Q$, see Eq.\,(\ref{barQQQCD2}). As far as the first subleading, dimension two, four-fermion 
operator $O_{4q}=\bar{Q}\Gamma_1 Q\bar{q}\Gamma_2q$ - $\Gamma_{1,2}$ denoting color and spinor matrices - 
is concerned\footnote{In contrast to (3+1) dimensional QCD, 
where the operator $g^2\bar{Q}\sigma^{\mu\nu}G_{\mu\nu}Q$ is the first subleading operator in 2D 
it has dimension four and thus is not the first subleading operator.} 
it can be shown that its Wilson coefficient $c_{4q}$ does not develop an 
imaginary part at leading order in $g$ and at one loop. Thus this operator does not contribute 
to the total width up to $O(1/m_b^3)$. There is a contribution of $O(1/m_b^5)$ at two loop, 
and hence we have up to $O(1/m_b^4)$
\eqb
\label{totalwidthOPEQCD2}
\Gamma_{H_Q}=\frac{G^2}{4\pi}\frac{m_Q^2-m_q^2}{\sqrt{m_Q^2-\beta^2}}\left[
\frac{\la H_Q|\bar{Q}Q|H_Q\ra}{2M_{H_Q}}+O\left(\frac{1}{m_Q^5}\right)\right]\,.
\eqe
Eq.\,(\ref{totalwidthOPEQCD2}) can be compared with the following 
result of a calculation obtained by using the 't Hooft equation (now with $m_Q$ and the mass of the spectator 
quark $m_{sp}\not=m_Q$) \cite{BigiShifmanUraltsevVainshtein1999}
\eqb
\label{tHtotalwidthOPEQCD2}
\Gamma_{H_Q}=\frac{G^2}{4\pi}\frac{m_Q^2-m_q^2}{m_Q}\left[\frac{m_Q}{M_{H_Q}}
\int_0^1 \frac{dx}{x}\,\phi^2_{H_Q}(x)+O\left(\frac{1}{m_Q^5}\right)\right]\,.
\eqe
Writing the operator $\bar{Q}Q$ in terms of quark-components in the 
light-cone formalism and absorbing renormalization factors, 
the matrix element in Eq.\,(\ref{totalwidthOPEQCD2}) 
can be expressed as a functional of $\phi_{H_Q}$ such that Eq.\,(\ref{tHtotalwidthOPEQCD2}) 
is reproduced. Up to $O(1/m_Q^4)$ the OPE prediction matches the expansion of the 
exact result. Also, the absence of a $1/m_b$ correction, found by analyzing the OPE, 
can directly be verfied using the HQE of the 't Hooft equation in the 
approach of Ref.\,\cite{Burkardt1992D}. 

\begin{figure}[tb]
\begin{center}
\hspace{-2cm}
\begin{minipage}[t]{8 cm}
\epsfig{file=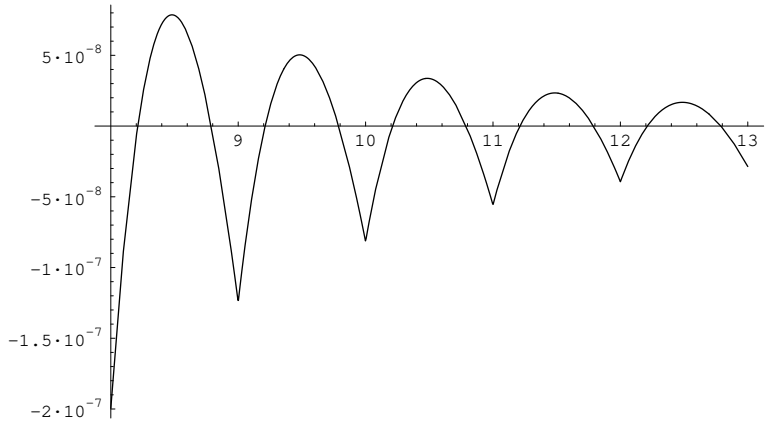,scale=0.7}
\end{minipage}
\begin{minipage}[t]{16.5 cm}
\caption{The oscillatory part of $\Gamma_{H_Q}$ represented as the 
ratio $\Delta\Gamma^{osc}/(G^2\beta)$ versus the ratio $M^2_{H_Q}/(\pi^2\beta^2)$. 
Taken from \cite{BigiShifmanUraltsevVainshtein1999}.\label{Fig-Delta-osc}}
\end{minipage}
\end{center}
\end{figure}

To derive Eq.\,(\ref{tHtotalwidthOPEQCD2}) a sum over exclusive widths $\Gamma_n$ 
for the decay into the $n$th meson had to be performed from which the off-shell 
part ($m_n>M_{H_Q}$) had to be subtracted
afterwards. It can be estimated 
that the analytic part of the latter term is $O(1/m_Q^5)$.
If there is a nonanalytic violation of local duality $\Delta\Gamma^{osc}$ it must 
reside in the sum with $m_n>M_{H_Q}$. An estimate of $\Delta\Gamma^{osc}$ was made in 
\cite{BigiShifmanUraltsevVainshtein1999} for $m_{sp}\le\beta$. 
In Fig.\,\ref{Fig-Delta-osc} $\Delta\Gamma^{osc}$ is shown 
as a function of the heavy meson mass. 
The relative amplitude of the oscillatory part can be estimated as
\eqb
\label{relAmQCD2}
\left|\frac{\Delta\Gamma^{osc}}{\Gamma_Q}\right|_{\tiny\mbox{max}}\sim \frac{3\pi^4}{2}
\left(\frac{\beta}{M_{H_Q}}\right)^9\,,
\eqe
and hence it is strongly power-suppressed in the weak-coupling limit $M_{H_Q}\gg\beta$. 

To conclude, there is definitely an oscillatory, duality violating component 
in the inclusive decay width for the semileptonic or hadronic 
decay of a heavy meson in QCD$_2$ considered in the limit 
$N_c\to\infty$, $g^2 N_c$ fixed. It is strongly power-suppressed for $M_{H_Q}\gg\beta$.

\section{OPE and nonperturbative nonlocality\label{sec:PhenNN}} 

In the last section we used field-theory models, which are not too 
far from first-principle, realistic QCD, to pin down the terms 
absent in practical OPEs that lead to a violation of 
local quark-hadron duality. This section approaches 
the problem from a more phenomenological side. By relating moments of hadronic 
light-cone wave functions (also called distribution amplitudes (DAs)) to the OPE representation 
of an appropriate vacuum correlator, thereby gainig experimentally testable 
information on the wave functions in terms of perturbative QCD and condensates 
\cite{ChernyakZhitnitsky1977,ChernyakZhitnitsky1981}, a phenomenological analysis of the validity of the OPE can be performed. 
Allowing for deviations from the locality of condensates - a possibility 
discussed since the early days of QCD sum rules\cite{Gromes1982,
DoschSimonov1988,BalitskyBraun1990,MikhailovRadyushkin1992,Hofmann12001,Hofmann22001,
Hofmann32001,BakulevMikhailov2002,HoangHofmann2003,BakulevMikhailovStefanis2003} - strongly 
suggests that a truncated, practical OPE is insufficient to reproduce the experimentally 
favored shapes of DAs \cite{MikhailovRadyushkin1992}. The same conclusion can be reached by a comparison of a re-ordering of the OPE, 
which sums up an infinite series of operators of 
increasing powers in covariant derivatives, with experimental data. 

\subsection{\it Distribution amplitudes and photon-photon annihilation}

Before going into the details of the sum-rule approaches of \cite{ChernyakZhitnitsky1981} and 
\cite{MikhailovRadyushkin1992} to the determination of DAs 
we will set the stage by explaining and applying them in (3+1) dimensions. In the 2D case 
we have already encountered them in our discussion of the 't Hooft model.    

\begin{figure}[tb]
\begin{center}
\hspace{-2.5cm}
\begin{minipage}[t]{8 cm}
\epsfig{file=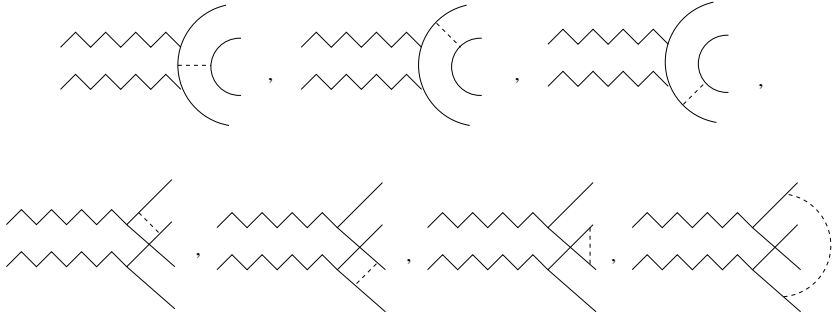,scale=0.4}
\end{minipage}
\begin{minipage}[t]{16.5 cm}
\caption{Diagrams contributing to $T_{\lambda\lambda^\prime}$ of Eq.\,(\ref{ggpp}) to lowest order in
$\alpha_s$. Wiggly lines denote photons, a solid line a quark or an 
antiquarks, and a dashed line is a gluon.\label{Fig-HSC2}}
\end{minipage}
\end{center}
\end{figure}

In the calculation of exclusive processes with a large momentum 
transfer - like the hadronic 
decay of heavy mesons or the photon-photon annihilation into hadrons - mesonic 
DAs appear quite naturally. They were introduced to separate 
nonperturbative large-distance physics from 
the perturbatively accessible small-distance physics in the calculation of 
transition amplitudes \cite{LepageBrodsky1979,EfremovRadyushkin1980,LepageBrodsky1980} 
(an exhaustive discussion is given in \cite{EfremovRadyushkin1980}). In general, the DA of a hadron decomposes 
into a part describing the valence quark and/or 
gluon content and into parts associated with the 
contribution of higher, color-singlet Fock states. For example, the $\pi^+$  
wave function $\vp(x,\mu^2)$ associated with 
the lowest-twist, nonlocal operator
 and with the lowest Fock state 
is defined as
\eab
\label{pionWF}
&&\la 0\left|\bar{d}(z)\gamma_\nu\gamma_5{\cal P}
\exp\left[ig\int_{-z}^{z}dy^\mu\,A_\mu\right]u(-z)\right|\pi^+(q)\ra_{\mu^2}\nonumber\\ 
&=&\sum_{n=0}^\infty\frac{(-1)^n}{n!}
\la 0\left|\bar{d}(z)\gamma_\nu\gamma_5\left(z^\mu[
\overrightarrow{D}_\mu-\overleftarrow{D}_\mu]\right)^n u(0)\right|\pi^+(q)\ra_{\mu^2}\nonumber\\ 
&=&iq_\nu\,\phi(z\cdot q,\mu^2)+\cdots\,,\nonumber\\ 
\phi_\pi(z\cdot q,\mu^2)&=&\int_{-1}^{+1}d\xi\,\e^{i\xi(z\cdot q)}\vp_\pi(\xi,\mu^2)\,
\eae
where the separation $2z$ is light-like, $z^2=0$, 
the sum is over even $n$, and ${\cal P}$, $\mu$ denote path ordering and the normalization point of 
the operators, respectively. 
In Eq.\,(\ref{pionWF}) only the function $\phi_\pi(z\cdot q,\mu^2)$ associated with 
the {\sl valence} quark distribution is explicitly indicated, 
paranthesis refer to higher Fock states. Writing $\xi=2x-1$, 
the quantity $x$ or $(x-1)$ 
is the fraction of {\sl longitudinal} pion momentum 
carried by the quark or antiquark, respectively. The DA $\vp_\pi(\xi,\mu^2)$ is interpreted 
as the integral of the full Bethe-Salpeter wave function over {\sl transverse} momenta $k_{\perp}$ 
down to $k_{\perp}\sim\mu$. The dependence on the cutoff $\mu$ in $\vp_\pi(\xi,\mu^2)$ 
can most conveniently be obtained by expanding into eigenfunctions of the evolution kernel obtained 
by considering one-gluon-exchange ladder diagrams which contribute perturbatively to the wave function \cite{LepageBrodsky1979}. 
The expansion is given as
\eqb
\label{expandWF}
\vp_\pi(\xi,\mu^2)=
(1-\xi^2)\sum_{n=0}^\infty a_n C_n^{3/2}(\xi)\left[\log\left(\frac{Q^2}{\Lambda_{QCD}^2}\right)\right]^{-\gamma_n}\,
\eqe
where the eigenvalue $\gamma_n$ is given as 
$\gamma_n=\frac{4}{3b}\left[1-2/((n+1)(n+2))+4\sum_{l=2}^{n+1}1/l\right]$, and $b=11-\frac{2}{3}$ is the
first coefficient of the QCD beta function. Due to the orthonormality (with weight $(1-\xi^2)$) of the Gegenbauer
polynomials $C_n^{3/2}$ on [-1,1] the expansion coefficients 
are simply given as $a_n=\int_{-1}^{1}d\xi\,C_n^{3/2}(\xi)\vp(\xi,\mu_0^2)$. 
\begin{figure}[tb]
\begin{center}
\hspace{2cm}
\begin{minipage}[t]{8 cm}
\epsfig{file=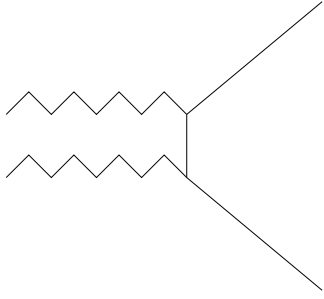,scale=0.6}
\end{minipage}
\begin{minipage}[t]{16.5 cm}
\caption{Diagram contributing to the hard scattering amplitude in $\gamma\gamma^*\to\pi^0$ to lowest order in
$\alpha_s$. Wiggly lines denote photons, a solid line a quark or an 
antiquark.\label{Fig-HSC1}}
\end{minipage}
\end{center}
\end{figure}
Let us now discuss two applications of DAs. 
We first consider the annihilation of two photons 
into hadrons, $\gamma\gamma\to X_h$. One of the photons either is 
highly virtual or it is real but incident
under a large angle $\theta_{c.m.}$ in the center-of-mass frame.  This is described in terms of a convolution of a 
hard scattering amplitude 
$T_H$, calculable in perturbation theory (see Figs.\,\ref{Fig-HSC1},\ref{Fig-HSC2}), 
with the DAs of the hadrons produced. For example, for the process 
$\gamma\gamma^*\to\pi^0$ 
this arises by considering a partial, perturbative contraction of quark fields in the 
amplitude $\la\pi^0|j_\mu(y) j_\mu(0)|0\ra$, yielding propagator lines to which the external 
photons couple, and a soft 
remainder which in lowest-order twist 
is given by Eq.\,(\ref{expandWF}). The vertex $\Gamma_\mu$ 
for this process is
\eqb
\label{vgaga}
\Gamma_\mu=-ie^2F_{\pi\gamma}(Q^2)\,\ep_{\mu\nu\rho\sigma}p^\nu\epsilon^\rho q^\sigma\,
\eqe
where $p$ is the pion's momentum, $\epsilon^\rho$ the 
polarization vector of the real photon, $q$ (with $q^2=-Q^2$) 
the momentum of the virtual photon, and $F_{\pi\gamma}(Q^2)$ 
the photon-pion transition form-factor. The latter can be written as 
the convolution of hard scattering amplitude (see Fig.\,\ref{Fig-HSC1}) 
with the pion wave function. To zeroth order in $\alpha_s$ 
we have \footnote{Abusing the notation, we denote the different
functional dependences of the wave function on $\xi$ and $x$ by the same symbol $\vp_\pi$.}
\eqb
\label{ggp}
F_{\pi\gamma}(Q^2)=\frac{2}{\sqrt{3} Q^2}\int_0^1 dx\,\frac{\vp_\pi(x,Q^2_{x})}{x(1-x)}\,
\eqe
where $Q_x=\mbox{min}(x,1-x)\,Q$. The amplitude ${\cal M}_{\lambda\lambda^\prime}$ 
for the process $\gamma\gamma\to \pi^+\pi^-$ at large angle $\theta_{c.m.}$ 
is given as \cite{BrodskyLepage1981}
\eqb
\label{ggpp}
{\cal M}_{\lambda\lambda^\prime}=\int_0^1 dx_1\,\int_0^1\,dx_2\,\vp_\pi(x_1,Q^2_{x_1})\vp_\pi(x_2,Q^2_{x_2})
\times T_{\lambda\lambda^\prime}(x_1,x_2,s,\theta_{c.m.})\,
\eqe
where $\lambda,\lambda^\prime$ denote the helicities $+$ or $-$ of 
the photons and $Q_{x_1}\sim\mbox{min}(x_1,1-x_1)
\sqrt{s}|\sin\theta_{c.m.}|$ (accordingly for $Q_{x_2}$).
In the case of equal helicities one has ${\cal M}_{++}={\cal M}_{--}$ which can be expressed in terms 
of the pion's electromagnetic form factor $F_{\pi^{\pm}}$ as
\eqb
\label{hff}
{\cal M}_{++}(s)=\frac{16\pi\alpha}{1-\cos^2\theta_{c.m.}} F_\pi^{\pm}(s)\,
\eqe
where
\eqb
\label{ffpion}
F_\pi^{\pm}(s)=\frac{16\pi\alpha_s}{3s}\int_0^1 dx_1\,\int_0^1 dx_2\,
\frac{\vp_\pi(x_1,Q^2_{x_1})\vp_\pi(x_2,Q^2_{x_2})}{x_1(1-x_1)x_2(1-x_2)}\,.
\eqe
Notice that the $\alpha_s$ expansion of the 
form factor $F_\pi$ in Eq.\,(\ref{ffpion}) starts at 
linear order due to the kinematic situation shown in Fig.\,\ref{Fig-HSC2} - 
in contrast to the process $\gamma\gamma^*\to\pi^0$ (see again Fig.\,\ref{Fig-HSC1}).  

If the respective form factors are measured accurately - like the CLEO collaboration 
\cite{CLEO1998} did in the case of $F_{\pi\gamma}(Q^2)$ - then 
Eqs.\,(\ref{ggp}),(\ref{ffpion}) can be used to constrain the form of the 
pion DA. For $Q\to\infty$ it should 
approach the asymptotic parton-model form $\vp_\pi(x,Q^2_{x}\to\infty)=6f_\pi\,x(1-x)$.  

\subsection{\it OPE and DAs}

\subsubsection{Pion DA and local condensates}

Armed with the concept of DAs and the conviction of its usefulness 
we are now in a position to relate them to the OPE parameters characterizing the 
nonperturbative QCD vacuum - the gauge invariant condensates. 
This idea was pioneered by Chernyak and Zhitnitsky \cite{ChernyakZhitnitsky1977,ChernyakZhitnitsky1981} (for a review see 
\cite{ChernyakZhitnitsky1984}). We will mostly follow their presentation which 
focuses in particular on pion wave functions. 

As one can derive from Eq.\,(\ref{pionWF}), the following 
relation holds for the moments of the wave function, 
$\la \xi^N\ra_{\mu^2}\equiv\int_{-1}^{+1}d\xi\, \xi^N\bar{\vp}(\xi,\mu^2)$,
\eqb
\label{momentsder}
if_\pi(z\cdot q)^{N+1}\la \xi^N\ra_{\mu^2}=\la0|\bar{d}(0)\gamma_\nu z^\nu\gamma_5\left(iz^\mu[
\overrightarrow{D}_\mu-\overleftarrow{D}_\mu]\right)^Nu(0)|\pi^+(q)\ra_{\mu^2}\,
\eqe
where $\bar{\vp}$ is a dimensionless wave function, $\bar{\vp}=\frac{1}{f_\pi}\vp$. 
According to Eq.\,(\ref{momentsder}) we have for the zeroth moment: 
\eqb
\label{zeroWF}
\la\xi^0\ra=1\,.
\eqe
To make contact with vacuum condensates the central object to consider 
is the correlator 
\eab
\label{corrvacOO}
T^{N_1N_2}(q^2,z\cdot q)&=&i\int d^4x\,\e^{iqx}\la T O_{N_1}(x)O_{N_2}(0) \ra\nonumber\\ 
&=&(z\cdot q)^{N_1+N_2+2} I_{N_1N_2}(q^2)\,
\eae
of the local operators 
\eqb
\label{OnCZ}
O_N\equiv \sqrt{\frac{1}{2}}\bar{u}\gamma_\nu z^\nu\gamma_5\left(iz^\mu[
\overrightarrow{D}_\mu-\overleftarrow{D}_\mu]\right)^Nu-(u\leftrightarrow d)\,.
\eqe
Expanding $I_{N_1N_2}(q^2)$ into an OPE  in the euclidean region $q^2=-Q^2<0$, we have
\eab
\label{OPEINN}
I_{N_1N_2}(q^2)&=&
-\frac{3}{(\bar{N}+1)(\bar{N}+3)}\frac{1}{4\pi^2}\log\left(\frac{Q^2}{\mu^2}\right)
+\frac{1}{12(Q^2)^2}\la\frac{\alpha_s}{\pi}G^2\ra\times\nonumber\\ 
&&\frac{1}{\bar{N}+1}\left(1+\frac{3}{\bar{N}-1}\left[(N_1-N_2)^2-\bar{N}\right]\right)+
\frac{32}{81(Q^2)^3}\pi\alpha_s\la\bar{u}u\ra^2_{Q^2}\left(11+4\bar{N}\right)\,
\eae
where $\bar{N}\equiv N_1+N_2$, and SU(3)$_F$ symmetry 
and vacuum saturation for the four-quark condensates 
have been assumed. The index $Q^2$ at the dimension-six condensate refers to the normalization point. 
The intermediate states in the correlators $I_{00}$ and $I_{N0}$ are the same, 
the resonances to which the ``currents'' $O_0$ and $O_N$ both couple are the 
$\pi^+$ and spin-one excitations such as the $A_1$, $\cdots$. We assume the first to be explicit, and we 
suppose that the rest can be treated as a part of the
perturbative, spectral continuum which starts at some 
threshold $s_{N0}$. In the chiral limit $m_\pi\to 0$ we obtain the following 
Borel sum rule for the correlator $T^{N0}$
\eab
\label{BSRmoments}
f_\pi^2\la\xi^N\ra_{M^2}&=&\frac{3M^2}{4\pi^2(N+1)(N+3)}\left(1-\e^{-s_{N0}/M^2}\right)+
\nonumber\\ 
&&\frac{1}{12M^2}\frac{3N+1}{N+1}\la\frac{\alpha_s}{\pi}G^2\ra+\frac{16}{81M^4}\,
\alpha_s\pi(4N+11)\la\bar{u}u\ra_{M^2}^2\,.
\eae
Evaluating the sum rule in Eq.\,(\ref{BSRmoments}) for the first three moments $N=2,4,6$ by  
using the SVZ values for the condensates\footnote{Anomalous dimensions 
in the four-quark operators have been 
neglected, for a discussion of the evaluation of the sum rules see \cite{ChernyakZhitnitsky1981}, 
the SVZ value for the condensates it taken at $\mu=1\,$GeV.}, $\la\frac{\alpha_s}{\pi}G^2\ra=1.2\times10^{-2}\,$
GeV$^4$ and $\alpha_s \la\bar{u}u\ra=1.83\times10^{-4}\,$GeV$^6$ the following results at
$\mu^2_0=(0.5\,\mbox{GeV})^2$ were obtained in \cite{ChernyakZhitnitsky1981}
\eqb
\label{momentsCZ}
\la{\xi}^2\ra_{(0.5\,{\tiny \mbox{GeV})^2}}=0.46\,,\ \ \ \ \ \ 
\ \la{\xi}^4\ra_{(0.5\,{\tiny \mbox{GeV})^2}}=0.30\,,\ \ \ \ \ \ 
\ \la{\xi}^6\ra_{(0.5\,{\tiny \mbox{GeV})^2}}=0.21\,.
\eqe
Using the convex asymptotic form $\bar{\vp}_{as}=\frac{3}{4}(1-\xi^2)$ for $\mu^2\to\infty$ - the contribution 
at $n=0$ in Eq.\,(\ref{expandWF}) -  one arrives at too low moments 
in comparison with Eq.\,(\ref{momentsCZ}): 
$\la\xi^2\ra_{as}=0.20\,,\la\xi^4\ra_{as}=0.086\,,$ and $\la\xi^6\ra_{as}=0.048\,$. 
This form of the wave function seems to be excluded by the sum rules for the moments.
Therefore, Chernyak and Zhitnitsky proposed the following pion DA
\eqb
\label{CZWF}
\bar{\vp}_{CZ}(\xi,\mu_0^2=(0.5\,\mbox{GeV})^2)=\frac{15}{4}\,\xi^2(1-\xi^2)
\eqe
which is not convex anymore, strongly concentrated at the endpoints $\xi=\pm1$. This DA 
gives moments that are in reasonable agreement with Eqs.\,(\ref{zeroWF}),(\ref{momentsCZ})
\eqb
\label{momentsCZmod}
\la{\xi}^0\ra^{CZ}_{(0.5\,{\tiny \mbox{GeV})^2}}=1\,,\ \ \ \ 
\la{\xi}^2\ra^{CZ}_{(0.5\,{\tiny \mbox{GeV})^2}}=0.43\,,\ \ \ \ 
\ \la{\xi}^4\ra^{CZ}_{(0.5\,{\tiny \mbox{GeV})^2}}=0.24\,,\ \ \ \ 
\ \la{\xi}^6\ra^{CZ}_{(0.5\,{\tiny \mbox{GeV})^2}}=0.15\,.
\eqe
Working with $\bar{\vp}_{CZ}$ to describe branching ratios of pionic decays 
of charmonium levels, experimentally obtained values can well be matched \cite{ChernyakZhitnitsky1981}. 
In addition, it can be demonstrated using $\bar{\vp}_{CZ}$ that the annihilation contributions 
to mesonic $D$ decays are comparable to the direct ones leading to a better 
agreement between theory and experiment.

\subsubsection{Pion DA and nonlocal condensates\label{sec:PionDAsNLC}}

The approach of Chernyak and Zhitnitsky, which relates the moments of DAs to 
the {\sl local} condensates appearing in the OPE of a suitable 
current correlator (see Eq.\,(\ref{corrvacOO})), implicitly assumes that 
quarks and gluons in the nonperturbative QCD vacuum 
have zero momentum. If this was the case for quarks, 
the {\sl nonlocality parameter} $\Lambda_q^2$, defined as 
\eqb
\label{nonlocpara}
\Lambda_q^2\equiv\frac{\la\bar{q}(0)D^2q(0)\ra}{\la\bar{q}(0)q(0)\ra}\,,
\eqe
$q$ denoting a light-quark field, would vanish
\footnote{The parameter$\Lambda_q^2$ is a measure for the deviation from perfect, 
gauge invariant field correlation at different points in spacetime. It appears in the 
first nontrivial term when expanding the corresponding gauge invariant two-point function 
$\la\bar{q}(z){\cal P}\exp\left[ig\int_0^{z} dy^\mu\,A_\mu(y)\right]q(0)\ra$ 
in powers of $z^2$.}. QCD sum rules, however, predict in the chiral limit 
a value of $\Lambda_q^2\sim 0.4-0.5$\,GeV$^2$ with an error of 10-20\%\,
\cite{BelyaevIoffe1983,OvchinnikovPivovarov1988}. An estimate using the single-instanton approximation 
of the instanton liquid yields the slightly higher value 
$\Lambda_q^2\sim 0.6$\,GeV$^2$ \cite{PolyakovWeiss1996,DorokhovEsaibegianMikhailov1997} while an 
unquenched lattice computation with four fermion flavors yields 
values close to the sum-rule estimate 
\cite{D'EliaDiGiacomoMeggiolaro1999,DoschEidemullerJaminMeggiolaro2000}. 
A more recent investigation of lattice estimates and the 
extrapolation to the chiral limit thereof was performed in \cite{BakulevMikhailov2002}. As 
a result, a window $0.37\,\mbox{GeV}^2\le\Lambda_q^2\le 0.55\,\mbox{GeV}^2$ 
was obtained. vPhenomenologically, it is thus an established fact that gauge invariant 
correlations between quark fields are of finite range. Moreover, 
the mass scale governing the fall-off of the quark correlator in 
Euclidean spacetime is larger than the usually accepted condensate 
scale $\sim\Lambda_{QCD}\sim 0.3-0.4\,$GeV which makes the OPE 
philosophy of an expansion in powers of the ratio $\Lambda_{QCD}/Q$ 
questionable. In relation to the moments of DAs the problem can be quantified as follows. 
According to Eq.\,(\ref{BSRmoments}) the ratio of nonperturbative 
contributions to the perturbative part in the OPE is (at large $N$) 
cubically growing with $N$. This means that even in a truncated OPE, effectively, the nonperturbative 
scales in the vacuum, which determine the $N$th moment of the DA, 
appear to be rapidly growing with $N$. To compensate for this effect higher and 
higher values of the continuum threshold $s_{0N}$ have to be adopted to reach 
stability of the sum rule for the $N$th moment \cite{ChernyakZhitnitsky1981}. 
But this amounts to neglecting 
higher resonances in the spectrum that would be related to the OPE contributions 
governed by large, nonperturbative mass scales. The quality of such a 
cancellation between neglected parts of the integrated spectrum and 
neglected condensates of a large mass scale in the OPE 
is obscure. It is very reasonable though to expect it to become worse with growing 
$N$. Considerations of this type lead Mikhailov, Radyushkin, and 
Bakulev to postulate a modification of the OPE with 
built-in nonlocal effects \cite{MikhailovRadyushkin1986,
MikhailovRadyushkin1989,BakulevRadyushkin1991,MikhailovRadyushkin1992}. 
An approach, which expands nonlocal condensates into complete sets of 
resolution dependent, finite-width functions in position space, 
was proposed in \cite{HoangHofmann2003} and will be discussed 
in detail in Sec.\,\ref{sec:DOE} with respect to the question 
of local duality violation. Instead of writing sum rules for the moments $\la\xi^N\ra$ 
as in Eq.\,(\ref{BSRmoments}) one can use the truncated OPE to directly write a 
sum rule for the wave function $\bar{\vp}_\pi(x)$ 
\cite{MikhailovRadyushkin1989,MikhailovRadyushkin1992}
\eab
\label{srpiWF}
f_\pi^2\bar{\vp}_\pi(x,M^2)&=&
\frac{M^2}{4\pi^2}\left(1-\e^{-s_0/M^2}\right)\bar{\vp}_\pi(x,\mu^2\to\infty)+\frac{1}{24M^2}
\la\frac{\alpha_s}{\pi}G^2\ra\left[\delta(x)+\delta(1-x)\right]+\nonumber\\ 
&&\frac{8}{81M^4}\pi\alpha_s\la\bar{q}q\ra^2\left(11\left[\delta(x)+\delta(1-x)\right]+
2\left[\delta^\prime(x)+\delta^\prime(1-x)\right]\right)\,
\eae
where the prime denotes differentiation with respect to $x$. Taking into account 
radiative corrections in order $\alpha_s$, 
it was obtained in \cite{MikhailovRadyushkin1992} that the followinf substitution should be performed 
\eqb
\label{radcorrWF}
\bar{\vp}_\pi(x,\mu^2\to\infty) \to
\bar{\vp}_\pi(x,\mu^2\to\infty)
\left[1+\frac{4}{3}\frac{\alpha_s}{4\pi}\left(5-\frac{\pi^2}{3}+\log^2\frac{1-x}{x}\right)\right]\,.
\eqe
Note that this form of the DA, which arises from a 
truncated local expansion of the correlator of {\sl nonlocal}, gauge invariant currents, 
violates the boundary condition 
\eqb
\bar{\vp}(x,\mu^2)\le Kx^\epsilon\ \ \ \ \mbox{for} \ \ \ x\to 0\ \ \ \ (\epsilon>0)\,,
\eqe
which assures the convergence of the expansion in Eq.\,(\ref{expandWF}) 
\cite{LepageBrodsky1979}. Hence Eq.\,(\ref{radcorrWF}) is incomplete. The complete form of 
Eq.\,(\ref{srpiWF}) is obtained by substituting $5\to 5+2\log(M^2/\mu^2)$ 
in Eq.\,(\ref{radcorrWF}) \cite{BakulevMikhailov1998,BakulevMikhailovStefanis2001}. 

The starting point in \cite{MikhailovRadyushkin1986,
MikhailovRadyushkin1989,BakulevRadyushkin1991,MikhailovRadyushkin1992,BakulevMikhailov1998} is a parametrization 
of the gauge invariant bilinear scalar or vector quark condensate (or correlator) of the form
\eqb
\label{alpharep}
\la\bar{q}(0)q(z)\ra=\int_0^\infty\e^{\nu z^2/4}f_S(\nu)\,,\ \ \ 
\ \la\bar{q}(0)\gamma^\mu q(z)\ra=iz^\mu\frac{2}{81}\pi\alpha_s(M^2)\la\bar{q}q\ra^2\,
\int_0^\infty\e^{\nu z^2/4}f_V(\nu)\,
\eqe
where $f_S$ and $f_V$ denote the distribution function for quark 
momenta in the vacuum associated to the scalar and vector 
correlation on the left-hand sides, respectively. In writing Eq.\,(\ref{alpharep}) the 
Fock-Schwinger gauge together with a straight-line connection
between the point $0$ and $z$ were assumed. Higher $n$-point functions as well 
as the nonlocal gluon condensate 
are parametrized in a similar manner, see 
\cite{MikhailovRadyushkin1992,Mikhailov1993} for an extended discussion
\footnote{In the case of the gluon correlator 
$\la A^a_\mu(z)A^b_\nu(y)\ra$ (with straight lines connecting the origin with $z$ and $y$) 
one would in principle not only 
have a dependence on $(z-y)^2$ but also on $z^2$ and $y^2$. However, 
a local expansion reveals that the dependences on $z^2$ and $y^2$ are
much weaker than the one on $(z-y)^2$, and therefore one may neglect them.}.

The functions $f_S$ and $f_V$ have to be modelled. For condensates without covariant derivatives 
in the OPE a formal expansion of the functions $f_S, f_V$ into the 
set $\delta^{(n)}(\nu)\,,(n=0,1\cdots)$ is truncated at $n=0$. Operators in the OPE 
with $n$ powers of 
covariant derivatives are associated with the $\delta^n(\nu)$ part 
of the expansion. We have already seen that there 
are problems with this local truncation. Mikhailov and Radyushkin 
proposed to model $f_S$ and $f_V$ by simply 
shifting the argument of the delta function of the local expansion 
to generate finite-width correlations in position space
\eqb
\label{shiftdelta}
\delta(\nu) \to \delta(\nu-\mu_{S,V}^2)\,.
\eqe
In the case of higher $n$-point functions the $f$ functions are modelled 
as products of delta functions centered at nonzero values of $\nu_i$. As a 
consequence, the expression for the pionic DA is smooth at $x=0$ 
and the moments of the nonperturbative terms {\sl decrease} with $N$ in such a way that the ratio to the
perturbative terms increases only moderately - much in contrast to the case 
of a truncated OPE, see Eq.\,(\ref{BSRmoments}). Other ans\"atze than the 
one in Eq.\,(\ref{shiftdelta}), which are better suited for sum rules with nondiagonal 
correlators, were proposed and discussed 
in \cite{Radyushkin1994,BakulevMikhailov1995,BakulevMikhailov1996,BakulevMikhailov2002}. 
Using $\Lambda_q^2=0.4\,$GeV$^2$, the following values for the first 
three moments were obtained
\eqb
\label{momentsMrad}
\la{\xi}^2\ra=0.25\,,\ \ \ \ \ \ 
\ \la{\xi}^4\ra=0.12\,,\ \ \ \ \ \ 
\ \la{\xi}^6\ra=0.07\,,
\eqe
which are close to the asymptotic values, see text below Eq.\,(\ref{momentsCZ}). 
For recently obtained small alterations of these values see 
\cite{BakulevMikhailovStefanis2001}. 
These values can be well reproduced by using the 
model DA $\bar{\vp}=\frac{8}{\pi}\sqrt{x(1-x)}$ which differs quite drastically from the 
one in Eq.\,(\ref{CZWF}) obtained by Chernyak and Zhitnitsky and 
is qualitatively similar to the asymptotic one. 

\subsubsection{Direct experimental evidence for nonlocal condensates from meson DAs}

The only way we can decide whether at a given order in twist and $\alpha_s$ the 
prediction of an endpoint concentrated pionic DA, which rest 
on a truncated OPE (Chernyak and Zhitnitsky), or a prediction 
close to the asymptotic, convex form, which is 
based on nonlocal condensates (Mikhailov, Radyushkin, and Bakulev), is true 
is to compare them with precise and independent experimental data on 
exclusive quantities. A very well suited observable is 
the pion-photon transition form factor $F_{\pi\gamma}(Q^2)$ of Eq.\,(\ref{ggp}). 
It was measured rather precisely by the CLEO 
collaboration for $Q^2$ from 1.5 to 9 GeV$^2$ \cite{CLEO1998}. 
Together with the data taken by the CELLO collaboration \cite{CELLO1991} 
a range of $Q^2$ values from 0.7 to 9 GeV$^2$ is covered. 

\begin{figure}[tb]
\begin{center}
\hspace{-1.5cm}
\begin{minipage}[t]{8 cm}
\epsfig{file=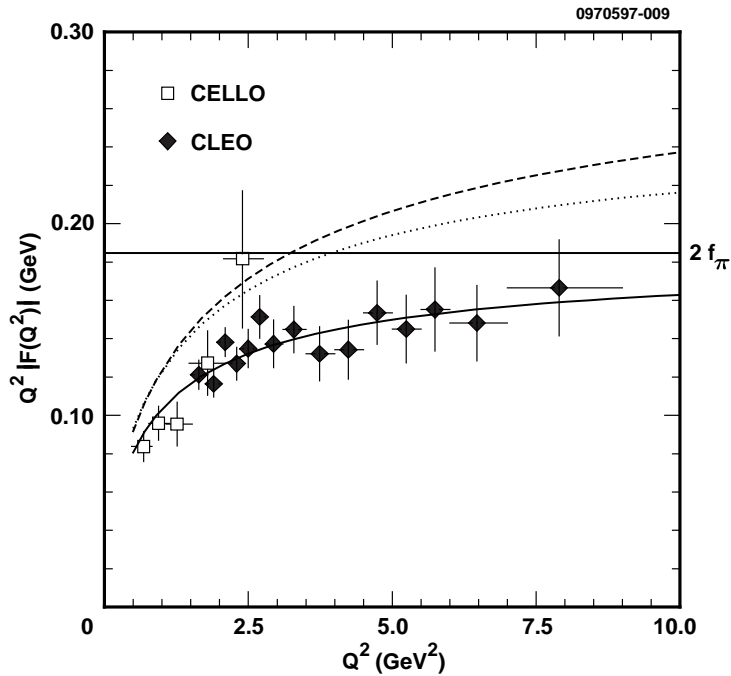,scale=0.6}
\end{minipage}
\begin{minipage}[t]{16.5 cm}
\caption{Comparison of the CLEO-results (points) for $Q^2F_{\pi\gamma}(Q^2)$ with 
the theoretical predictions made by \cite{Kroll1996} using the asymptotic DA (solid curve) 
and the Chernyak-Zhitnitsky DA of Eq.\,(\ref{CZWF}) (dashed curve). 
The dotted curve shows the prediction made with 
the Chernyak-Zhitnitsky DA when its QCD evolution is 
taken into account. The solid line at 
$2f_\pi$ indicates the asymptotic value. Plot taken from \cite{CLEO1998}.\label{Fig-TFF1}}
\end{minipage}
\end{center}
\end{figure}

\begin{figure}[tb]
\begin{center}
\hspace{-1.5cm}
\begin{minipage}[t]{8 cm}
\epsfig{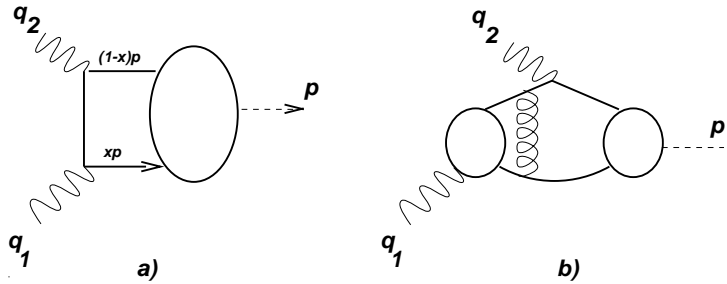}
\end{minipage}
\begin{minipage}[t]{16.5 cm}
\caption{Lowest-order diagrams in perturbative QCD for the three-point 
correlator $2\pi i\int d^4x d^4y\, \e^{-iqx}\e^{ipx}\la T j^5_\lambda(y) j_\mu(x) j_\nu(0)\ra$ 
where $j^5_\lambda=\bar{u}\gamma_5\gamma_\lambda u-\bar{d}\gamma_5\gamma_\lambda d$, 
and $j_\mu$ denotes the electromagnetic current, $j_\mu=\frac{2}{3}\bar{u}\gamma_\mu u-
\frac{1}{3}\bar{d}\gamma_\mu d$. Taken from \cite{RadyushkinRuskov1996}.\label{Fig-PTFF}}
\end{minipage}
\end{center}
\end{figure}

\begin{figure}[tb]
\begin{center}
\hspace{-1.5cm}
\begin{minipage}[t]{8 cm}
\epsfig{file=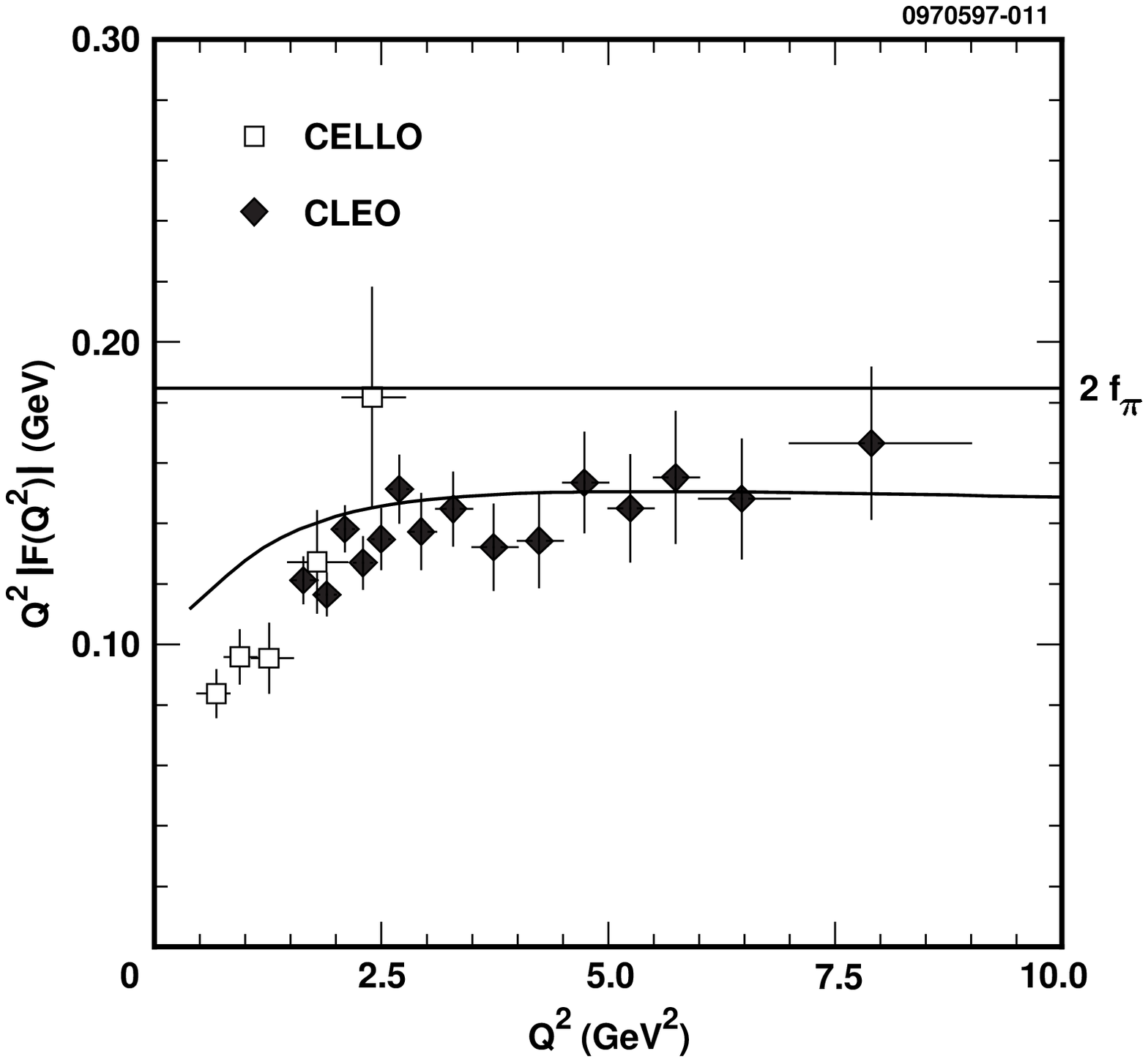,scale=0.6}
\end{minipage}
\begin{minipage}[t]{16.5 cm}
\caption{Comparison of the CLEO-results (points) 
for $Q^2F_{\pi\gamma}(Q^2)$ with the prediction of 
\cite{RadyushkinRuskov1996} which is based on a 
QCD sum rule for $F_{\pi\gamma^*}(Q^2=-q^2,{Q^\prime}^2=-{q^\prime}^2)$ with two virtual photons. 
The limit ${Q^\prime}^2\to 0$ was taken after applying a factorization procedure for the 
infrared singularities which arise perturbatively in this limit. 
Taken from \cite{CLEO1998}.\label{Fig-TFF2}}
\end{minipage}
\end{center}
\end{figure}

\begin{figure}
\begin{center}
\leavevmode
\leavevmode
\vspace{6cm}
\includegraphics{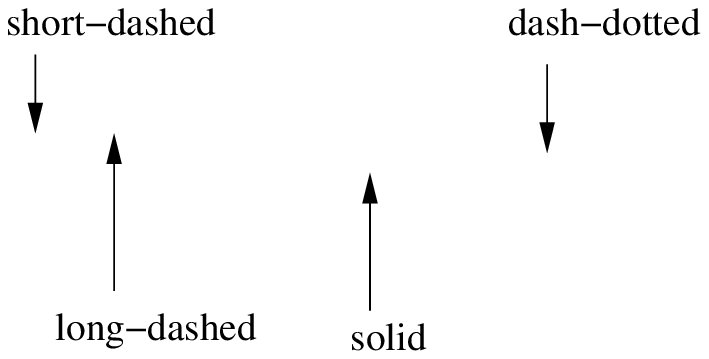}
\end{center}
\caption{Comparison of $Q^2F_{\pi\gamma}(Q^2)$ as obtained 
from the analysis using a light-cone sum rule \cite{Khodjamirian1999} and 
the experimental data. The solid line is associated with 
the asymptotic DA, the long-dashed line with the Chernyak-Zhitnitsky DA, the 
dash-dotted line with the simple interpolation formula 
$F_{\pi\gamma}(Q^2)=\frac{\sqrt{2}f_\pi}{4\pi^2f_\pi^2+Q^2}$ in \cite{LepageBrodsky1979}, 
and the short-dashed line with a DA obtained in \cite{BraunFilyanov1989}. 
Plot taken from \cite{Khodjamirian1999}.\label{Fig-Kho}}      
\end{figure}

In \cite{Kroll1996} a perturbative approach was applied to the calculation of the 
transition form factor $F_{\pi\gamma}(Q^2)$ which takes into 
account transverse momentum effects and the Sudakov suppression of longitudinal 
momenta close to the endpoints $x=0,1$. 
More precisely, the (perturbatively undressed) 
valence Fock state pion wave function $\Psi_0(x,\vec{b},\mu_F^2)$ 
is assumed to factorize into the DA $\bar{\vp}(x,\mu^2)$ for the 
longitudinal momentum fraction and a portion $\hat{\Sigma}(\sqrt{x(1-x)}\vec{b})$
containing information about the distribution in two dimensional transverse position space
\eqb
\label{Kroll}
\Psi_0(x,\vec{b},\mu^2)=\frac{f_\pi}
{2\sqrt{6}}\,\bar{\vp}(x,\mu^2)\,\hat{\Sigma}(\sqrt{x(1-x)}b)\,
\eqe
where $\mu$ is the scale for the factorization of hard scattering 
and soft momentum distribution, and the 2d vector $\vec{b}$ 
denotes the quark-antiquark separation. Certain constraints on $\hat{\Sigma}$ 
can be derived from duality arguments \cite{ChibisovZhitnitsky1995} 
which are minimally satisfied by the following Gaussian ansatz
\eqb
\label{Gausstr}
\hat{\Sigma}(\sqrt{x(1-x)}b)=4\pi\exp\left[-\frac{x(1-x)b^2}{4a^2}\right]\,
\eqe
where $a$ denotes the transverse size parameter. It 
can be fixed by the requirement that the wave 
function $\Psi_0(x,\vec{b},\mu^2)$ be 
normalized to $\sqrt{6}/f_\pi$ (due to $\pi^0\to\gamma\gamma$). For models with a 
power-law ansatz for the dependence of the wave function on transverse momentum see 
\cite{FredericoPauli2001,PauliMukherjee2001,Pauli2002,MukherjeeMusatovPauliRadyushkin2003}.
Taking into account Sudakov suppressions in 
the hard scattering amplitude $T_H$ when retaining its dependence on 
transverse momentum $k_\perp$ \cite{BottsSternman1989,LiSternman1992,MusatovRadyushkin1997}, 
the expression for $F_{\pi\gamma}(Q^2)$ is altered in comparison to Eq.\,(\ref{ggp}) as
\eqb
\label{ggpa}
F_{\pi\gamma}(Q^2)=\int_0^1dx\,
\frac{d^2b}{4\pi}\Psi_0(x,\vec{b},\mu^2=1/b^2)\,T_H(x,\vec{b},Q^2)\,
\exp[-S(x,\vec{b},Q^2)]\,
\eqe
where $S$ denotes the Sudakov exponent which accounts 
for gluonic radiative corrections 
not contained in the evolution of
the DA $\bar{\vp}$. Note that the factor $\Psi_0$ in Eq.\,(\ref{ggpa}) represents 
an {\sl ansatz} designed to take effects of intrinsic transverse 
momentum into account, it is not derived from first
principles. The accuracy of this ansatz was 
questioned in \cite{StefanisSchroersKim19992000}. 
Also, subleading logarithms in the Sudakov corrections were taken into
account in \cite{StefanisSchroersKim19992000}. 
Still, using the {\sl asymptotic} form $\bar{\vp}=6x(1-x)$, very good agreement 
between the data and the numerical result based 
on Eq.\,(\ref{ggpa}) is obtained at $\mu^2=1$\,GeV$^2$ in \cite{Kroll1996}, see Fig.\,(\ref{Fig-TFF1}). 
A perturbative investigation of the transition form factor 
\eqb
\label{TFFgamma*}
F_{\pi\gamma^*}(Q^2,{Q^\prime}^2)=\frac{4f_\pi}{\sqrt{3}}\int_0^1dx\,
\frac{\bar{\vp}(x)}{xQ^2+(1-x){Q^\prime}^2}
\eqe
for the process $\gamma^*\gamma^*\to\pi^0$ 
using the same approach as in \cite{Kroll1996} was performed in \cite{Ong1995}. 
At $Q^2\sim{Q^\prime}^2$ this calculation provides at test of the hard scattering 
amplitude since the transition form factor is essentially independent of 
the pionic DA. The result of \cite{Kroll1996} that the DA of 
Chernyak and Zhitnitsky is excluded and the asymptotic DA is favored 
by experimental data was confirmed in \cite{Ong1995} when taking the 
limit ${Q^\prime}^2\to 0$. However, a more recent light-cone sum-rule analysis of the 
CLEO data \cite{SchmeddingYakovlev2000} indicates
\footnote{The analysis in \cite{SchmeddingYakovlev2000} obtains constraints on the first 
two Gegenbauer polynomials.} that the 
neither the asymptotic DA (although favored) nor 
the Chernyak-Zhitnitsky model are compatible with the data. This was qualitatively confirmed in 
\cite{BakulevMikhailovStefanis2004} where an incompatibility with the CLEO data on the 
3 $\sigma$ and 4 $\sigma$ level was obtained when using the asymptotic DA and the Chernyak-Zhitnitsky model,
respectively. 
In \cite{RadyushkinRuskov1996} the transition form factor $F_{\pi\gamma^*}(Q^2,{Q^\prime}^2)$ 
was calculated using a QCD sum rule for the three-point correlator associated with 
the process $\gamma^*\gamma^*\to\pi^0$, see Fig.\,\ref{Fig-PTFF} 
for the lowest-order perturbative diagrams contributing to it. 
Applying a factorization procedure to separate off the infrared singularities emerging 
in the limit ${Q^\prime}^2\to 0$, subsequently absorbing 
these singularities into bilocal correlation functions, and 
finally assuming certain forms for these correlation functions, the $Q^2$ 
dependence of the transition form factor $F_{\pi\gamma}(Q^2)$ was studied and compared with the data 
from CELLO \cite{CELLO1991}, see Fig.\,\ref{Fig-TFF2}. As a result, again evidence was provided by 
this approach to $F_{\pi\gamma}(Q^2)$ that $\bar{\vp}(x,Q^2)$ is rather 
close to the asymptotic form $\bar{\vp}(x,Q^2\to\infty)=6x(1-x)$.
The same conclusion was reached in \cite{Khodjamirian1999} where a light-cone 
sum rule (see next section for its foundations) for the two-point correlator 
\eqb
\label{twopcK}
\int d^4x\,\e^{-iqx}\,\la\pi^0(p)|Tj_\mu(x)j_\nu(0)|0\ra\
\eqe
of the electromagnetic currents was used in the variable\footnote{Photon 
virtualities are again $Q^2=q^2$ and ${Q^\prime}^2=-(p-q)^2$.} ${Q^\prime}^2$. In the chiral limit 
the light-cone OPE for the transition form factor $F_{\pi\gamma}(Q^2)$ can 
then be written in terms of a dispersion integral over 
the imaginary part of the transition form factor 
$\mbox{Im}\,F_{\pi\gamma^*}(Q^2,s)$ which, in turn, is given by the pion DA up to twist four as
\eqb
\label{dispFFWF}
\frac{1}{\pi}\mbox{Im}
F_{\pi\gamma^*}(Q^2,s)=\frac{\sqrt{2}f_\pi}{3}\left.\left(\frac{\bar{\vp}_\pi(u,Q^2)}{s+Q^2}-
\frac{1}{Q^2}\frac{d\bar{\vp}^{(4)}(u,Q^2)}{ds}\right)\right|_{u=\frac{Q^2}{s+Q^2}}\,
\eqe
where $\bar{\vp}^{(4)}$ denotes the twist-four part. 
As a result, even at a higher order in twist 
the asymptotic DA is again favored by the CLEO and CELLO data, see Fig.\,\ref{Fig-Kho}. Notice, however, 
that a very recently performed analysis of the CLEO and CELLO data using a light-cone sum rule 
with the inclusion of twist four {\sl and} $O(\alpha_s)$ perturbative 
corrections to the pion DA seems to exclude both the asymptotic 
as well as the Chernyak-Zhitnitsky form of 
the DA at lowest twist \cite{BakulevMikhailovStefanis2003,BakulevMikhailovStefanis20032}. 
Still, in the determination of the pion DA from a 
nonlocal sum rule at the same order in twist and $\alpha_s$ 
a nonlocality parameter $\Lambda_q^2=0.4\,$GeV$^2$ is compatible with the result 
obtained from the light-cone sum rule. Seemingly, the inclusion of radiative corrections 
has a greater effect on the deviation from the asymptotic DA than the inclusion of the 
twist-four correction. This fits well with the result of 
\cite{BakulevMikhailovStefanis2001} where the sum rule for the moments $\la\xi^N\ra$ with nonlocal 
condensates (including the $A_1$ resonance explicitly into the spectral function) 
{\sl and} $O(\alpha_s)$ corrections but no twist-four corrections was analyzed.  
A number of calculations of the twist-two pion DA $\bar{\vp}$ were recently performed 
using instanton models \cite{PetrovPolyakovRuskovWeissGoeke1999,Dorokhov2002} 
or in an instanton motivated model \cite{PrazalowiczRostworowski}. 
As compared to the asymptotic DA all of these investigations found 
a slight depletion of the model DA within the central 
region but no qualitative deviation from the asymptotic form. 

We may thus conclude that the compatibility (noncompatibility) of the 
pion DA directly extracted from the experimentally well 
measured transition form factor $F_{\pi\gamma}(Q^2)$ and the pion 
DA obtained from a sum rule involving a truncated, 
practical OPE with nonlocal (local) condensates 
provides rather strong phenomenological evidence that 
a local expansion of nonperturbative gauge invariant correlation functions 
is not sufficient to describe the hard and 
exclusive process $\gamma\gamma^*\to\pi^0$. 

The technically more involved and theoretically more shaky case of 
$\rho$ meson DAs will not be discussed in detail in this review\footnote{Instead of one DA there are 
four DAs $\bar{\vp}_{\perp}, \bar{\vp}_{\parallel}, g^v_{\perp}$, and $g^a_{\perp}$ 
at the leading twist two \cite{AliBraunSimma1994,BallBraun1996} and valence Fock state. The DAs 
$\bar{\vp}_{\perp}$ and $\bar{\vp}_{\parallel}$ describe the distribution of 
longitudinal quark momentum in a transversely and longitudinally polarized $\rho$ meson, 
respectively. The twist-two DAs $g^v_{\perp}$ and 
$g^a_{\perp}$ are connected to $\bar{\vp}_{\parallel}$ \cite{BallBraun1996} 
by Wandzura-Wilczek type relations \cite{WandzuraWilczek1977} 
whose status as a dynamical statement 
is still under debate, see for 
example \cite{Lazar2001,BallLazar2001}.}. 
We only mention here the QCD sum-rule analysis of Refs.\,\cite{BallBraun1996,BakulevMikhailov1998} using 
a truncated, practical OPE with local and nonlocal 
condensates $(\Lambda_q^2\sim0.4$\,GeV$^2)$, respectively. In both cases only leading twist 
was considered and, in contrast to the old calculation of 
Refs.\,\cite{ChernyakZhitnitsky1981,ChernyakZhitnitsky19832,
ChernyakZhitnitskyZhitnitsky1983}, radiative corrections to the perturbative 
parts of the sum rules were taken into account. In \cite{BallBraun1996} a sign 
error in the four-quark contribution to
the OPE for the second moment of $\bar{\vp}_{\perp}$ was pointed 
out leading to a drastic change in the prediction of the 
shape of this DA as compared to the prediction in \cite{ChernyakZhitnitsky1981,ChernyakZhitnitsky19832,
ChernyakZhitnitskyZhitnitsky1983}. A comparison between $\bar{\vp}_{\parallel}$ at a 
normalization scale $\mu=1$\,GeV, obtained from the sum rules for the zero-helicity state 
reveals no essential difference in both approaches. The treatment with nonlocal condensates, 
however, gives a much better stability of the 
Borel curves of the moments $\la\xi^N\ra$ up to large $N$ 
than it does in the local case where a large-$N$ 
prediction is unreliable, see Eq.\,(\ref{BSRmoments}) 
for a similar situation in the pion case. The shape of $\bar{\vp}_{\parallel}$ 
obtained in \cite{BallBraun1996,BakulevMikhailov1998} is rather 
close to the result in \cite{ChernyakZhitnitsky1981,ChernyakZhitnitsky19832,
ChernyakZhitnitskyZhitnitsky1983} which is not too far from the asymptotic one; 
for $\bar{\vp}_{\perp}$, however \cite{BallBraun1996} 
get a much wider distribution than Chernyak and Zhitnitsky. 
Let us also mention that a computation of $\bar{\vp}_{\perp}$ was performed 
in the framework QCD sum rules with nonlocal condensates in 
\cite{BakulevMikhailov2001}.

To summarize, we have seen that the use of 
nonlocal condensates in the OPEs of current 
correlators which determine the moments of mesonic DAs is superior 
to the treatment with local condensates in the following ways: 
(i) the prediction of high moments $\la\xi^N\ra$ is feasible, (ii) 
Borel sum rules for a given moment exhibit a wide window 
of stability, (iii) the prediction of close-to-asymptotic 
behavior of the lowest-twist pion DA using nonlocal condensates in 
spectral sum rules for the first few moments agrees well 
with an independent determination in terms of 
light-cone sum rules using the experimental data on the transition form factor 
$F_{\pi\gamma}(Q^2)$ as a phenomenological input. 

\subsection{\it Reshuffling the OPE\label{sec:ResOPE}}

Motivated by the phenomenological results in the last section we will in this section 
review the theoretical foundations for a light-cone expansion 
of a current-current correlator into gauge invariant, nonlocal operators. 
Taking the hadron-to-hadron (hadron-to-vacuum) 
matrix elements of this expansion allows to relate experimentally 
measurable deep inelastic scattering (transition) cross sections 
to structure functions (distribution amplitudes). 

A phenomenological expansion of (2-point) nonlocal operator averages and Wilson coefficients 
into nonlocal objects, which captures more information about the 
nonlocal operator average than its truncated local expansion, 
is proposed for the vacuum-to-vacuum case. 
This approach introduces a so-called 
resolution parameter. The nonperturbative evolution of the 
nonlocal objects in this parameter will be derived. Several 
applications will be discussed, and a ``running'' 
gluon condensate will be extracted from 
experimental data in two channels. The implications of our 
nonlocal modification of the OPE in view of the nature of this expansion and an OPE-based realization 
of local quark-hadron duality are investigated 
phenomenologically.

\subsubsection{Light-cone expansion into string operators of a current-current product}

We have already mentioned in the last section the use of light-cone sum rules as a means 
to extract information on the twist-two pion DA from the experimental data for the transition form factor
$F_{\pi\gamma}$ \cite{Khodjamirian1999}. In this section we will give some 
theoretical background for this expansion. 

\begin{figure}[tb]
\begin{center}
\hspace{-0.5cm}
\begin{minipage}[t]{8 cm}
\epsfig{file=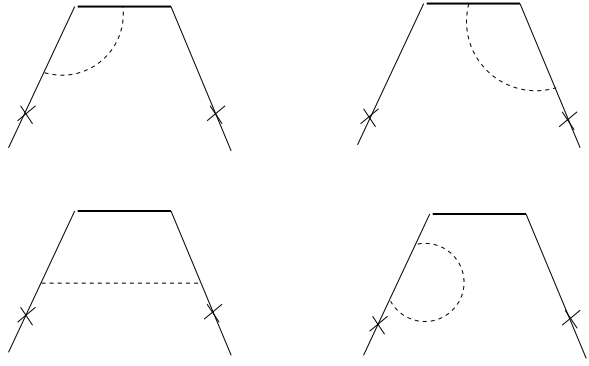,scale=0.4}
\end{minipage}
\begin{minipage}[t]{16.5 cm}
\caption{Diagrams contributing to the one-loop renormalization of the flavor-nonsinglet 
string operator of Eq.\,(\ref{stringops}). 
Thick lines denote the classical gauge string, crossed lines classical 
fermion fields and dashed lines are fluctuating gluon fields. \label{Fig-SOren}}
\end{minipage}
\end{center}
\end{figure}

The light-cone expansion of an object like $iTj_\mu(x)j_\nu(-x)$, $j_\mu$ 
being the electromagnetic current, into nonlocal string operators \cite{Karchev1983,Balitsky1883,BalitskyBraun1988} 
and the know-how about the 
perturbative renormalization of the latter 
under a change of scale are of paramount importance 
for the experimental determination of hadronic structure 
functions (hadron-to-hadron matrix element of nonlocal operator) 
and DAs (hadron-to-vacuum matric element of nonlocal operator). 
Within perturbation theory the proof for the existence of the light-cone expansion 
of a current-current correlator in terms of quark twist-two string operators like\footnote{We consider the
flavor nonsinglet string operator first since for this case we do not have to consider the mixing with 
the gluon string operator.} 
\eqb
\label{stringops}
\bar{\psi}(x)\lambda^a\hat{x}{\cal P}\exp\left[ig\int_0^x dz^\mu\,A_\mu(z)\right]\psi(0)\equiv 
\bar{\psi}(x)\lambda^a\hat{x}[x,0]\psi(0)\,
\eqe
was given in \cite{AnikinZavialov1978}. In Eq.\,(\ref{stringops}) $\lambda^a$ is a 
flavor Gell-Mann matrix and 
$\hat{x}\equiv\gamma_\mu x^\mu$. Some time later the one-loop dependence 
of gauge-string operators on the renormalization scale $\mu$ was calculated in \cite{BalitskyBraun1988} 
by integrating over fluctuations $q^q$ and $A_\mu^q$ about classical backgrounds $q^c$ and $A_\mu^c$. 
The relevant diagrams are shown in Fig.\,\ref{Fig-SOren}.

Using the Fock-Schwinger 
gauge $x^\mu A^c_\mu=0$, the background gauge-string $[\alpha x,0]$ is unity and thus 
is omitted in some of the expressions to follow below. Evaluating the diagrams 
in Fig.\,\ref{Fig-SOren} and keeping only the first 
logarithmic term in $x^2$, one then obtains \cite{BalitskyBraun1988}
\eqb
\label{renSOe}
\left.\bar{\psi}(x)\lambda^a\hat{x}\psi(0)\right|_{\mu_2^2}=\int_0^1 d\alpha 
\int^\alpha_0 d\beta \,
\left[\delta(1-\alpha)\delta{\beta}-
\frac{2\alpha_s}{3\pi}\log\frac{\mu_2^2}{\mu_1^2}\,K(\alpha,\beta)
\right]\left.\bar{\psi}(\alpha x)\lambda^a\hat{x}\psi(\beta x)
\right|_{\mu_1^2}\,
\eqe
where 
\eqb
\label{Kreg}
K(\alpha,\beta)=\frac{1}{2}\,\delta(1-\alpha)\delta({\beta})-\delta(\alpha)\left[\frac{1-\beta}{\beta}
-\delta({\beta})\int d\beta^\prime\frac{1-\beta^\prime}{\beta^\prime}\right]-\delta(\beta)
\left[\frac{\alpha}{1-\alpha}
-\delta({\alpha})\int d\alpha^\prime\frac{\alpha^\prime}{1-\alpha^\prime}\right]-1\,.
\eqe
The leading logarithms in Eq.\,(\ref{renSOe}) can be summed by 
solving a one-loop renormalization-group equation. In analogy to an expansion 
of a {\sl local} operator into an eigenbasis (conformal operators, multiplicatively renormalized) 
of the evolution kernel one may also expand a {\sl nonlocal} 
string operator into a nonlocal eigenbasis. The result is\cite{BalitskyBraun1988} 
\eab
\label{decnlSO}
&&\left.\bar{\psi}\left(\frac{\alpha+\beta}{2}x\right)\lambda^a\hat{x}
\left[\frac{\alpha+\beta}{2}x,\frac{\alpha-\beta}{2}x\right]\psi
\left(\frac{\alpha-\beta}{2}x\right)\right|_{\mu_2^2}\nonumber\\ 
&=&\int\frac{dk}{4\pi}\,\e^{-ik\alpha}\int_{-1/2-i\infty}^{-1/2+i\infty}dj\,(j+\frac{1}{2})
\,\beta^{-3/2}J_{j+1/2}(k\beta)
\int_{-\infty}^{\infty}d\alpha^\prime\,\e^{ik\alpha^\prime}\int_0^\infty d\beta^\prime\,
\sqrt{\beta^\prime}H^2_{j+1/2}(k\beta^\prime)\left(\frac{\alpha(\mu_1^2)}{\alpha(\mu_2^2)}
\right)^{-\gamma_j}\nonumber\\ 
&&\times\left.\bar{\psi}\left(\frac{\alpha^\prime+\beta^\prime}{2}x\right)\lambda^a\hat{x}
\left[\frac{\alpha^\prime+\beta^\prime}{2}x,\frac{\alpha^\prime-\beta^\prime}{2}x\right]\psi
\left(\frac{\alpha^\prime-\beta^\prime}{2}x\right)\right|_{\mu_1^2}\,
\eae
where $J_{j+1/2}(H_{j+1/2})$ is a Bessel(Hankel) function, 
\eqb
\label{anomdim}
\gamma_j=\frac{8}{3b}\int_0^1du\,u^{j-1}\left[\frac{1}{2}\,\delta(1-u)-(1-u)-2\left(
\frac{u}{1-u}-\delta(u)\int du^\prime\frac{u^\prime}{1-u^\prime}\right)\right]\,,
\eqe
and the integrations over $\alpha^\prime,\beta^\prime$ and $k,j$ indicate the projection onto and 
the (continuous) decomposition into renormalization-group covariant, conformal string operators, 
respectively.

\begin{figure}[tb]
\begin{center}
\hspace{-0.5cm}
\begin{minipage}[t]{8 cm}
\epsfig{file=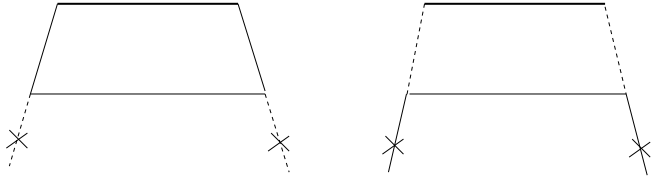,scale=0.4}
\end{minipage}
\begin{minipage}[t]{16.5 cm}
\caption{Diagrams responsible for the mixing of twist-two, flavor singlet quark string operators
and flavor singlet gluon string operators.
\label{Fig-mixQG}}
\end{minipage}
\end{center}
\end{figure}

In addition to the quark string operators of Eq.\,(\ref{stringops}) there is 
a twist-two flavor singlet gluon string operator which enters the light-cone 
expansion of a current-current correlator. It is 
given as
\eqb
\label{gluonstring}
G(u,v)=x_\alpha G_{\mu\alpha}^a(ux)\,[ux,vx]_{ab}\,G_{\mu\beta}^b(vx)x_\beta\,
\eqe
where the straight-path Wilson line $[ux,v,x]_{ab}$ is now 
in the {\sl adjoint} representation! The flavor singlet quark string operator 
\eqb
\label{fsquarkstring}
\tilde{Q}(u,v)=\frac{i}{2}\left[\bar{\psi}(ux)\hat{x}[ux,vx]\psi(vx)-
\bar{\psi}(vx)\hat{x}[vx,ux]\psi(ux)\right]
\eqe
and the gluon string operator of Eq.\,(\ref{gluonstring}) mix 
under renormalization, see Fig.\,(\ref{Fig-mixQG}).
Formulas simular to the one in Eq.\,(\ref{decnlSO}) relate 
$\tilde{Q}$ and $G$ at a renormalization point $\mu_2^2$ to a linear combination 
of their respective conformal expansions at some other renormalization point 
$\mu_1^2$, details can be found in \cite{BalitskyBraun1988}. 
The derivation for the evolution of string operators containing three Wilson 
lines is similar except for the fact that conformal symmetry is 
not sufficient to determine the solution explicitly.  

It was shown in \cite{BalitskyBraun1988,Balitsky1883} how the light-cone expansion of a $T$ product like 
$Tj_\mu(x)j_\nu(-x)$ into nonlocal string operators instead of local operators (in the latter case 
the choice of a suitable operator bases at a given nonleading twist 
is obscure because there are relations between the
local operators dictated by the equations of motion \cite{Jaffe1983,JaffeSoldate1982}) 
leads to a compact and much better managable series in singularities 
in deviations from the light-cone $x^2=0$ than in the local case. The leading term 
$-\frac{1}{16\pi^2x^4}\bar{\psi}(x)\gamma_\mu\hat{x}\gamma_\nu \bar{Q}^2\psi(-x)$, where 
$\bar{Q}=\frac{1}{2}\lambda^3+\frac{1}{2\sqrt{3}}\lambda^8$ is the charge matrix, 
has no definite twist. To project onto the leading twist, one can perform a symmetrization and trace
subtraction in the local expansion of this string operator, which can be reassembled 
into a nonlocal expression. Higher twist contributions are technically quite involved. 

The nonlocal light-cone expansion of \cite{BalitskyBraun1988} 
was used extensively in phenomenology. 
Besides the applications already addressed in the last section let us just mention 
two more examples: 
Light-cone sum rules were employed 
in \cite{BelyaevBraunKhodjamirianRuckl1995} to predict the heavy-mesons couplings to 
pions in nonleptonic $D$ and $B$ decays in terms of two- and three- particle 
DAs up to twist four. The pion form factor at 
intermediate momentum $Q\sim 1\,$GeV was estimated 
in \cite{BraunKhodjamirianMaul2000} using light-cone sum rules with higher 
twist distributions and radiative corrections. 

\subsubsection{Delocalized operator expansion\label{sec:DOE}}

After having discussed the expansion of a $T$ product of electromagnetic 
currents into nonlocal string operators in 
view of applications involving a nonlocal {\sl hadron-to-vacuum matrix element} or 
a nonlocal {\sl hadron-to-hadron matrix element} we turn now to the case of a 
vacuum-to-vacuum average which leads to the occurrence of 
nonlocal {\sl condensates}. The discussion of \cite{HoangHofmann2003}, which we follow, 
will be on a more phenomenological level. 
We have already introduced the approach of 
\cite{MikhailovRadyushkin1986,MikhailovRadyushkin1989,BakulevRadyushkin1991,
MikhailovRadyushkin1992,BakulevMikhailov1998} to nonlocal 
condensates in Euclidean spacetime in Sec.\,\ref{sec:PionDAsNLC}. 

Let us first discuss the case of two-point condensates. They are associated with nonlocal versions of 
the quark and gluon condesates in the conventional OPE. 
The parametrization of the nonlocal scalar quark condensate in Eq.\,(\ref{alpharep}) 
is to a good approximation also applicable to the nonlocal gluon condensate in Euclidean spacetime: 
In Fock-Schwinger gauge we have 
\cite{MikhailovRadyushkin1992}
\eqb
\label{nonGC}
\la A_\mu^a(z) A_\nu^b(y)\ra=\delta^{ab}\left(y_\mu z_\nu-\delta_{\mu\nu}(z\cdot y)\right)\frac{\la G^2\ra}{384}
\times g((z-y)^2,z^2,y^2)+\cdots
\eqe
where the local expansion of $g((z-y)^2,z^2,y^2)$ reads
\eqb
\label{gfunctionexp}
g((z-y)^2,z^2,y^2)=1-\frac{\la GD^2G\ra-\frac{2}{3}\la j^2 \ra}{18\la G^2\ra}
\left((y-z)^2+\frac{y^2+z^2}{8}\right)+\cdots\approx g((z-y)^2)\,.
\eqe
In Eq.\,(\ref{nonGC}) the paranthesis denote contributions to $\la A_\mu^a(z) A_\nu^b(y)\ra$ 
which are not captured by the {\sl scalar} function $g((z-y)^2,z^2,y^2)$. 
Clearly, the dependence on $(y-z)^2$ is much stronger than the 
dependence on $y^2$ or $z^2$. The dependence on the latter two variables thus 
can be omitted on the level of accuracy at which the local condensates are known. 
In the expansion of the scalar part $T(q^2)$ of a given current-current 
correlator $i\int d^4x\,\la Tj_\mu(x)j_\nu(0)\ra$ the Lorentz, color, 
and flavor indices in front of the nonlocal condensate get contracted and integrations over $x,y,$ and $z$ 
are performed. The nonperturbative correction to $T_{pert}(q^2)$ arising from a 
two-point condensate $g(x^2)$ can thus be written as
\eab
\label{twopointcond}
\int_{-\infty}^\infty d^4x\,f(x_1,\ldots, x_4)\,g(x^2) 
& = & 
\sum_{n_1,\ldots,n_4=0}^{\infty} f_{n_1,\ldots,n_4}(\Omega) \, 
g_{n_1,\ldots,n_4}(\Omega)
\,,\quad
\label{multDOEddim}
\end{eqnarray} 
where
\begin{eqnarray}
f_{n_1,\ldots,n_4}(\Omega) & \equiv & 
\int d^4x \, f(x_1,\ldots, x_4) \, 
\bigg[\,
\prod_{i=1}^4 \frac{H_{n_i}(\Omega x_i)}{(2 \, \Omega)^{n_i}}
\,\bigg]
\nonumber\\[2mm]
& = &
\left.
\bigg[\,
\prod_{i=1}^4 \, \frac{1}{(2 \, \Omega)^{n_i}}\,
H_{n_i}\bigg(\Omega\bigg(i\frac{d}{dk_i}\bigg)\bigg) 
\,\bigg]
\,\tilde f(k_i,\ldots ,k_4) \right|_{k_1,\ldots ,k_4=0}
\,,
\label{fmultidim}
\\[2mm]
g_{n_1,\ldots,n_4}(\Omega) & \equiv & 
\int
d^dx\, 
\bigg[\,
\prod_{i=1}^4 \frac{\Omega^{n_i+1}}{\sqrt{\pi}\,n_i!}\,H_{n_i}(\Omega x_i)
\,\bigg]
 \,\e^{-\Omega^2 x^2}
\,g(x_1,\ldots, x_4)
\nonumber\\[2mm]
& = &
\int\frac{d^4k}{(2\pi)^4}\,
\bigg[\,
\prod_{i=1}^4 \,\frac{(i k_{n_i})^{n_i}}{n_i!}
\,\bigg]
\,\e^{-\frac{k^2}{4\Omega^2}} \,\tilde g(k_1,\ldots ,k_4)
\,. 
\label{fandgddim}
\eae
A number of comments are in order. In Eq.\,(\ref{twopointcond}) the function $f(x_1,\ldots, x_4)$ 
denotes the strongly-peaked at $x=0$, perturbatively calculable, 
short-distance dependence as it arises 
in Euclidean spacetime after separating off the condensate part in an 
application of the background-field method\footnote{We have notationally suppressed 
the dependence of $f$ on the external
momentum $Q$.}. It corresponds to 
the Wilson coefficient for the local condensate in position space 
and has a width $\Delta_f$ which is comparable to 
the inverse external momentum $Q^{-1}$, see Fig.\,(\ref{Fig-fg}). 
\begin{figure}[tb]
\begin{center}
\hspace{-2.5cm}
\begin{minipage}[t]{8 cm}
\epsfig{file=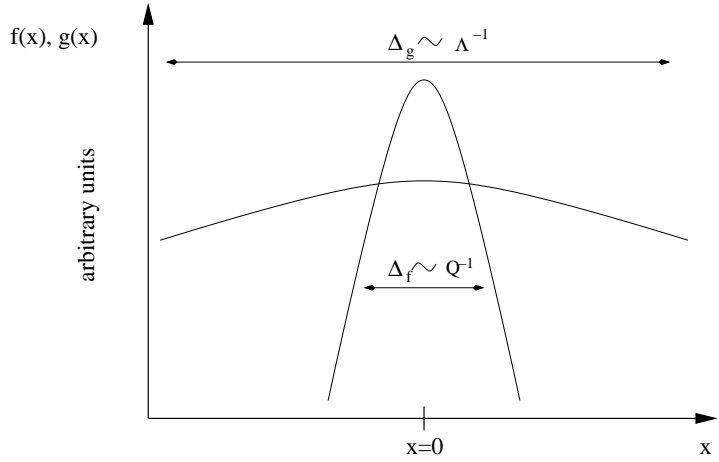,scale=0.4}
\end{minipage}
\begin{minipage}[t]{16.5 cm}
\caption{Schematic drawing of the short-distance function $f(x)$ and
the 2-point condensate $g(x)$ illustrating the scale hierarchy
$\Delta_f/\Delta_g\sim\Lambda/Q\ll 1$. Plot taken from \cite{HoangHofmann2003}.
\label{Fig-fg}}
\end{minipage}
\end{center}
\end{figure}
To arrive at the second line a complete set of functions 
$e^\Omega_{n_1\cdots n_4}(x)$ and $\tilde{e}^\Omega_{n_1\cdots n_4}(x)$, 
which span a {\sl dual space} with respect to the bilinear form
\begin{equation}
\label{bilinear1}
(f,g) \, \equiv \, \int d^4x\, f(x_1,\cdots,x_4) \,g(x^2)\, 
\end{equation}
and which are normalized to 
\begin{equation}
\label{ortho1}
\left(\,e^\Omega_{n_1,\cdots,n_4}\,,\,\tilde{e}^\Omega_{m_1,\cdots,m_4}\,\right)
\, = \, 
\delta_{n_1m_1}\cdots\delta_{n_4m_4}\,,
\end{equation}
has been inserted inbetween $f$ and $g$. In the limit, where the {\sl resolution parameter} $\Omega$ 
goes to infinity, $\Omega\to\infty$, the set 
$e^\Omega_{n_1\cdots n_4}(x)$ and $\tilde{e}^\Omega_{n_1\cdots n_4}(x)$ reproduces 
the local expansion of the nonlocal condensate, and we have 
\begin{equation}
e^{\Omega\to\infty}_{n_1\cdots n_4}(x) \, \equiv \, \frac{(-1)^{n_1+\cdots +n_4}}{n_1!\cdots n_4!}\,
\delta^{(n_1)}(x_1)\cdots\delta^{(n_4)}(x_4) 
\qquad\mbox{and}\qquad
\tilde{e}^{\Omega\to\infty}_{n_1\cdots n_4}(x) \, \equiv \, x_1^{n_1}\cdots x_4^{n_4} 
\,.
\label{deltabasis}
\end{equation}
In Eq.\,(\ref{deltabasis}) $\delta^{(n_i)}$ denotes the $n_i$th derivative of the delta function. 
For definiteness, we have chosen in Eqs.\,(\ref{fmultidim}),(\ref{fandgddim}) the following set of 
products of Hermite polynomials and their duals
\begin{eqnarray}
e_{n_1,\cdots,n_4}^\Omega(x) & \equiv &
\frac{\Omega^{(n_1+1)+\cdots+(n_4+1)}}{\pi^2\,n_1!\cdots n_4!}\,H_{n_1}(\Omega x)\cdots 
H_{n_4}(\Omega x)\,\e^{-\Omega^2 (x_1^2+\cdots+x^2_4)} 
\,,
\nonumber 
\\[2mm]
\tilde e_{n_1,\cdots,n_4}^\Omega(x) & \equiv & \frac{H_{n_1}(\Omega x)\cdots 
H_{n_4}(\Omega x)}{(2 \, \Omega)^{n_1+\cdots+n_4}}
\,.
\label{hermitebasis}
\end{eqnarray}
Obviously, the basis (\ref{hermitebasis}) reduces to the basis (\ref{deltabasis}) in 
the limit $\Omega\to\infty$. Different choices of one-parameter 
basis functions are equally well possible. 
Let us emphasize at this point that the delocalization of the OPE, 
which is achieved in this manner differs from 
\cite{MikhailovRadyushkin1986,MikhailovRadyushkin1989,BakulevRadyushkin1991,
MikhailovRadyushkin1992,BakulevMikhailov1998} by the fact that with a truncation of the expansion 
(\ref{twopointcond}) at $n_1=0,\cdots,n_4=0$ a resolution 
parameter, which eventually will be chosen to be equal to the applied external momentum, 
cuts off the irrelevant long-distance contribution of the nonlocal condensate 
$g(x^2)$ to the integral. The Wilson coefficient in momentum space 
is equal to the one of the local expansion in this truncation, and 
this leads effectively to the running of the 
condensate with resolution. In Sec.\,\ref{sec:EPCO} an 
alternative approach to running condensates will be discussed and applied.

The properties of Hermite polynomials dictate 
the following evolution equations for $f_{n_1,\ldots,n_4}(\Omega)$ and 
$g_{n_1,\ldots,n_4}(\Omega)$ 
\begin{eqnarray}
\frac{d}{d\,\Omega}\,f_{n_1,\ldots,n_4}(\Omega) & = &
\frac{(n_1-1)\,n_1}{2\,\Omega^3}\,f_{n_1-2,\ldots,n_4}(\Omega)
+ \ldots +
\frac{(n_4-1)\,n_4}{2\,\Omega^3}\,f_{n_1,\ldots,n_4-2}(\Omega)
\,,
\nonumber\\[2mm]
\frac{d}{d\,\Omega}\,g_{n_1,\ldots,n_4}(\Omega) & = &
-\,\frac{(n_1+1)(n_1+2)}{2\,\Omega^3}
\,g_{n_1+2,\ldots,n_4}(\Omega)
- 
\nonumber\\ & & \hspace{1cm}
\ldots -
\frac{(n_4+1)(n_4+2)}{2\,\Omega^3}
\,g_{n_1,\ldots,n_4+2}(\Omega)
\,.
\label{fandgevolutionddim}
\end{eqnarray}
The parametric counting of the short-distance 
coefficients $f_{n_1,\ldots,n_4}$ and the 
long-distance functions $g_{n_1,\ldots,n_4}$ is similar 
as in the local expansion for $\Omega>Q$, the leading local terms are corrected 
by a {\sl finite} sum over powers in $Q/\Omega$ in the former case and 
an infinite sum over powers of $\Lambda/\Omega$ in the latter case. Here $\Lambda$ denotes the 
mass scale associated with the fall-off of $g(x^2)$ in Euclidean spacetime, 
for details see Sec.\,(\ref{sec:GFC}). 

The consideration of 2-point condensates outlined above can be generalized to 
$(N>2)$-point condensates by an auxiliary increase to $N-1$ spacetime dimensions on which 
the functions $f$ and $g$ depend. 

\subsubsection{The gluonic field strength correlator and bilocal quark condensate 
on the lattice\label{sec:GFC}} 

Here we review what is known
from lattice simulations about the gauge invariant, gluonic field
strength correlator which bears information on the nonlocal 
gluon condensate of Eq.\,(\ref{nonGC}). Appealing to Poincar\'{e}, parity and time inversion 
invariance in Euclidean spacetime, the following parametrization was introduced in \cite{DoschSimonov1988}: 
\begin{eqnarray}
g_{\mu\nu\kappa\lambda}(x) &\equiv& G^a_{\mu\nu}(x)\,[x,0]_{ab}\,G^b_{\kappa\lambda}(0)\nonumber\\ 
&=&(\,\delta_{\mu\kappa}\delta_{\nu\lambda}-\delta_{\mu\lambda}\delta_{\nu\kappa}\,)
\,\Big[\,D(x^2)+D_1(x^2)\,\Big]
\nonumber\\[2mm]  & &
+\,(\,x_\mu x_\kappa\delta_{\nu\lambda}-x_\mu x_\lambda\delta_{\nu\kappa}+
x_\nu x_\lambda\delta_{\mu\kappa}-x_\nu x_\kappa\delta_{\mu\lambda}\,)
\,\frac{\partial D_1(x^2)}{\partial x^2}
\,.
\label{pargfc}
\end{eqnarray}
To separate perturbative from nonperturbative contributions the scalar
functions $D$ and $D_1$ are usually \cite{DoschEidemullerJaminMeggiolaro2000,BaliBrambillaVairo1998,
D'EliaDiGiacomoMeggiolaro1997} fitted as
\begin{eqnarray}
D(x^2) & = & 
A_g\,\exp\bigg(\!-\frac{|x|}{\lambda_g}\bigg)
+\frac{a_g}{|x|^4}\,\exp\bigg(\!-\frac{|x|}{\lambda_g}\bigg)\,,
\nonumber\\[2mm] 
D_1(x^2) & = & 
A_1\,\exp\bigg(\!-\frac{|x|}{\lambda_g}\bigg)+
\frac{a_1}{|x|^4}\,\exp\bigg(\!-\frac{|x|}{\lambda_g}\bigg)
\,. 
\label{gfcfit}
\end{eqnarray}
The power-like behavior in Eqs.\,(\ref{gfcfit}) at small $|x|$ is
believed to catch most of the perturbative physics, although it is
known that partially summed perturbation theory may generate less divergent 
renormalon contributions at $|x|=0$ as well, see Sec.\,(\ref{sec:Renorm}). As we have seen, the coefficients 
of these contributions are ambiguous. We ignore this subtlety and assume that the 
purely exponential terms in Eqs.\,(\ref{gfcfit}) exclusively carry the 
nonperturbative information\footnote{This is in contrast to 
\cite{MikhailovRadyushkin1986,MikhailovRadyushkin1989,BakulevRadyushkin1991,
MikhailovRadyushkin1992,BakulevMikhailov1998} where a Gaussian behavior was assumed.}. 
We also note that due to the cusp of the exponential ansatz in
Eq.\,(\ref{gfcfit}),  derivatives of the nonperturbative part have nonlogarithmic
UV singularities at $x=0$ that can only be defined in a nonperturbative UV regularization
scheme. This is obviously in conflict with the well-tested idea that UV singularities, when 
treated in renormalization-group improved perturbation theory, lead to a logarithmic, 
anomalous scaling of averages over local operators. It does then make no sense to assume 
the exponential behavior of Eq.\,(\ref{gfcfit}) down to arbitrarily small
distances - after all the lattice resolution at which the field strength correlator 
is measured is finite, $a^{-1}\sim 2\,$GeV. We will re-address this concern 
in Secs.\,\ref{sec:MCHQ} and \ref{sec:EPCO}. We also would like to mention that 
in the framework of a nonrelativistic effective-field-theory 
formulation of QCD a decomposition 
of the correlator (\ref{pargfc}) into color magnetic and 
color electric correlations yields an electric 
correlation length, which is smaller than the magnetic, 
one as a result of lattice measurements \cite{MorningstarJugeKuti1998,
BrambillaPinedaSotoVairo2000,BaliPineda2003}. 
See also \cite{BrambillaEirasPinedaSotoVairo2003} for a relation between the 
decay widths of heavy quarkonia on the one hand and electric and 
magnetic gluon correlators on the other hand.

In a recent unquenched lattice simulation~\cite{D'EliaDiGiacomoMeggiolaro1997} 
(see also Ref.\,\cite{BaliBrambillaVairo1998}), where the gluon field strength correlator was measured with a
resolution of $a^{-1}\approx (0.1\,\mbox{fm})^{-1} \approx 2$\,GeV
between $3$ and $8$ lattice spacings, it was found that  
\begin{equation}
\label{lat}
\frac{A_g}{A_1} \, \approx \, 9
\qquad\mbox{and}\qquad
\lambda_g^{-1} \, \approx \, 0.7\,\mbox{GeV} 
\,.
\end{equation}
Notice that the inverse correlation length is somewhat
larger than the typical hadronization scale $\Lambda_{\rm QCD}$. 
While the actual size of $A_g$ and $A_1$ depend quite strongly on
whether quenched or unquenched simulations are carried out and also on the values for the 
light quark masses assumed, the ratio $A_g/A_1$
and the correlation length $\lambda_g$ were found to be 
quite stable \cite{D'EliaDiGiacomoMeggiolaro1997}. 
The $D_1$ contribution can be neglected since $A_g\gg A_1$ (see Eq.\,(\ref{pargfc})), and one may then 
write 
\begin{equation}
g_{\mu\nu\kappa\lambda}^{\rm non-pert}(x) \, = \, 
\frac{1}{12}\, 
(\,\delta_{\mu\kappa}\delta_{\nu\lambda}-\delta_{\mu\lambda}\delta_{\nu\kappa}\,)\,.
\,g(|x|)
\,
\label{pargfcsimple}
\end{equation}
This has the same tensor structure as the local condensate 
$g_{\mu\nu\kappa\lambda}^{\rm non-pert}(0)$. It is in agreement 
with the local expansion of $g_{\mu\nu\kappa\lambda}$ which reads 
\cite{NikolaevRadyushkin1983}
\begin{eqnarray}
\label{2dG^2}
-\la g^2 G^a_{\mu\nu}\pd_\rho\pd_\rho G^a_{\alpha\beta}\ra 
& = & 
8\,O^-\left(\delta_{\mu\beta}\delta_{\alpha\nu}-
\delta_{\mu\alpha}\delta_{\nu\beta}\right)
\nonumber\\[2mm]    
& & 
+ \,O^-\left(\delta_{\mu\beta}\delta_{\alpha\nu}+
\delta_{\alpha\nu}\delta_{\mu\beta}-\delta_{\alpha\mu}\delta_{\nu\beta}-
\delta_{\beta\nu}\delta_{\mu\alpha}\right)
\nonumber\\[2mm] 
& &
+\,O^+\left(\delta_{\mu\beta}\delta_{\alpha\nu}+
\delta_{\mu\beta}\delta_{\alpha\nu}- 
\delta_{\mu\alpha}\delta_{\nu\beta}-
\delta_{\nu\beta}\delta_{\mu\alpha}\right)
\,,
\end{eqnarray}
where
\begin{equation}
\label{O+-}
O^{\pm}
\, \equiv \,
\frac{1}{72}\la g^4 j^a_\mu j^a_\mu\ra\pm
\frac{1}{48}\la g^3f_{abc}G^a_{\mu\nu}G^b_{\nu\lambda}G^c_{\lambda_\mu}\ra\,,
\end{equation} 
and $j^a_\mu$ denotes a light flavor-singlet current. 
Using exact vacuum saturation for the four-quark operator, $\alpha_s(\mu=0.7\,\mbox{GeV})=0.7$, 
and the following instanton-calculus determined value of the condensate 
$\la g^3 f_{abc}G^a_{\mu\nu}G^b_{\nu\lambda}G^c_{\lambda\mu}\ra=0.045$\,GeV$^6$, and 
$\langle \bar{q}q\rangle=-(0.24\,\,\mbox{GeV})^3$, it can 
be shown that the tensor structure belonging to the 
local condensate dominates the remainder in 
$\la g^2 G^a_{\mu\nu}\pd_\rho\pd_\rho G^a_{\mu\nu}\ra$ by a 
factor of about nine \cite{HoangHofmann2003}. 

In the case of the scalar, gauge invariant 
quark correlator the situation is similar \cite{D'EliaDiGiacomoMeggiolaro1999}. 
For the longitudinal vector quark correlator defined as
\eqb
\label{vecqurk}
-\frac{x^\mu}{|x|}\bar{q}(x)\gamma_\mu[x,0]q
\eqe
an about ten times smaller correlation length as 
compared to the scalar case was obtained in \cite{D'EliaDiGiacomoMeggiolaro1999} 
with a four-flavor simulation (staggered fermions) at a lattice resolution of $a^{-1}\sim 2\,$GeV.

\subsubsection{Model calculation of the nonperturbative shift in 
the ground-state energy of heavy quarkonium\label{sec:MCHQ}}

In this section we demonstrate in a model calculation 
for the nonperturbative shift of the ground-state ($n^{\,2s+1}L_j=1^{\,3}S_1$) 
energy $E^{np}$ of the 
heavy quarkonia, where the ``exact'' 
result can be calculated, in dependence on the heavy-quark mass $m$ 
how much better the delocalized operator 
expansion converges as compared to the local 
expansion \cite{HoangHofmann2003}. Our investigation is not 
intended to represent a phenomenological study
of non-perturbative effects in heavy quarkonium energy levels.

Among the early applications of the OPE in QCD was the analysis of
non\-per\-tur\-ba\-tive effects in heavy quarkonium
systems \cite{Voloshin1979,Leutwyler1981,Voloshin1982}. 
Heavy quarkonium systems are non\-re\-la\-ti\-vis\-tic quark-antiquark bound
states for which the following hierarchy of the relevant  
physical scales $m$ (heavy quark mass), $m v$
(relative momentum), $m v^2$ (kinetic energy) and $\Lambda_{\rm QCD}$ holds:
\begin{equation}
\label{condition1}
m \, \gg \, m v \, \gg \, m v^2 \, \gg \, \Lambda_{\rm QCD}
\,.
\end{equation}
The spatial size of the quarkonium system $\sim (m v)^{-1}$ is
much smaller than the typical dynamical time scale $\sim (m
v^2)^{-1}$. In practice the last of the conditions in
Eq.\,(\ref{condition1}), which relates the vacuum correlation length
with the quarkonium energy scale, is probably not satisfied for any
known quarkonium state, not even for $\Upsilon$ mesons \cite{Gromes1982}.
Only for top-antitop quark threshold production
condition (\ref{condition1}) may be a viable assumption \cite{Hoang2000}. 

We adopt the local version of the
multipole expansion (OPE) for the expansion in the ratios of the
scales $m$, $m v$ and $m v^2$. The resolution dependent expansion
(DOE) is applied with respect to the ratio of  the scales $m v^2$ and 
$\Lambda_{\rm QCD}$.   
The former expansion amounts to the usual treatment of the dominant
perturbative dynamics by means of a nonrelativistic two-body
Schr\"odinger equation. The interaction with the
nonperturbative vacuum is accounted for by two insertions of the local 
$\bmx\bmE$ dipole operator, $\bmE$ being the chromoelectric
field.\,\cite{Voloshin1979} The chain of VEV's of the two gluon operator
with increasing numbers of covariant derivatives 
times powers of quark-antiquark octet propagators \cite{Voloshin1979},
i.e. the expansion in  $\Lambda/m v^2$, is treated in the DOE.
 
At leading order in the local multipole expansion with respect to the
scales $m$, $m v$, and $m v^2$ the expression for the nonperturbative
corrections to the ground state energy reads
\begin{equation}
E^{np} 
\, = \,
\int_{-\infty}^\infty\! dt \, f(t) \, g(t)
\,,
\label{Enonpertdef}
\end{equation}
where
\begin{equation}
f(t) \, =  \, \frac{1}{36}\,
\int \frac{dq_0}{2\pi}\,\e^{i q_0(it)}\,
\int\! d^3\bmx \int\! d^3\bmy\,
\phi(x)\,(\bmx\bmy)\,
G_O\bigg(\bmx,\bmy,-\frac{k^2}{m}-q_0\bigg)\,\phi(y)
\,,
\label{Enonpertfdef}
\end{equation}
with
\begin{eqnarray}
G_O\bigg(\bmx,\bmy,-\frac{k^2}{m}\bigg)
& = &
\sum_{l=0}^\infty\,(2l+1)\,P_l\bigg(\frac{\bmx\bmy}{x y}\bigg)\,
G_l\bigg(x,y,-\frac{k^2}{m}\bigg)
\,,
\nonumber\\[2mm]
G_l\bigg(x,y,-\frac{k^2}{m}\bigg) 
& = &
\frac{m k}{2\pi}\,(2kx)^l\,(2ky)^l\,\e^{-k(x+y)}\,
\sum_{s=0}^\infty \,\frac{L_s^{2l+1}(2kx)\,L_s^{2l+1}(2ky)\,s!}
          {(s+l+1-\frac{m\,\alpha_s}{12\,k})\,(s+2l+1)!}
\,,
\nonumber\\[2mm]
\phi(x) & = & \frac{k^{3/2}}{\sqrt{\pi}}\,\e^{-k x}
\,,\qquad
\nonumber\\[2mm]
k & = & \frac{2}{3}\,m\alpha_s
\end{eqnarray}
The term $G_O$ is the quark-antiquark octet
Green-function~\cite{Voloshin1982},
and $\phi$ denotes the ground state wave function of the quarkonium system. The functions $P_n$ and
$L_n^k$ are Legendre and Laguerre polynomials, respectively.
Since we neglect the
spatial extension of the quarkonium system with respect to the
interaction with the nonperturbative vacuum, the insertions of the
$\bmx\bmE$ operator probe only the temporal correlations in the
vacuum. This effectively renders the problem one-dimensional. Notice 
that $t$ is the Euclidean time. 

Since the spatial
extension of the quarkonium system is neglected and the average time
between interactions with the vacuum is of the order of the inverse
kinetic energy, the characteristic width of the function $f$ 
in Eq.\,(\ref{Enonpertfdef}) is of order $(m v^2)^{-1}\sim(m\alpha_s^2)^{-1}$. 
The values of the first few local multipole moments $f_n(\infty)$, which
correspond to local Wilson coefficients, read
\begin{equation}
\begin{array}{ll}
f_0(\infty) \, = \, 1.6518\,\frac{m}{36\,k^4}\,, \qquad 
&  
f_2(\infty) \, = \, 1.3130\,\frac{m^3}{36\,k^8}\,, 
\\[3mm]
f_4(\infty) \, = \, 7.7570\,\frac{m^5}{36\,k^{12}}\,,  
&  
f_6(\infty) \, = \, 130.492\,\frac{m^7}{36\,k^{16}}\,, 
\\[3mm]
f_8(\infty) \, = \, 4474.1\,\frac{m^9}{36\,k^{20}}\,,
&  
f_{10}(\infty) \, = \,262709.3\,\frac{m^{11}}{36\,k^{24}}\,,\ldots\,.
\end{array}
\label{fns}
\end{equation}
The term $f_0(\infty)$
agrees with Ref.\,\cite{Leutwyler1981,Voloshin1982} and $f_2(\infty)$ with
Ref.\,\cite{Pineda1997}. The results for $f_{n>2}(\infty)$ are new. 

We use a lattice-inspired model function $g(t)$ for the nonperturbative gluonic field strength
correlator of the form
\begin{eqnarray}
g(t) & = & 12\,A_g \,\exp\Big(-\sqrt{t^2+\lambda_g^{2}}/\lambda_g + 1\Big)
\,,
\nonumber\\[2mm]
A_g & = & 0.04~\mbox{GeV}^{4}\,,
\qquad
\lambda_g^{-1} \, = \, 0.7~\mbox{GeV}
\,,
\label{gmodel}
\end{eqnarray}
This model has an exponential large-time behavior according to
Eq.\,(\ref{gfcfit}) and a smooth behavior for small $t$. The 
local dimension four gluon condensate in this model is
\begin{equation}
\Big\langle\frac{\alpha_s}{\pi}\,G_{\mu\nu}^a G_{\mu\nu}^a\Big\rangle
\, = \, 
\frac{6\,A_g}{\pi^2} 
\, = \,  0.024\,\,\mbox{GeV}^4
\,.
\end{equation}
The exact form of the
model for the nonperturbative gluonic field strength
correlator is not important for our purposes as long as the derivatives of
$g(t)$ at $t=0$ are well defined, 
see e.g. Refs.\,\cite{Gromes1982,Balitsky1985} for different model choices. 
Tab.\,\ref{tabquarkonium} shows the exact result in the
model (\ref{gmodel}) and
the first four terms of the DOE of $E^{np}$ for quark 
masses $m=5,25,45,90,175$\,GeV and for $\Omega=\infty$
and $\Omega=k^2/m$. For each value of the quark mass the strong coupling has
been fixed by the relation $\alpha_s = \alpha_s(k)$. 
\begin{table}[t!]  
\begin{center}
\begin{tabular}{|c|c|c|c||r|r|r|r|r|} \hline
\multicolumn{5}{|c|}{} & \multicolumn{2}{|c|}{$\Omega=\infty$} 
                  & \multicolumn{2}{|c|}{$\Omega=k^2/m$}
 \\ \hline
$m$ &  & $k^2/m$ & $E^{np}$ & & $f_ng_n$ & $\sum^n_{i=0}f_ig_i$ 
                              & $f_ng_n$ & $\sum^n_{i=0}f_ig_i$
 \\ 
(GeV) & \raisebox{1.5ex}[-1.5ex]{$\alpha_s$} & (MeV) & (MeV)
   & \raisebox{1.5ex}[-1.5ex]{$n$} & (MeV) & (MeV) & (MeV) & (MeV)
 \\ \hline\hline
$5$ &   $0.39$ & $0.338$ & $24.8$ & 
           0 & $38.6$      &  $38.6$    & $24.2$    & $24.2$
\\ \cline{5-9}
& & & &    2 & $-65.7$     & $-27.2$    & $-3.9$    & $20.3$
\\ \cline{5-9}
& & & &    4 & $832.7$     & $805.5$    & $12.1$    & $32.4$
\\ \cline{5-9}
& & & &    6 & $-35048.0$  & $-34242.4$  & $-43.1$  & $-10.8$
\\ \hline\hline
$25$ &  $0.23$ & $0.588$ & $12.6$ & 
           0 & $16.0$    &  $16.0$    &  $12.8$ &  $12.8$
\\ \cline{5-9}
& & & &    2 & $-9.0$    &  $7.0$     &  $-1.2$ &  $11.5$
\\ \cline{5-9}
& & & &    4 & $37.8$    &  $44.8$    &  $2.6$  &  $14.1$
\\ \cline{5-9}
& & & &    6 & $-526.6$  &  $-481.8$  &  $-6.7$ &  $7.4$
\\ \hline\hline
$45$ &  $0.19$ & $0.722$ & $4.9$ & 
           0 & $5.9$    &  $5.9$    &  $5.0$   &  $5.0$
\\ \cline{5-9}
& & & &    2 & $-2.2$   &  $3.7$    &  $-0.4$  &  $4.6$
\\ \cline{5-9}
& & & &    4 & $6.1$    &  $9.8$    &  $0.7$   &  $5.3$
\\ \cline{5-9}
& & & &    6 & $-56.4$  &  $-46.6$  &  $-1.4$  &  $3.8$
\\ \hline\hline
$90$ &  $0.17$ & $1.156$ & $1.05$ & 
           0 & $1.15$   &  $1.15$  &  $1.07$   &  $1.07$
\\ \cline{5-9}
& & & &    2 & $-0.17$  &  $0.98$  &  $-0.04$  &  $1.02$
\\ \cline{5-9}
& & & &    4 & $0.18$   &  $1.16$  &  $0.04$   &  $1.07$
\\ \cline{5-9}
& & & &    6 & $-0.65$  &  $0.51$  &  $-0.07$  &  $1.00$
\\ \hline\hline
$175$ & $0.15$ & $1.750$ & $0.245$ & 
           0 & $0.258$   &  $0.258$  &  $0.249$   &  $0.249$
\\ \cline{5-9}
& & & &    2 & $-0.016$  &  $0.242$  &  $-0.005$  &  $0.244$
\\ \cline{5-9}
& & & &    4 & $0.008$   &  $0.250$  &  $0.003$   &  $0.247$
\\ \cline{5-9}
& & & &    6 & $-0.012$  &  $0.237$  &  $-0.003$  &  $0.244$
\\ \hline
\end{tabular}
\caption{\label{tabquarkonium} 
Nonperturbative corrections $E_{np}$ to the heavy quarkonium ground state level
at leading order in the multipole expansion with respect to the scales
$m$, $mv$ and $mv^2$ for various quark masses $m$ based on the model
in Eq.\,(\ref{gmodel}). Displayed are the exact result and the first
few orders in the DOE for $\Omega=\infty$ and $\Omega=k^2/m$.
The numbers are rounded off to units of $0.1$, $0.01$ or $0.001$\,MeV. 
}
\end{center}
\vskip 3mm
\end{table}
Notice that the series all appear to be asymptotic, i.e.\,\,they are not
convergent for any resolution. The local expansion ($\Omega=\infty$)
is badly behaved for small quark masses because for $k^2/m<\lambda_g^{-1}$ any
local expansion is meaningless. In particular, for $m=5$\,GeV the
subleading dimension-six term is already larger than the parametrically
leading dimension-four term. For quark masses, where $k^2/m\ge\lambda_g^{-1}$, the local
expansion is reasonably good. However, at the finite resolution
$\Omega=k^2/m$, the size of higher order terms is considerably smaller
than in the local expansion for all quark masses, and the series is apparently 
much better behaved. The size of the order-$n$ term is
suppressed by approximately a factor $2^{-n}$ as compared to the order-$n$ term 
in the local expansion. We see explicitly that terms in
the series with larger $n$ increase more quickly in magnitude in the local expansion
scale as compared to expansion at finite resolution. One also observes that even
in the case $k^2/m < \lambda_g^{-1}$, where the leading term of the
local expansion overestimates the exact result, the leading
term in the delocalized expansion for $\Omega=k^2/m$ agrees with the 
exact result within a few percent. 

For a realistic treatment of the nonperturbative contributions in the heavy quarkonium
spectrum a model-independent analysis should be carried out. In
addition, also higher orders in the local multipole expansion with
respect to the  ratios of scales $m$, $m v$ and $m v^2$ should be
taken into account, which have been neglected here. These corrections
might be substantial, in particular for smaller quark masses. 

\subsubsection{Running gluon condensate from the $\bar{c}c$ spectrum \label{sec:CSRRGC}}

In this section we review an extraction of the running gluon condensate 
from the charmonium spectrum using moment sum rules \cite{HoangHofmann2003}. 
We have already discussed this approach 
in Sec.\,(\ref{sec:J/psi}) where an OPE was assumed (for an analysis involving operators at $n=8$ see \cite{Zyablyuk2003}. 
The $n$th moment ${\cal M}_n$ reads 
\cite{NikolaevRadyushkin1983}
\begin{eqnarray}
{\cal M}_n 
& = &
\frac{3}{4\pi^2}\frac{2^n(n+1)(n-1)!}{(2n+3)!!}\,\frac{1}{(4m_c^2)^n}\,
\bigg\{\,
1 \, + \, \mbox{[pert. corrections]}
\nonumber\\[2mm]
 & & \hspace{1cm}
\,+\,
\delta^{(4)}_n\,\langle g^2 G^2\rangle 
\,+\,
\Big[\,
\delta^{(6)}_{G,n}\,\langle g^3 f G^3 \rangle
\,+\,
\delta^{(6)}_{j,n}\,\langle g^4 j^2 \rangle
\,\Big] \,+\,\ldots
\,\bigg\}
\,,
\nonumber\\[2mm]\mbox{where}
\nonumber\\[2mm]
\delta^{(4)}_n 
& = &
-\frac{(n+3)!}{(n-1)!\,(2n+5)}\,\frac{1}{9\,(4m_c^2)^2}
\,,
\nonumber\\[2mm]
\delta^{(6)}_{G,n} 
& = &
\frac{2}{45}\frac{(n+4)!\,(3n^2+8n-5)}{(n-1)!\,(2n+5)\,(2n+7)}\,
\frac{1}{9\,(4m_c^2)^3}
\,,
\nonumber\\[2mm]
\delta^{(6)}_{j,n} 
& = &
-\frac{8}{135}\,\frac{(n+2)!\,(n+4)\,(3n^3+47 n^2+244 n+405)}
      {(n-1)!\,(2n+5)\,(2n+7)}\,\frac{1}{9\,(4m_c^2)^3}
\,,
\label{Mexplicit}
\end{eqnarray}
and $\langle g^3 f G^3\rangle \equiv 
\langle g^3 f^{abc} G^a_{\mu\nu} G^b_{\nu\lambda}
G^c_{\lambda\mu}\rangle=0.045\,\,\mbox{GeV}^6$ (instanton gas approximation \cite{SVZ19791}), 
$\langle g^4 j^2 \rangle \equiv \langle g^4 j^a_\mu j^a_\mu\rangle=
-\rho\, 4/3(4\pi)^2\alpha_s^2\langle \bar{q}q\rangle^2$ ($\alpha_s(\mu=0.7\,\mbox{GeV})=0.7$ and 
$\langle \bar{q}q\rangle=-(0.24\,\,\mbox{GeV})^3$, $\rho=1$ $\rightarrow$ exact vacuum saturation), 
$j^a_\mu$ being the light flavor singlet current. We work with the ratio
\begin{equation}
r_n 
\, \equiv \,
\frac{{\cal M}_n}{{\cal M}_{n-1}}\,.
\label{mr}
\end{equation}
The extraction of the running of the gluon condensate is reliable\footnote{Recall, 
that the gluon condensate perturbatively is a renormalization-group invariant and 
thus does not scale logarithmically.} provided that subleading, 
dimension-six power corrections in Eq.\,(\ref{Mexplicit}) 
are much smaller than the dimension-six power correction
\begin{equation}
\frac{1}{4\Omega^2}\,\delta^{(4)}_n\,
\langle g^2 G D^2 G \rangle 
\label{runningdim6est}
\end{equation}
stemming from the local expansion of the 
dimension-four running gluon condensate. 
Table \ref{tabcharmonium} shows that in the OPE for $r_n$ 
this is indeed the case provided that $n$ 
is sufficiently large (in practice $n\ge 4$). As expected, Table \ref{tabcharmonium} 
indicates that for a value of the 
four-quark condensate twice the value obtained from 
exact vacuum saturation the convergence of the dimension-six part of the running 
gluon condensate towards that of the full OPE is slower than in the case of 
exact vacuum saturation.
\begin{table}[t!]  
\begin{center}
\begin{tabular}{|c|c|c|c|c|c|c|c|c|} \hline
$n$ & 1 &  2 &  3 &  4 &  5 &  6 &  7 &  8  
\\ \hline
$\rho=1$ & 0.13 & 0.36 & 0.56 & 0.73 & 0.87 & 0.99 & 1.08 & 1.16
\\ \hline
$\rho=2$ & 0.09 & 0.25 & 0.42 & 0.57 & 0.69 & 0.80 & 0.89 & 0.97
\\ \hline
\end{tabular}
\caption{Ratio of the local dimension-six contributions contained in
the running gluon condensate and in the full OPE for $r_n$ as a
function of $n$.\label{tabcharmonium} 
}
\end{center}
\vskip 3mm
\end{table}
On the experimental side of the sum rules we have used the spectrum 
as compiled in \cite{KuhnSteinhauser2001}, 
see this reference for details. 
To compare the running gluon condensate as extracted from the data 
with the model expression (using the lattice fit of Eq.\,(\ref{gfcfit}) and 
the $n_i=0$ expression in Eq.\,(\ref{fandgddim})) we have to make a choice for the 
resolution scale $\Omega$. The following physical arguments apply: 
For large $n$, i.e.\,\,in the nonrelativistic
regime, the width of the short-distance function $f$ 
is of the order of the quark c.m.\,kinetic energy 
$mv^2$, which scales like $m/n$ because the average quark velocity in
the $n$-th moment scales like $1/\sqrt{n}$ \cite{Hoang1999}. 
For small $n$, on the other hand, the relevant short-distance scale
is just the quark mass. We take this as a guideline to use $\Omega=2m_c/n$ 
as the relevant resolution scale in the model expression 
\begin{eqnarray}
\lefteqn{
\Big\langle\frac{\alpha_s}{\pi}G^2\Big\rangle_{\rm lat}(\Omega)
\, = \,
\frac{6\,A_g}{\pi^2}\,
\bigg\{\,
1 \, + \, \frac{1}{4\,\Omega^2\,\lambda_g^2} 
}
\nonumber\\[2mm]
& & \qquad
-\,
\frac{3\,\sqrt{\pi}}{4\,\Omega\,\lambda_g}\,
\bigg(1 + \frac{1}{6\,\Omega^2\,\lambda_g^2}\bigg)
\,\e^{1/(4\,\Omega^2\,\lambda_g^2)}
 \,\bigg(\,1 - {\rm erf}\Big(\frac{1}{2\,\Omega\,\lambda_g}\Big)
\,\bigg)
\,\bigg\}
\,.\qquad
\label{runningG2lattice}
\end{eqnarray}
for the running gluon condensate ($\lambda_g^{-1}=0.7$~GeV and $A_g=0.04$ which corresponds to 
$\langle(\alpha_s/\pi)G^2\rangle_{\rm lat}(\infty)=0.024\,\,\mbox{GeV}^4$). 
In Fig.\,(\ref{Fig-RGC}) the results of the data extraction 
and the model calculation are shown.
\begin{figure}[tb]
\begin{center}
\hspace{-4.5cm}
\begin{minipage}[t]{8 cm}
\epsfig{file=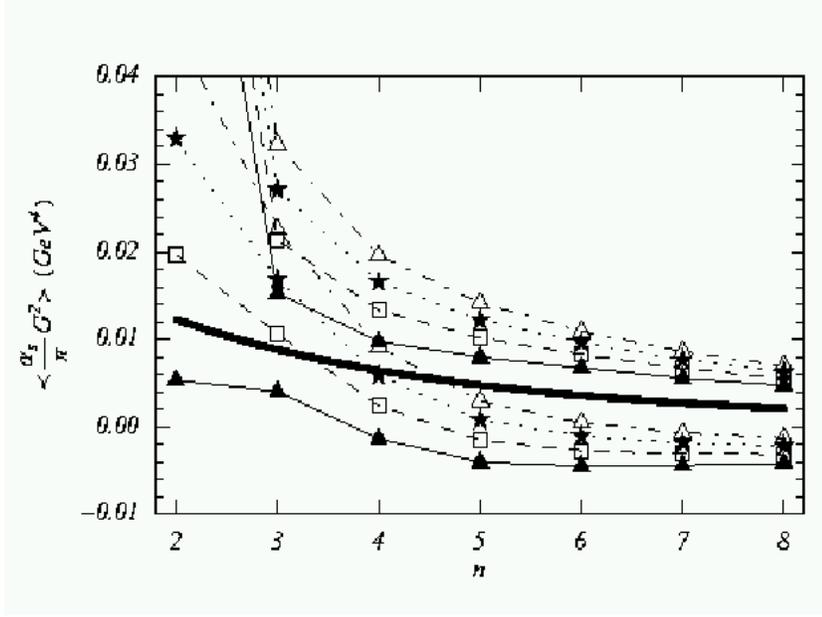,scale=0.5}
\end{minipage}
\begin{minipage}[t]{16.5 cm}
\caption{The running gluon condensate as a function of $n$ when extracted from
the ratio of moments $r_n$ for $\overline{\mbox{MS}}$ charm-quark masses 
$\overline{m}_c(\overline{m}_c)=1.23$ (white triangles), $1.24$ 
(black stars), $1.25$ (white squares) and $1.26$~GeV (black
triangles). The area between the upper and lower symbols represents the
uncertainties. 
The thick solid line indicates the running gluon condensate as it is 
obtained from the lattice-fitted ansatz in Eqs.\,(\ref{gfcfit}). Plot taken from \cite{HoangHofmann2003}.
\label{Fig-RGC}}
\end{minipage}
\end{center}
\end{figure}
The $n$-dependence of the lattice-inspired running gluon
condensate\footnote{Here and in Sec.\,\ref{sec:ccbarru} a 
steepening of the model curve (a stronger running) at large 
momenta (small $n$) will occur if the lattice fit of 
Eq.\,(\ref{gfcfit}) is considered as the result of a finite-resolution measurement, 
see Sec.\,\ref{sec:EPCO}.} and the result obtained from the charmonium moment sum rules are 
consistent for larger $n$. For small $n$ no conclusive statement can be made, 
recall Table\,\ref{tabcharmonium} and the fact that for small $n$ the sensitivity to 
the error in the continuum region of the spectrum is enhanced.

\subsubsection{Running gluon condensate from the $\tau$ decay spectrum\label{sec:ccbarru}}

In this section we review an independent extraction of the running 
gluon condensate from the spectral function of the $V+A$ channel 
\cite{HoangHofmann2003} as it was measured in 
$\tau$ decays by Aleph \cite{Aleph1998} and Opal \cite{Opal1999} at LEP. In this channel 
the OPE of the associated current-current correlator 
$i\int d^4\!x\, \e^{iqx}\la T j_\mu^{L}(x)j_\nu^{R}(0)\ra$ 
(with currents $j_\mu^{L/R}=\bar{u}\gamma_\mu(1\pm\gamma_5)d$) is dominated 
by the gluon condensate \cite{BraatenNarisonPich1992,Le DiberderPich19921,Le DiberderPich19922}, 
the dimension-six power corrections that are not 
due to the local expansion of the running gluon condensate 
are suppressed. We use a sum rule for the cutoff independent Adler 
function
\begin{equation}
\label{Adler}
D(Q^2) 
\, \equiv \, 
-Q^2\,\frac{\partial\,\Pi^{\rm V+A}(Q^2)}{\partial\, Q^2}
\, = \,
\frac{Q^2}{\pi}\,\int_0^\infty ds\, 
 \frac{\mbox{Im}\,\Pi^{\rm V+A}(s)}{(s+Q^2)^2}
\,.
\label{Adlerdef}
\end{equation}
For the $V+A$ spectral function we have used
the Aleph measurement \cite{Aleph1998} in the resonance region up to 
$2.2$~GeV$^2$. For the continuum region above $2.2$~GeV$^2$ 
3-loop perturbation theory, which we also used on the OPE side, was employed (with $\alpha_s(M_Z)=0.118$), and we
have set the renormalization scale $\mu$ equal to $Q$. 
The Wilson coefficient for $\Big\langle\frac{\alpha_s}{\pi}G^2\Big\rangle$ 
was taken into account up to order $\alpha_s$ \cite{ChetyrkinGorishnySpiridonov1985}. 
It can be read off from
\begin{equation}
T^{\rm V+A}_{\rm np}(Q^2) 
\, = \,
\frac{1}{6\,Q^4}\,
\bigg(\,1-\frac{11}{18}\frac{\alpha_s}{\pi}\,\bigg)\,
\Big\langle\frac{\alpha_s}{\pi}G^2\Big\rangle
\, + \, \ldots
\,.
\label{VpAG2}
\end{equation}  
With the same values for the local condensates as in Sec.\,\ref{sec:CSRRGC} 
the evaluation of the dimension-six contribution in the OPE and the 
dimension-six contribution of the local 
expansion of the running gluon condensate yields a comparable 
magnitude and equal signs which gave us a phenomenological justification for 
the extraction of the running gluon condensate in the $V+A$ channel. 
Fig.\,\ref{Fig-RGCVA} shows the result. 
\begin{figure}[tb]
\begin{center}
\hspace{-4.5cm}
\begin{minipage}[t]{8 cm}
\epsfig{file=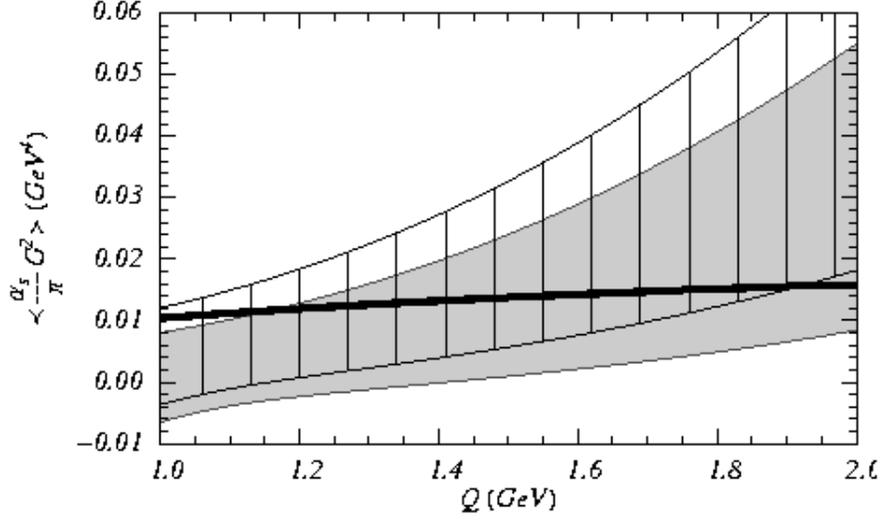,scale=0.7}
\end{minipage}
\begin{minipage}[t]{16.5 cm}
\caption{The running gluon condensate as a function of $Q$ when extracted from
the Adler function.
The grey area represents the allowed region using perturbation
theory at ${\cal O}(\alpha_s^3)$ and the striped region using
perturbation theory at  ${\cal O}(\alpha_s^2)$.
The thick solid line denotes the running gluon condensate as obtained from the
lattice-fitted ansatz in Eqs.\,(\ref{gfcfit}) for $\Omega=Q$. 
Plot taken from \cite{HoangHofmann2003}.
\label{Fig-RGCVA}}
\end{minipage}
\end{center}
\end{figure}
The uncertainties are due to 
the experimental errors in the spectral function and a variation of the
renormalization scale $\mu$ in the range $Q\pm 0.25$\,GeV. The analysis is 
restricted to the range $1\,\mbox{GeV}\le Q\le 2\,\mbox{GeV}$ because 
for $Q<1$\,GeV perturbation theory becomes unreliable and for $Q>2$\,GeV the 
experimentally unknown part of the spectral function at $s\ge 2.2$\,GeV$^2$ 
is being probed. The thick black line in Fig.\,\ref{Fig-RGCVA} shows the
lattice-inspired model for the running gluon condensate of Eq.\,(\ref{runningG2lattice}) 
for $\Omega=Q$. It is consistent with the phenomenological extraction and an 
increasing function of $Q$. Since the $Q$ dependence of the 
model is rather weak for $1\,\mbox{GeV}\le Q\le 2\,\mbox{GeV}$ and the error 
band of the extraction rather large it is not possible to draw a more quantitative conclusion 
at present.  

\subsubsection{Euclidean position-space $V\pm A$ correlators 
at short distance\label{sec:EPCO}}

In Sec.\,\ref{sec:DOE} a delocalized version of the OPE was obtained by projections 
of a perturbatively calculable short-distance function $f(x)$ and a 
nonperturbative, long-distance function $g(x^2)$ 
on a resolution dependently ``rotated'' basis in dual space, 
see Eq.\,(\ref{multDOEddim}). Thereby, the 
function $g(x^2)$ is assumed to be determined at infinite resolution. 
However, the only so-far available first-principle approach to $g(x^2)$ is a lattice calculation 
which can only be performed at a finite 
resolution, $\Omega\sim a^{-1}$. 

An alternative approach to running condensates than the one in Sec.\,\ref{sec:DOE} is to 
view the OPE as the usual local expansion but now involving averages over 
local composite operators at a resolution $\Omega\sim Q$. These operator averages are, 
besides their usual perturbative evolution, nonperturbatively 
coarse-grained in a gauge invariant way \cite{Hofmann12001}. In the case of a composite with two fundamental
operators like the quark or the gluon condensate nonperturbative coarse graining is performed 
by integrating out length scales between 
$\Omega^{-1}$ and $(\Omega-d\Omega)^{-1}$, which are 
associated with a loss in resolution of $\Omega^2/d\Omega$, 
in an $\Omega$ {\sl dependent}, gauge invariant and nonperturbative 
correlation function $g(x,\Omega)$, using a sharply cut off spherical well 
\eqb
\label{sphwell}
\frac{2\,(d\Omega)^{4}}{\pi^2}\,\theta\left(1/d\Omega-|x|\right)\,
\eqe
as a weight function. If we assume self-similarity of $g(x,\Omega)$, 
that is, an exponential form 
\eqb
\label{sesim}
g(x,\Omega)=A(\Omega)\exp[-|x|/\lambda_g]\,,
\eqe
where only the coefficient\footnote{Interpreting $\lambda_g^{-1}$ as 
the mass of the lowest intermediate hadronic state reached 
by the gauge invariant correlation, see for example \cite{DoschEidemullerJamin1999}, the 
correlation length $\lambda_g$ is viewed as an observable 
which does not depend on the resolution.} $A(\Omega)$ depends on $\Omega$, 
then the following evolution equation for $A(\Omega)$ is easily 
derived \cite{Hofmann12001}
\eqb
\label{evequ}
\frac{\pd}{\pd\Omega}A(\Omega)=\frac{4}{5\lambda}\Omega^{-2}A(\Omega)\,.
\eqe
The solution of Eq.\,(\ref{evequ}) corresponding to the 
running gluon condensate would then take the form
\eqb
\label{run}
\langle\frac{\alpha_s}{\pi}G^2\Big\rangle(\Omega)=\frac{6A_g}{\pi^2}
\exp\left[-\frac{4}{5\lambda_g}\left(\frac{1}{\Omega}-\frac{1}{\Omega_{lat}}\right)\right]\ .
\eqe
when the lattice fit of Eq.\,(\ref{gfcfit}) is used as a model with 
$\Omega_{lat}=a^{-1}=2\,$GeV. 

Following \cite{Hofmann32001} we now 
use dependences like Eq.\,(\ref{sphwell}) to 
investigate the effect 
of running condensates on the $|x|$ dependence 
of the Euclidean $V\pm A$ position-space correlators when they are expanded into a DOE 
in the chiral limit \cite{DeGrand2001}
\eab
\label{OPEps}
R_{V-A}(|x|\sim 1/\Omega)&\equiv&\frac{T^{V}(|x|)-T^{A}(|x|)}{2T_0(|x|)}=
\frac{\pi^3}{9}\,\alpha_s(\Omega)\la\bar{q}q\ra_{\Omega}^2\,\log[(|x|\Omega)^2]\,|x|^6\, ;\nonumber\\ 
R_{V+A}(|x|\sim 1/\Omega)&\equiv&\frac{T^{V}(|x|)+T^{A}(|x|)}{2T_0(|x|)}=
1-\frac{\pi^2}{96}\la\frac{\alpha_s}{\pi}(F_{\mu\nu}^a)^2\ra_{\Omega}|x|^4
-\nonumber\\ 
& &\frac{2\pi^3}{81}\,\alpha_s(\Omega)\la\bar{q}q\ra_{\Omega}^2\,\log[(|x|\Omega)^2]\,|x|^6\,
\eae
in the $V\pm A$ channel. There is no perturbative and gluon-condensate contribution 
in $R_{V-A}$, the correlator is extremely sensitive to chiral symmetry breaking. 
For the treatment of the four-point quark correlators at $D=6$ 
we assume a factorization into scalar two-point quark correlators. In analogy to the case of the 
field strength correlator a lattice measurement of the scalar two-point quark correlator 
\cite{D'EliaDiGiacomoMeggiolaro1999} with $N_F=4$ 
staggered fermions and a quark mass $am=0.01$ 
at $a^{-1}\sim2\,$GeV yields
\eqb
\label{para}
\lambda_q=3.1\, \mbox{GeV}^{-1}\ ,\ \ \ 
\ A_q(a^{-1}=\Omega_{lat}\sim 2\,\mbox{GeV})=(0.212\,\mbox{GeV})^3\,. 
\eqe
Substituting $\Omega=1/|x|$ in Eq.\,(\ref{OPEps}) and working with the parameters of Eqs.\,
(\ref{para}), (\ref{gmodel}), and a fixed value\footnote{The running of $\alpha_s$ almost cancels the 
log-powers at $D=6$ in Eq.\,(\ref{OPEps}).} 
of $\alpha_s(\Omega_{lat})=0.2$ \cite{Bode2001}, an $|x|$ dependence as depicted in Fig.\,
\ref{Fig-EV-+A} is obtained.
\begin{figure}[tb]
\begin{center}
\hspace{-6.cm}
\begin{minipage}[t]{8 cm}
\epsfig{file=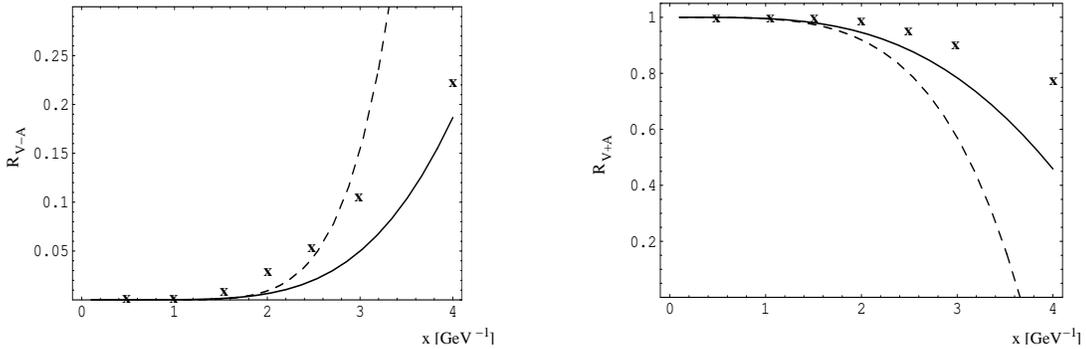,scale=0.65}
\end{minipage}
\begin{minipage}[t]{16.5 cm}
\caption{$R_{V\pm A}$ as a function of distance $x$ in the chiral limit. 
The solid line corresponds to the DOE with an assumed factorization of four-point 
quark correlators at $D=6$ into two-point, scalar quark correlators, 
the dashed line to the OPE. Crosses depict the 
result of the instanton liquid 
calculation of \cite{SchaferShuryak2001} which is taken from \cite{DeGrand2001}. 
Plot taken from \cite{Hofmann32001}.
\label{Fig-EV-+A}}
\end{minipage}
\end{center}
\end{figure}
In addition to a much better agreement (in comparison to the OPE) 
with the result obtained in the instanton liquid \cite{SchaferShuryak2001} the result of the DOE 
calculation almost perfectly agrees with the result of a quenched 
lattice calculation obtained using an overlap action \cite{DeGrand2001}. 
The result of an extraction of $R_{V\pm A}(|x|)$ 
from the $\tau$-decay data, see for example \cite{SchaferShuryak2001}, slightly overshoots the DOE and lattice 
results which were obtained in the 
chiral limit see Fig.\,\ref{Fig-Aleph-Inst}. 
\begin{figure}[tb]
\begin{center}
\hspace{1.cm}
\begin{minipage}[t]{8.5 cm}
\epsfig{file=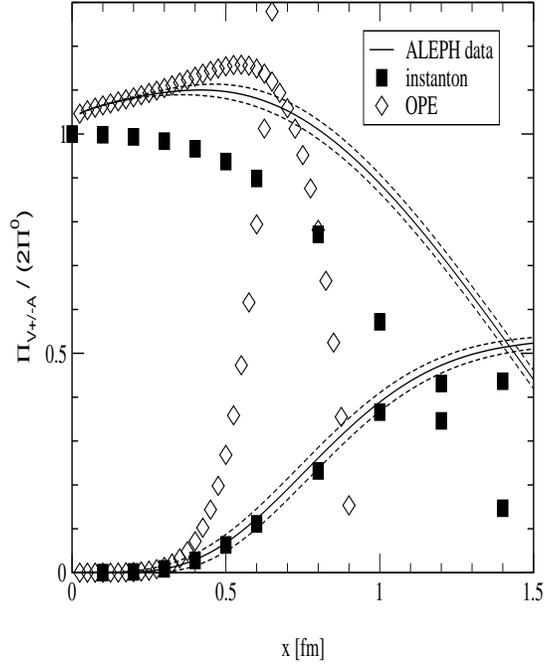,scale=0.55}
\end{minipage}
\begin{minipage}[t]{10.5 cm}
\caption{$R_{V\pm A}$ as a function of distance $x$. 
The dashed lines indicate the errors in the experimental determination of the respective 
spectral functions by Aleph. The squares show the results obtained in 
the model of a random instanton liquid. The diamonds are the OPE prediction using 
realistic light-quark masses. Plot taken from \cite{SchaferShuryak2001}.
\label{Fig-Aleph-Inst}}
\end{minipage}
\end{center}
\end{figure}

\subsubsection{Mesonic spectra from the DOE?}

The good agreement that we have found in Sec.\,\ref{sec:EPCO} 
between Euclidean $V\pm A$ correlators when 
expanded into an OPE with running condensates and when computed on a lattice lead us 
to investigate to what extent one can predict properties of 
the spectral functions of light-quark channels within the 
resonance region from the (practical) OPE with running condensates by analytical 
continuation to time-like momenta, $Q^2=-(s+i\ep)$ or $Q=-i\sqrt{s}$, $(s>0)$, 
and by taking the imaginary part afterwards \cite{Hofmann22001}. For a contribution 
of the form 
\eqb
\label{pc}
\frac{A_4(\Omega_{lat})}{(Q^2)^2}\,\exp\left[-\frac{4}{5\lambda_4}\left(\frac{1}{Q}-
\frac{1}{\Omega_{lat}}\right)\right]\,
\eqe
the associated contribution to the spectral function reads
\eqb
\label{sf}
\frac{A_4(\Omega_{lat})}{s^2}\exp\left[\frac{4}{5\lambda_4}\frac{1}{\Omega_{lat}}\right]
\sin\left[-\frac{4}{5\lambda_4 \sqrt{s}}\right]
\eqe
Notice that oscillations only start if $\sqrt{s}$ becomes smaller than the inverse 
correlation length $\lambda_4^{-1}$. As in Sec.\,\ref{sec:EPCO} 
a factorization of four-quark correlators into 
scalar two-quark correlators within the usual vacuum saturation hypothesis of the local case 
is assumed. All running condensates thus have the form as in Eq.\,(\ref{sf}), 
the effective correlation length for running 
condensates of $D=6$ is $\lambda_q/2$.
\begin{figure}[tb]
\begin{center}
\hspace{-6.cm}
\begin{minipage}[t]{8 cm}
\epsfig{file=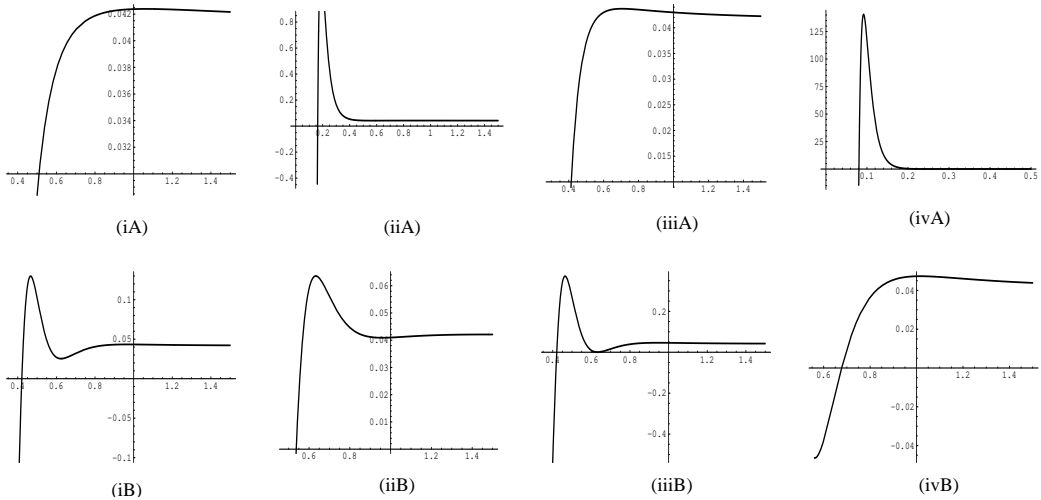,scale=0.4}
\end{minipage}
\begin{minipage}[t]{16.5 cm}
\caption{The spectral functions $\mbox{Im}\Pi(Q^2=-s-i\ep)$, $(s>0)$ 
for sets $(A)$ and $(B)$, where $(i)$, $(ii)$, $(iii)$, and $(iv)$ correspond 
to the $\rho$, $a_1$, $\pi$, and $\phi$ channels, respectively. 
The unit of $s$ is (GeV)$^2$. Taken from \cite{Hofmann22001}.
\label{Fig-SFDOE}}
\end{minipage}
\end{center}
\end{figure}
The $\rho$, $a_1$, $\pi$, and $\phi$ meson channels where 
investigated in \cite{Hofmann22001}, and the following two 
sets of parameters were used:
\eab
\label{paraA}
\underline{\mbox{Set}\, A:}& & \nonumber\\ 
\lambda_q&=&3.1\, \mbox{GeV}^{-1}\ ,\ \ \ \ A_q(a^{-1}=\Omega_{lat}\sim 2\,\mbox{GeV})=(0.212\, \mbox{GeV})^3\,,\nonumber\\ 
\lambda_g&=&1.7\, \mbox{GeV}^{-1}\ ,\ \ \ \ A_g(a^{-1}=\Omega_{lat}\sim 2\,\mbox{GeV})=0.015\, (\mbox{GeV})^4\,\nonumber\\ 
\underline{\mbox{Set}\, B:}& & \nonumber\\ 
\lambda_q&=&0.3\, \mbox{GeV}^{-1}\ ,\ \ \ \ A_q(a^{-1}=\Omega_{lat}\sim 2\,\mbox{GeV})=(0.212\, \mbox{GeV})^3\,,\nonumber\\ 
\lambda_g&=&1.7\, \mbox{GeV}^{-1}\ ,\ \ \ \ A_g(a^{-1}=\Omega_{lat}\sim 2\,\mbox{GeV})=0.015\, (\mbox{GeV})^4\,.
\eae
The quark correlation length $\lambda_q$ in Set A corresponds to 
that of the scalar quark correlator, the one 
in Set B to the longitudinal vector quark correlator as 
obtained in \cite{D'EliaDiGiacomoMeggiolaro1999}. Judging 
from Fig.\,\ref{Fig-SFDOE}, the result obtained with the 
small fermionic correlation length in Set B is more realistic but still far off 
the experimentally observed behavior. The most disturbing feature of the spectra in 
Fig.\,\ref{Fig-SFDOE} is the fact that they become negative at 
low values of $s$. For a discussion of the moments of 
the spectra in Fig.\,\ref{Fig-SFDOE} see \cite{Hofmann22001}. 
This is unacceptable and certainly has to do with the incompleteness 
and hence poor practical convergence of the expansion at low momenta.

\subsubsection{Speculations on convergence 
properties of expansions with nonperturbatively 
running operators} 

How does the partial resummation of operators involving powers 
of covariant derivatives, which leads to the nonperturbative 
running of the condensates in the usual OPE, possibly affect the ``convergence'' properties 
of the modified operator expansion? Let us give some (admittedly speculative) 
arguments.   

First of all, one notices the difference between the exponential(-like) dependences of 
Eq.\,(\ref{run}) (or Eq.\,(\ref{runningG2lattice})) with the situation in the instanton model 
in Sec.\,\ref{sec:QPIB} where exponentially 
small terms of the form $\exp[\frac{Q}{\Lambda_{QCD}}]$ 
were observed. As it was argued in \cite{ChibisovDikemanShifman1996} the latter exponentials 
may occur in the exact result to cure the ambiguities of the $1/Q^2$ 
expansion which may arise due to the factorially-in-$D$ rising coefficients where $D$ 
roughly refers to the power in $1/Q$. A possibility is that this 
factorial divergence of coefficients would already be present in a modification $\overline{\mbox{OPE}}$ of the 
OPE where operators containing powers of covariant derivatives are 
omitted\footnote{We do not want to apply the equations of motion and Bianchi identities which 
reduce these to operators without powers of covariant derivatives.} 
-- the factorially-in-$D$ rising coefficients would then simply arise 
from the combinatorial variety of composites involving powers of 
quark and gluon field operators. So already the $\overline{\mbox{OPE}}$ and, more generally, 
each subseries of the OPE with a given, fixed power of covariant
derivatives would be an asymptotic expansion which could be made unambiguous by adding 
exponentially small terms of the form $\exp[\frac{Q}{\Lambda_{QCD}}]$.

Second, in the OPE we may resum a part of the 
powers-of-covariant-derivative series associated with each 
operator in $\overline{\mbox{OPE}}$ by applying our methods of Secs.\,\ref{sec:DOE} or 
\ref{sec:EPCO}. For an operator of dimension $D$ the associated contribution after partial 
resummation is roughly of the form 
\eqb
\label{pca}
\frac{A_n(\Omega_{lat})}{Q^n}\,\exp\left[-\lambda^{-1}_n 
\left(\frac{1}{Q}-\frac{1}{\Omega_{lat}}\right)\right]\,
\eqe
where $\lambda_n$ denotes an effective correlation length. The expression in Eq.(\ref{pca}) 
has a maximum at $Q=(n\,\lambda_n)^{-1}$ with value 
\eqb
P_n\equiv\left(\Lambda_n\,\lambda_n\,\exp[-1]\,n\right)^n \,
\exp\left[\frac{\lambda^{-1}_n}{\Omega_{lat}}\right]\,
\eqe
where $\Lambda_n\equiv (A_n(\Omega_{lat}))^{(1/n)}$. 
As can be motivated from the example of factorizing the four-quark 
correlations into two-quark correlations in Sec.\,\ref{sec:EPCO} it is 
likely that $\lambda_n$ is a {\sl decreasing} function of $n$. If we assume a 
fall-off as $\lambda_n=\lambda/n^{(1-\ep)}$, $(1>\ep>0)$ then the position of 
the maximum of the expression in Eq.(\ref{pca}) decreases as 
$Q_n=\lambda^{-1} n^{-\ep}$. 
The value of the maximum in this case is
\eqb
P_n=\left(\Lambda_n\,\lambda\, n^\ep\,
\exp[-1+\frac{n^{-\ep}}{\lambda \Omega_{lat}}]\right)^n\,. 
\eqe
Assuming that $\Lambda_n$ does not depend on $n$, 
$\Lambda_n\equiv \Lambda_{QCD}$, as suggested by the phenomenology of 
practical OPEs, the critical mass dimension $n_c$ from which on maxima explode 
is given in implicit form as
\eqb
\Lambda_{QCD}\,\lambda\, n_c^\ep\,\exp[-1+\frac{n_c^{-\ep}}{\lambda \Omega_{lat}}]=1\,.
\eqe
A truncation of the OPE with running condensates at some $n<n_c$ would mean that physics 
below the maximum $Q_n$ can not be described, a truncation at $n\ge n_c$ does not make 
sense since due to the rapid increase of maxima at $n\ge n_c$ 
the expansion does not approximate anymore. Provided our above assumptions are met, 
another source of asymptotic behavior is identified. 
Let us give some numerical examples. For $\ep=0.5$ and $\Lambda_{QCD}=0.4$ GeV (the rounded) value of $n_c$ was 
calculated as a function of $\lambda$ in \cite{Hofmann22001} as
\eab
\lambda&=&3\,\mbox{GeV}^{-1}\ \ \rightarrow\ \ n_c=4\ ;\ \ \ \ \ \ \,
\lambda=2\,\mbox{GeV}^{-1}\ \ \ \ \ \rightarrow\ \ \ n_c=10\ ;\nonumber\\ 
\lambda&=&1\,\mbox{GeV}^{-1}\ \ \rightarrow\ \ n_c=40\ ;\ \ \ \ \ 
\lambda=0.5\,\mbox{GeV}^{-1}\ \ \ \rightarrow\ \ \ n_c=158\,. 
\eae
In conclusion, we have argued that exponential-like behavior of the form 
\eqb
\label{expbeha}
\exp[\frac{Q}{\Lambda_{QCD}}]\ \ \ \ \mbox{and}\ \ \ \ \exp\left[-\lambda^{-1}_n 
\left(\frac{1}{Q}-\frac{1}{\Omega_{lat}}\right)\right]
\eqe
may coexist in an improved version of the OPE. The former may assure 
the uniqueness of the asymptotic $1/Q$ expansion involving operators 
with a given, fixed number of covariant derivatives \cite{ChibisovDikemanShifman1996} while the latter arises 
from a partial summation of the expansion in powers of covariant 
derivatives acting on a given, fixed composition of powers of 
quark and gluon operators.

\section{Summary}

In this article we have presented a review of theoretical 
approaches to the indentification of the mechanisms leading to the violation of 
local quark-hadron duality in QCD. 
We have started our discussion with a mini-review on QCD sum rules in (axial)vector-meson 
channels in vacuum, at finite temperature, and at finite baryon density. The status of
 renormalon singularities as possible 
sources for power corrections in the OPE was very briefly addressed. We then have 
reviewed two theoretical model approaches to the calculation of current correlators: 
the instanton-gas model and the 
't Hooft model. The largest part of the review was devoted to a discussion of 
phenomenlogical evidence for the violation of quark-hadron duality. 
We have gathered indications that nonperturbatively nonlocal effects 
are insufficiently incorporated in a truncated local expansion if 
nonlocal quantities like parton distribution amplitudes, meson 
form factors, and hadron transition amplitudes are to be predicted. Phenomenologically motivated 
solutions to this shortcoming were discussed. Possible implications for OPE-based predictions 
of hadron spectra were speculated upon.    
 
\section*{Acknowledgments}
I would like to thank A. Dutt-Majumdar, A. Faessler, T. Gutsche, A. Hoang, E. Marco, M. Pospelov, and 
W. Weise for their collaboration on topics discussed in this review. I have, in particular, much benefitted 
from intense collaborations with A. Hoang and M. Pospelov. I am indebted to V. I. Zakharov for 
many stimulating conversations and continuing encouragement during my work on 
nonperturbative nonlocality at MPI Munich. Many thanks go to H. Gies, A. Hoang, M. Jamin, and V. I. Zakharov 
for their very useful comments on the manuscript.

\end{document}